\documentclass[
aps,
prd,
reprint,
showpacs,
superscriptaddress,
longbibliography,
floatfix,
nobalancelastpage,
nofootinbib
]{revtex4-2}

\usepackage{amsmath} 
\usepackage{mathtools} 
\usepackage{comment}
\usepackage{amssymb}
\usepackage{multirow}
\usepackage{graphicx}  
\usepackage{dcolumn}   

\usepackage{bm}        
\usepackage{booktabs}
\usepackage{tabularx}
\usepackage{siunitx}

\usepackage{flushend}

\usepackage{enumitem}
\usepackage[super]{nth}
\usepackage[normalem]{ulem}
\usepackage{wasysym}
\usepackage{microtype}
\usepackage{makecell}

\usepackage[dvipsnames, svgnames, table]{xcolor}

\usepackage{autofigs12}

\usepackage{braket}
\usepackage{color} 

\usepackage[colorlinks=true, citecolor=RoyalBlue,
linkcolor=BrickRed, urlcolor=ForestGreen]{hyperref}

\usepackage[capitalise]{cleveref}

\usepackage{subfiles}

\usepackage[english]{babel}
\makeatletter
\@namedef{l@en}{\@nameuse{l@english}}
\@namedef{captionsen}{\@nameuse{captionsenglish}}
\@namedef{dateen}{\@nameuse{dateenglish}}
\makeatother

\usepackage[bottom]{footmisc}

\DeclareMathAlphabet{\mathcal}{OMS}{cmsy}{m}{n}

\graphicspath{{figures/}{figures_new/}}
\newif\ifshowcomments

\ifshowcomments
    \newcommand{\DG}[1]{{\color{teal}{\texttt{[DG: #1]}}}}
    \newcommand{\SV}[1]{{\color{magenta}{{ #1}}}}
    \newcommand{\IM}[1]{{\color{blue}{\texttt{[IM: #1]}}}}
    \newcommand{\LM}[1]{{\color{purple}{\texttt{[LM: #1]}}}}
\else
    \newcommand{\DG}[1]{}
    \newcommand{\SV}[1]{}
    \newcommand{\IM}[1]{}
    \newcommand{\LM}[1]{}
\fi

\newcommand{\substrate}{\mathrm{s}}
\newcommand{\coating}{\mathrm{c}}
\newcommand{\longitudinal}{\mathrm{L}}
\newcommand{\shear}{\mathrm{S}}
\newcommand{\bulk}{\mathrm{B}}
\newcommand{\thermal}{\mathrm{T}}
\newcommand{\particular}{\mathrm{P}}
\newcommand{\homogeneous}{\mathrm{H}}

\newcommand{\red}[1]{\ensuremath{\overline{#1}}}

\newcommand{\qo}[1]{\ensuremath{\hat{#1}}}

\newcommand{\K}[1][ ]{\ensuremath{\kappa}}
\newcommand{\rK}[1][ ]{\ensuremath{\red{\kappa}}}

\newcommand{\HqM}[1][ ]{\ensuremath{\qo{D}}}
\newcommand{\HpM}[1][ ]{\ensuremath{D}}
\newcommand{\Hpm}[1][ ]{\ensuremath{\tilde{d}}}

\newcommand{\PqE}[1][ ]{\ensuremath{\qo{E}}}
\newcommand{\PpE}[1][ ]{\ensuremath{E}}
\newcommand{\Ppe}[1][ ]{\ensuremath{\tilde{e}}}

\newcommand{\qoa}[1][ ]{\ensuremath{\qo{\mathfrak{a}}}}

\newcommand{\crefBSMN}{Appendix~\hyperref[sec:BSMN]{B.1}}                       \newcommand{\crefBSTE}{Appendix~\hyperref[sec:BSTE]{B.2}}  
\newcommand{\crefSMNSemi}{Appendix~\hyperref[sec:SMNSemi]{A.3}}
\newcommand{\crefCTOAvg}{Appendix~\hyperref[sec:CTO_Averaging]{C.3}}

\DeclareUnicodeCharacter{2009}{\,}

\allowdisplaybreaks

\newif\ifdcc

\begin{document}

\title{High-Frequency Thermal Noise in Michelson Interferometers}
\author{Daniel Grass}
    \email{dgrass@caltech.edu}
\author{Sander~M. Vermeulen}
\author{Ian~A.~O. MacMillan}
\author{Lee McCuller}
\affiliation{Division of Physics, Mathematics and Astronomy, California Institute of Technology, Pasadena, CA 91125}

\date{\today}

\begin{abstract}
Thermal noise in optics constrains the precision of optical experiments, including Michelson interferometers, optical clocks, and optomechanical sensors. While high-power optical experiments such as gravitational-wave detectors are currently limited mainly by quantum noise, new experiments incorporating recent insights from quantum metrology are being developed that evade the quantum shot noise background. In particular, Michelson interferometers that use photon-counting readout will look for weak, high-frequency signals. Since shot noise is no longer the dominant noise source with these readout schemes, it is important to accurately model thermal noise to characterize signals and design more sensitive experiments. However, previous modeling uses approximations that are no longer valid in these frequency regimes. In the MHz band, the quasistatic approximation does not apply. We therefore develop more general models of substrate and coating mechanical (Brownian) noise, substrate and coating thermoelastic noise, and coating thermorefractive noise. We validate the models with comparisons to previous low-frequency modeling and high-frequency spectra from an experiment that has already taken data, the Holometer. We then apply the new models to GQuEST, an experiment under construction.

\end{abstract}
\maketitle

\section{Introduction}

Recent theoretical work~\cite{VerlindeJHEP20SpacetimeFluctuations,ZurekPLB22VacuumFluctuations,KwonFP25PhenomenologyHolography} proposes that a small but detectable signal of quantum gravity is measurable at high frequencies in Michelson interferometers. The proposed quantum gravity signals are weak and stochastic, making them difficult to distinguish from quantum shot noise. To overcome this limitation, new experiments are being developed that use alternative readout schemes that avoid shot noise; these schemes have similarities to spectroscopy, and may enable discovery of these signals~\cite{VermeulenPRX25PhotonCountingInterferometry,VermeulenCQG21ExperimentObserving}. In the absence of a quantum shot noise background, the dominant noise at the frequencies of interest is classical, specifically, thermal noise from the interferometers' optics.

However, prior modeling of thermal noises~\cite{BonduPLA98ThermalNoiseb,LevinPRD98InternalThermal, LiuPRD00ThermoelasticNoiseb,HarryCQG02ThermalNoise,HongPRD13BrownianThermala,GurkovskyPLA10ThermalNoise,BraginskyPLA99ThermodynamicalFluctuationsa,Braginsky03ThermodynamicalFluctuations,FejerPRD04ThermoelasticDissipation,EvansPRD08ThermoopticNoise,SomiyaPRD09CoatingThermal,BraginskyPLA00ThermorefractiveNoisea,LevinPLA08FluctuationDissipation} uses approximations that are no longer valid in the MHz frequency band of the quantum gravity detectors under consideration. One common simplification is the quasistatic approximation, which is the assumption that the measurement frequency band is below the frequency of the first mechanical mode of the mirror. This assumption is invalid for MHz-band interferometers, even with millimeter-thick mirrors. We develop more general models of substrate and coating mechanical (Brownian) noise, and substrate and coating thermoelastic noise. Another common assumption in previous work is that the beam penetration depth into the interferometer end mirrors is smaller than the thermal diffusion length. This assumption is also no longer valid at high frequencies. We thus develop a new model of coating thermorefractive noise. We validate these models with comparisons to prior low-frequency models as well as high-frequency spectra from data taken by the Holometer experiment. We then apply the new models to a new quantum gravity experiment, Gravity from the Quantum Entanglement of Space-Time (GQuEST)~\cite{VermeulenPRX25PhotonCountingInterferometry}. 

We analytically derive a theory for the spectral features from bulk-acoustic mechanical mirror modes in the Holometer and find agreement with experimental data. We determine that the dominant noise in that experiment is not coating mechanical noise, as prior quasistatic models would predict, but is instead substrate mechanical noise. The GQuEST design is informed by our models and optimized to have frequency bands with low substrate and beamsplitter mechanical noise. In these bands, the amplitude of beamsplitter mechanical noise is approximately equal to that of the coating mechanical noise, while substrate mechanical noise is slightly subdominant.

In \cref{sec:ModelingBackground}, we briefly review the noise modeling methods we will use in the following sections.

In \cref{sec:MechNoise}, we calculate mechanical (Brownian) noise of the end mirror substrates and coatings when the quasistatic approximation is no longer appropriate. Previous literature~\cite{BonduPLA98ThermalNoiseb,LevinPRD98InternalThermal, LiuPRD00ThermoelasticNoiseb, HarryCQG02ThermalNoise, SomiyaPRD09CoatingThermal, GurkovskyPLA10ThermalNoise,HongPRD13BrownianThermala} has assumed the readout frequency to be much smaller than the first mechanical eigenfrequency. This section derives noise models that do not require this assumption. A surprising result is that the coating mechanical noise away from the mechanical eigenfrequencies does not depend on the substrate's material properties, unlike the quasistatic result.

In \cref{sec:STENoise}, we use the results of \cref{sec:MechNoise} to calculate the substrate thermoelastic noise of the end mirrors. This extends the quasistatic work of~\cite{BraginskyPLA99ThermodynamicalFluctuationsa, LiuPRD00ThermoelasticNoiseb} to frequencies near and above the first mechanical eigenfrequency.

In \cref{sec:CTONoise}, we use the results of \cref{sec:MechNoise,sec:STENoise} to evaluate the noise from thermal fluctuations in the end mirror coatings. We again depart from the quasistatic limit of previous literature~\cite{Braginsky03ThermodynamicalFluctuations,FejerPRD04ThermoelasticDissipation,EvansPRD08ThermoopticNoise,SomiyaPRD09CoatingThermal}; this departure modifies the coating thermoelastic (CTE) noise model. Furthermore, we no longer take the thermal diffusion length in the coating to be larger than the beam penetration depth~\cite{BraginskyPLA00ThermorefractiveNoisea, LevinPLA08FluctuationDissipation,EvansPRD08ThermoopticNoise, SomiyaPRD09CoatingThermal}, which improves the coating thermorefractive (CTR) noise model.

\ifdcc
In \cref{sec:Comparison_with_Experimental_Data}, we apply these models and present a noise budget for GQuEST, an experiment under construction. 
\else
In \cref{sec:Comparison_with_Experimental_Data}, we evaluate the accuracy of these models (together with previously modeled sources of classical noise) on the data of a previous experiment, the Holometer. We then take these models and present a noise budget for GQuEST, an experiment under construction.
\fi

The appendices contain further details and extensions of the work. Specifically, we go into additional mathematical detail of the noise sources in \cref{app:SMN Details,app:CTO Details}, we explore noise in the beamsplitter in \cref{app:BSN}, we discuss implications for km-scale interferometers in \cref{app:Implications}, and we provide tables of experimental parameters and material constants in \cref{app:Tables}.

\section{Modeling Background}\label{sec:ModelingBackground}
All of the noise sources modeled in this paper are thermal noises, i.e., noise from internal degrees of freedom in the optics. Random material or temperature fluctuations inside the mirror change the phase of the light, sourcing a noise contribution at the interferometer's readout.

The modeling in this paper relies on the fluctuation-dissipation theorem (FDT) of Callen and Welton~\cite{CallenPR51IrreversibilityGeneralizeda}. This theorem states that a system that dissipates energy to the thermal bath must also experience fluctuations from that bath. More precisely, it states that the susceptibility of a degree of freedom (DOF) of a system to a drive acting on this DOF determines the coupling of fluctuations in the system back into that DOF. We take the mean spatial displacement $\delta z$ of the Gaussian laser beam wavefront as our DOF, as variations of $\delta z$ are measured by the output of an interferometer.
    
This DOF of interest is an effective optical path length, which can be modulated through various mechanisms, including material deformation, thermal expansion, and thermally induced variation in the material refractive index. By considering any potential drive of the DOF, we can model
the dissipation of the energy driven through the system and fully determine the coupling of thermal fluctuations to the DOF. This works even if the drive is ``fictitious" and can be formulated without relying on a microphysical mechanism.

For the systems under consideration, this dissipation is directly related to the admittance or susceptibility to the drive. 

In terms of the admittance, the FDT states
\begin{align}
    S_z(f) = \frac{k_{\mathrm{B}} T}{\pi^2 f^2} |\mathrm{Re}\{\mathcal{Y}(f)\}| = \frac{4k_{\mathrm{B}} T}{\omega^2} |\mathrm{Re}\{\mathcal{Y}(\omega)\}|,
    \label{eq: fdt}
\end{align}
where $S_z$ is the one-sided noise power spectral density (PSD) of the observable $\delta z$, $k_{\mathrm{B}}$ is the Boltzmann constant, $T$ is the temperature of the system, $f$ is the frequency of interest, $\omega$ is the angular frequency of interest ($\omega = 2 \pi f$), $\mathrm{Re}\{\}$ denotes the real part, and $\mathcal{Y}(f)$ is the admittance, i.e., the inverse of impedance. The angular frequency $\omega$ is often more natural in calculations than $f$ and will be predominantly used in this paper. The real part of the admittance is the inverse of the resistance. The admittance can be expressed as
\begin{align}
    \mathcal{Y}(f) = \frac{v(f)}{F(f)} = 2\pi i f \chi(f),
\end{align}
where $F(f)$ is the drive (analogous to force) on $\delta z$, $v(f)$ is the resulting response (analogous to velocity) of $\delta z$. $\chi(f)$ is the susceptibility, defined as
\begin{align}
    \chi(f) = \frac{\delta z(f)}{F(f)}.
\end{align}
The susceptibility, admittance, velocity, and displacement are all referred to the DOF $\delta z$. We thus express the FDT using $\chi$ as
\begin{align}\label{eq:FDTxi}
    S_z(f) = \frac{2k_{\mathrm{B}} T}{\pi f} \left|\mathrm{Im}\left\{\chi(f)\right\}\right| = \frac{4k_{\mathrm{B}} T}{\omega} \left|\mathrm{Im}\left\{\chi(\omega)\right\}\right|,
\end{align}
where $\mathrm{Im}\{\}$ denotes the imaginary part.

Below, we first apply the FDT to find the noise induced in $\delta z$ by mechanical (bulk-motional) modes of the mirror. The FDT states that the mirror's mechanical susceptibility to a pressure in the shape of a wavefront gives the coupling of thermal mechanical fluctuations to the displacement of an incident wavefront.
Though this pressure is exerted by a beam in a real system, the dynamics induced by the laser on the mirror are not relevant here; we merely use the pressure as a modeling tool to calculate the coupling of mirror fluctuations into the reflected light. We will refer to this pressure as a ``fictitious force" for consistency with previous work.

An alternative method of calculating mechanical noise has been used by Saulson and then Gillespie and Raab~\cite{SaulsonPRD90ThermalNoise, GillespiePRD95ThermallyExcited}; they decompose the total admittance as a sum over the admittances of mechanical eigenmodes, and then use the FDT to find the noise. While this method has proven very useful, including for calculations in~\cite{VermeulenPRX25PhotonCountingInterferometry}, at higher frequencies eigenmodes become harder to calculate with out-of-the-box tools such as finite element analysis (FEA), which is limited by mesh size and increasingly complicated spatial modal structure. For this reason, it is advantageous to use analytic methods when working at high frequencies to avoid calculating every eigenmode. Furthermore, the model of Gillespie and Raab assumes a single material and therefore a single loss angle. Due to these limitations of the eigenmode method, we proceed with the susceptibility approach laid out above.

Another important approach for calculating noise through the FDT comes from Levin~\cite{LevinPRD98InternalThermal}: the ``direct" method, so called as it calculates the fluctuation directly from the drive energy dissipation, rather than obtaining this quantity through finding the real part of the admittance of the drive with respect to the DOF. For the systems under consideration, we have
\begin{align}
    |\mathrm{Re}\{\mathcal{Y}(f)\}| = 2\pi f|\mathrm{Im}\{\chi(f)\}| = \frac{2 W_{\text{diss}}(f)}{F_0^2},
\end{align}
where $W_{\text{diss}}(f)$ is the power dissipated from the drive and $F_0$ is the amplitude of the drive force. The FDT can be written directly in terms of the dissipated power as 
\begin{align}\label{eq:LevinFDT}
    S_z(f) = \frac{2 k_{\mathrm{B}} T}{\pi^2 f^2} \frac{W_{\text{diss}}(f)}{F_0^2} = \frac{8 k_{\mathrm{B}} TW_{\text{diss}}(\omega)}{\omega^2F_0^2}.
\end{align}

We rely on the calculation of $W_{\text{diss}}(f)$  (Levin's method~\cite{LevinPRD98InternalThermal}) to calculate the mechanical dissipation in mirror coatings and thus their thermal fluctuations. This calculation relies on the substrate admittance computations, but the method is simpler than modifying the admittance to include coating effects.

\section{Mechanical Noise in Mirror Substrates and Coatings}\label{sec:MechNoise}
\subsection{Mechanical Noise in the Mirror Substrates}\label{subsec:MechNoiseSub}
Mechanical noise arises from elastic modes in the mirrors, changing the path length of the interferometer arms. This noise source has previously been called ``Brownian noise," but we avoid this term as there is no mass diffusion in this noise process. One can think of this noise source as vibrations of the optic. The dissipative term in the susceptibility is often parameterized as the ``loss angle" $\phi$, defined as
\begin{align}
    Y = Y_0 (1+i\phi),
\end{align}
where $Y$ is complex Young's modulus and $Y_0$ is Young's modulus in the absence of mechanical loss. Note that a loss angle exists for other moduli, and we use this notation for them as well. This (small) imaginary component will cause many variables, such as the various moduli, speeds of sound, or wavenumbers, in the following sections to have an imaginary component. As with moduli, a subscript of $0$ indicates the magnitude (i.e., the absolute value) of a variable. For the entirety of this paper, we drop terms of order $\phi^2$ or smaller as $\phi$ is small for the materials considered here, around $10^{-5}$~\cite{GrasPRD17AudiobandCoating,PennPLA06FrequencySurface, RodriguezSR19DirectDetectiona,UchiyamaPLA99MechanicalQuality}. We assume an isotropic material and a single loss angle for ease of calculation\footnote{One can also conceptualize the loss in terms of the quality factor $Q$. When there is a single, constant-in-frequency loss angle, one can write $Q=1/\phi$. However, the quality factor is, in general, frequency- and mode-dependent, and we calculate spectra away from resonances, so we avoid $Q$ in this paper.}. Isotropic materials have two loss angles, and certain crystalline materials like AlGaAs have three~\cite{AbernathyPLA18BulkShear,PennJOSAB19MechanicalRingdown,LIGOP1800359v2Brownian}. The loss angle may have frequency dependence ($\phi(f)$); we drop notation for this explicit dependence for brevity.

First, we will evaluate the contribution of thermal noise from the substrate alone (which we will call substrate mechanical noise, or SMN). In~\cref{sec:CMN}, we will incorporate the coating, which makes a significant contribution to the total noise because of its large loss angle compared to the substrate's loss angle. We start by modeling the optical substrate geometry as an infinite slab (i.e., a disk or plate with infinite lateral extent) with a finite thickness $h$. The infinite slab is the simplest model that contains the salient Lamb-wave dynamics introduced by the finite mirror thickness; our model is a Rayleigh-Lamb model with an excitation due to the fictitious force at the front surface from the impinging Gaussian beam. We choose the infinite-slab geometry to model this noise because the mirror radius is much larger than the beam size, and often greater than or equal to the thickness. We model the mirror as a cylinder, i.e., a mirror with finite radial extent in addition to finite thickness, in~\cref{subsubsec:SMN_cylinder}. Although real interferometer mirrors are not infinite slabs, the infinite slab model can actually be more appropriate than the cylinder model due to an inability to measure individual mechanical modes of the cylinder; see~\cref{sec:Holometer}.

\begin{figure}[t]
\includegraphics[width=1\linewidth]{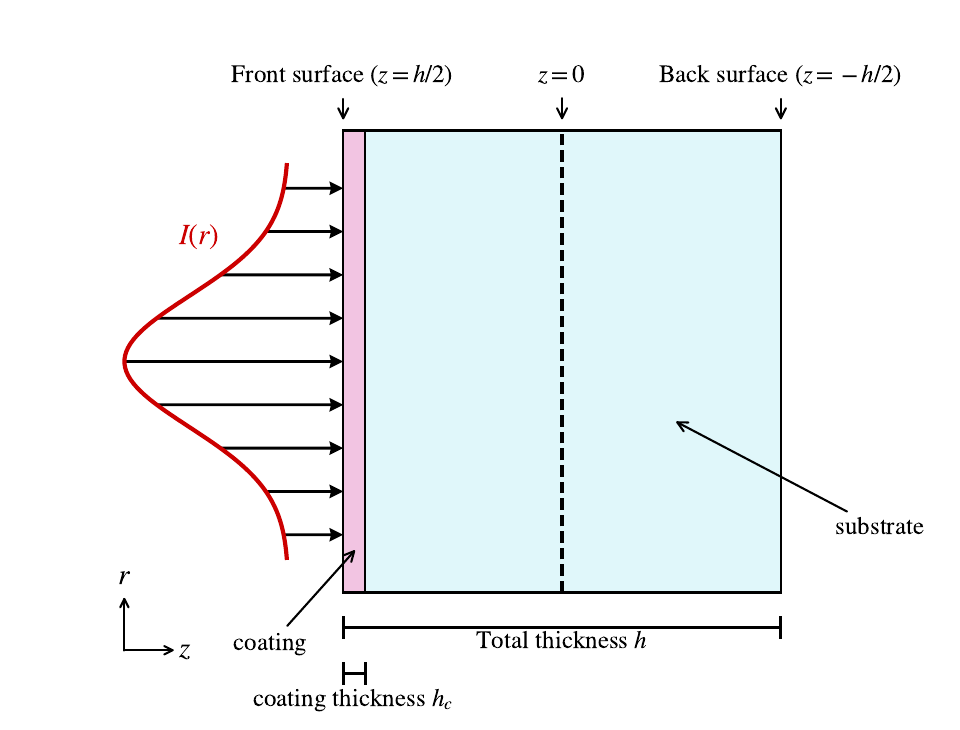}
  \caption{The cross-section of an example mirror made up of a thin High-Reflective (HR) coating and a substrate, $h_\mathrm{c}\ll h$. The total thickness of the mirror is $h$ along the $z$-axis. In this case, $z=0$ occurs at the center of the mirror, meaning that the front face is at $z=h/2$ and the back surface is at $z=-h/2$. The Gaussian beam weight factor, given in \cref{eq:GaussianBeamWeightFactor}, is depicted on the left incident on the mirror.}
  \label{fig:Mirror1}
\end{figure}

To apply the FDT, we seek to calculate the susceptibility to a mechanical drive acting on $\delta z$. We therefore consider a pressure drive with a Gaussian spatial profile $I(r)$ with peak amplitude $F_0$, 
\begin{align}\label{eq:periodic_drive}
    F(r,t) = F_0I(r)e^{i\omega t},
\end{align}
where $\omega$ is the angular frequency of interest and $t$ is time. $I(r)$ is the unit-normalized Gaussian beam weight factor; it is depicted on the mirror in \cref{fig:Mirror1},
\begin{align}\label{eq:GaussianBeamWeightFactor}
    I(r) = \frac{2}{\pi w^2}e^{-2r^2/w^2},
\end{align}
where $r$ is the radial coordinate and $w$ is the Gaussian beam spot radius ($1/e^2$ in intensity). The mean wavefront displacement $\delta z(t)$, due to the mechanical displacement of the mirror surface $u_z(z=h/2,r,t)$, is
\begin{align}
    \delta z(t) = \int u_z(z=h/2,r,t) I(r)\,d^2r,
\end{align}
where the integral is over the front surface of the mirror, located at $z = h/2$ (the mirror thickness is $h$), and $u_z(z, r,t)$ is the $z$ (beam propagation axis) displacement of the mirror resulting from the force $F(r,t)$.
Note that the Gaussian profile appears both in the driving force and in our definition of the DOF of interest, as required by the FDT. In future work, we will examine the generation of higher-order spatial modes and will use a different form of $I(r)$.

We first transform $\delta z$ to the spatial frequency and temporal frequency domains. We calculate in the spatial frequency domain because the problem has spatial symmetry, and in the temporal frequency domain because we assume stationarity to compute a spectrum and use a periodic drive to model the noise. We first transform the time coordinate, writing
\begin{align}\label{eq: delta z r integral}
    \widehat{\delta z}(\omega) = \int_0^\infty \hat{u}_z(z=h/2,r, \omega) I(r)\,d^2r.
\end{align}
For now, we will solve the problem for an infinite slab in cylindrical coordinates ($z \in [-h/2, h/2], r \in [0,\infty)$). See \cref{fig:Mirror1}. We transform the integral into the spatial frequency domain using a Hankel transform\footnotemark{} and write,
\begin{align}
    \tilde{I}(k) = \frac{1}{2\pi}e^{-k^2w^2/8},
\end{align}
where $k$ is the transverse wavenumber. The DOF of interest becomes
\begin{align}
    \widehat{\delta z}(\omega) = 2\pi\int_0^\infty \tilde{u}_z(z=h/2,k,\omega) \tilde{I}(k) k \,dk.
\end{align}
We now proceed to find $\tilde{u}_z(z,k,\omega)$ induced by the drive, evaluate it at the mirror's front surface, and integrate over $k$ to find the susceptibility. We start with the elastic wave equation in $z$, $r$, and $t$ coordinates,
\begin{align}
    \rho \frac{\partial^2 \bm{u}}{\partial t^2} = (\lambda + 2\mu)\nabla (\nabla \cdot \bm{u}) - \mu \nabla \times (\nabla \times \bm{u}),
\end{align}
where $\rho$ is the substrate density, $\bm{u}$ is the displacement vector (that specifies the displacement of a point relative to its stationary position), and $\lambda$ and $\mu$ are the two Lam\'e parameters~\cite{Bedford94IntroductionElastic} (these are complex as well). $\mu$ is also called the shear modulus. The boundary conditions are given by the stress $\sigma$, with units of pressure, at the boundaries $z=\pm h/2$. Some additional key parameters from the derivation in \cref{app:SMN Details} are the speeds of sound in the substrate,
\begin{align}
    v_\longitudinal = \sqrt{\frac{\lambda + 2\mu}{\rho}} = \sqrt{\frac{M}{\rho}};~v_\shear = \sqrt{\frac{\mu}{\rho}},
\end{align}
where $v_\longitudinal$ and $v_\shear$ are the longitudinal and shear (transverse) speeds of sound, respectively, and $M = \lambda + 2\mu$ is the P-wave modulus. We also require the wavenumbers,
\begin{align}\label{eqs:wavenumbers}
    k_{z,\longitudinal} = \sqrt{k^2 - k_\longitudinal^2};~k_{z,\shear} = \sqrt{k^2 - k_\shear^2};~k_\longitudinal=\frac{\omega}{v_\longitudinal};~k_\shear=\frac{\omega}{v_\shear}.
\end{align}
Please note that these speeds of sound and wavenumbers are complex since the elastic moduli are complex. We leave the details for finding $\tilde{u}_z(z,k,\omega)$ via the elastic wave equation to the appendix, see \cref{app:SMN Details}. We write the displacement of the mirror surface as
\begin{align}
    \tilde{u}_z(z = h/2,k) &= F_0\tilde{I}(k) \xi_z(k),
\end{align}

where $\xi_z(k)$ is a complex function, the ``receptance density" (displacement over force per unit area of $k$-space), which is related to the admittance and susceptibility as
\begin{align}\label{eq:rcptnc_to_admtnc}
    \mathcal{Y}(\omega)
    = i\omega\,\chi(\omega)
    = i\omega\,2\pi \int_0^\infty \xi_z(k)\,\tilde I(k)^2\,k\,dk
\end{align}
We find the receptance density is
\begin{widetext}
\begin{multline}\label{eq:xi_z}
    \xi_z(k) = -\frac{k_{z,\longitudinal} k_\shear^2}{2\mu} \bigg(\frac{1}{(2k^2-k_\shear^2)^2\coth(k_{z,\longitudinal} h/2) - 4k^2k_{z,\longitudinal}k_{z,\shear}\coth(k_{z,\shear} h/2)} +\\
    \frac{1}{(2k^2-k_\shear^2)^2\tanh(k_{z,\longitudinal} h/2) - 4k^2k_{z,\longitudinal}k_{z,\shear}\tanh(k_{z,\shear} h/2)}\bigg).
\end{multline}
\end{widetext} 
Note that there is a nontrivial frequency dependence in $\xi_z(k)$ as $k_{\longitudinal}$ and $k_\shear$ depend on $\omega$. Finally, plugging our result into the FDT \cref{eq:FDTxi}, we can express the substrate mechanical noise on an infinite slab as
\begin{align}\label{eq:SMN_Spectrum}
    S_{z,\text{SMN,Slab}}(\omega) = \frac{2k_{\mathrm{B}} T}{\pi \omega}\int_0^\infty\left|\mathrm{Im}\left\{\xi_z(k)\right\}\right|ke^{-k^2w^2/4}\,dk.
\end{align}

\begin{figure*}[t]
\includegraphics[width=1\linewidth]{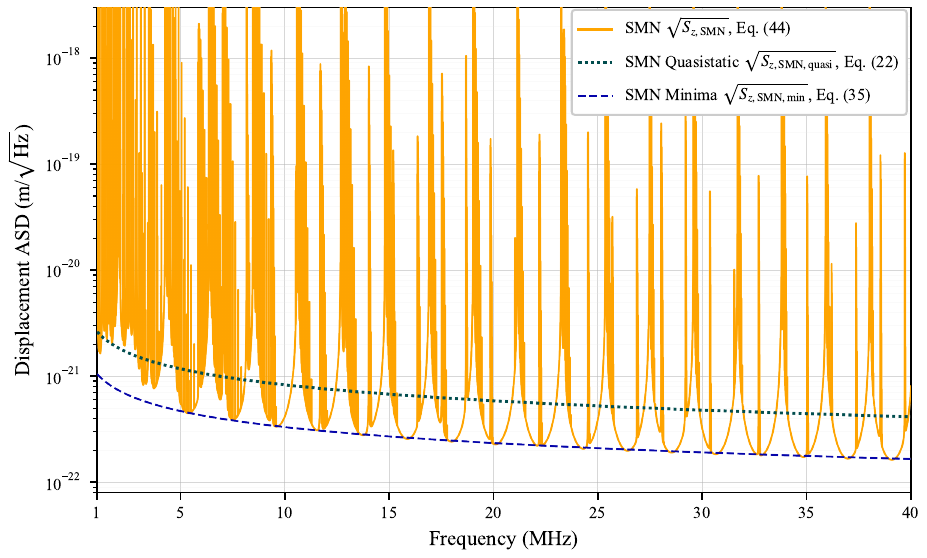}
  \caption{The Modeled Amplitude Spectral Density (ASD) of the Substrate Mechanical Noise (SMN) for the GQuEST End Mirrors. The orange-colored curve plots the substrate mechanical noise for a cylindrical end mirror (\cref{eq:SMN_Spectrum_Cyl}). We compare it to the quasistatic, infinite-half-space model (dotted line) from~\cite{BonduPLA98ThermalNoiseb} and the minima at high frequencies (\cref{eq:SMNMin}). Note this and the other ASD plots are the total noise for two end mirrors (or two beamsplitters for \cref{fig:BS_Noise}). The parameters used to generate this and subsequent plots can be found in~\cref{app:Tables}. A frequency sample width of $\Delta f_{\text{sample}} = 3.9$~Hz was used. Because this sample width is smaller than the $\sim 10$~Hz mode linewidth, the peaks are resolved; their heights on the graph are physical, and many have an amplitude density larger than $3\cdot10^{-18}~\text{m}/\sqrt{\text{Hz}}$ due to the small loss angle of the substrate.}
  \label{fig:SMN}
\end{figure*}

Physical and analytical features of this spectrum are explored in \cref{fig:SMN} and \cref{fig:SMNZoom}.

\footnotetext[2]{As this is an axisymmetric problem in two dimensions, we use the order-0 Hankel transform. This transformation goes from two coordinates, say $r$ and $\varphi$, to one, $k$.}

To compare to previous results, we take the quasistatic ($\omega \rightarrow 0$), infinite-$h$ limit. Previous authors have preferred to use Young's modulus, $Y$, and the Poisson ratio, $\nu$, so we express $\xi_z(k)$ in this limit as
\begin{align}\label{eq:xi_quasistatic}
    \xi_{z,\text{quasistatic}}(k) = \frac{2(1-\nu^2)}{Yk}.
\end{align}
The $k$-integral simplifies, and the spectrum becomes (under the assumption that there is a single loss angle)
\begin{align}
    S_{z,\text{SMN quasistatic}}(\omega) = \frac{4k_{\mathrm{B}} T (1-\nu_0^2) \phi}{\sqrt{\pi}Y_0 w \omega}.
\end{align}
This agrees completely with the quasistatic, infinite-half-space mirror calculation by~\cite{BonduPLA98ThermalNoiseb} (their Eq. (14)).\footnote{While this leading-order limit of $\xi_z(k)$ recovers the noise spectrum in that work, higher-order corrections in $h$ or $\omega$ lead to unphysical noise at low frequencies. The issue arises from allowing elastic waves to propagate into an infinite-half-space (previous literature did not include elastic waves). To solve this, we need a more physical model like in \cref{subsubsec:SMN_cylinder} or ultimately~\cite{HutchinsonJoAM80VibrationsSolid}.} This agreement is not shown in~\cref{fig:SMN} as the lower frequency bound of 1 MHz is above the lowest GQuEST end mirror mechanical eigenfrequency. The plot therefore does not show the quasistatic regime but includes the quasistatic regime's noise curve to highlight the difference between it and the high-frequency spectrum.

\subsubsection{Analytic Substrate Mechanical Noise Spectrum at High Frequencies}\label{subsubsec:SMN_Analytic}
The integral form of \cref{eq:SMN_Spectrum} provides an accurate noise spectrum for arbitrary frequencies. We have approximated this form in the quasistatic, infinite-$h$ limit, and found a closed-form expression (\cref{eq:xi_quasistatic}). Now we proceed to find a closed-form expression for the noise spectrum at high frequencies. Specifically, we want to calculate the noise spectrum for 
\begin{align}
    \frac{2}{w} \ll \frac{\omega}{|v_\longitudinal|} = |k_\longitudinal|
\end{align}
and
\begin{align}
    \frac{1}{h} \ll \frac{\omega}{|v_\longitudinal|} = |k_\longitudinal|.
\end{align}

For finding an approximate closed-form expression, we take the following ansatz for the imaginary part of the receptance density:

\begin{multline}\label{eq:xi_z_approx}
    \mathrm{Im}\left\{\xi_z(k)\right\} \approx \mathrm{Im}\left\{\xi^{(0)}_z\right\} +\\ \sum_{p=1}^\infty \mathrm{Re}\left\{\text{Res}\left\{\xi_z(k),k_p+i\Gamma_p\right\}\right\}\frac{\Gamma_p}{\Gamma_p^2+(k-k_p)^2},
\end{multline}
where $\xi^{(0)}_z$ is the zeroth-order coefficient of a Taylor expansion of $\xi_z(k)$ about $k = 0$,\footnote{The Taylor series does not converge to $\xi_z(k)$ for $k>k_1$, where $k_1$ is the pole with the lowest wavenumber. Therefore, the Taylor expansion should only be used in the regime $k \ll k_1$.} $k_p$ is the wavenumber corresponding to the $p$th Lorentzian peak of $\mathrm{Im}\left\{\xi_z(k)\right\}$, $\Gamma_p = \phi k_p/2$ is the Lorentzian half-width-at-half-maximum bandwidth, and $\text{Res}\left\{\xi_z(k),k_p+i\Gamma_p\right\}$ is the residue of $\xi_z(k)$ evaluated at the poles $k_p+i\Gamma_p$. This ansatz is motivated by a qualitative assessment of the numerically evaluated noise spectrum \cref{eq:SMN_Spectrum}, as we explain below.

Due to the Gaussian weighting factor in \cref{eq:SMN_Spectrum}, the natural scale for spatial frequencies $k$ that couple strongly is approximately $2/w$. We can therefore take $k\approx2/w$, which together with the high-frequency approximation $k_\longitudinal\gg2/w$ gives $k_\longitudinal\gg k$. Thus, in this regime, the $z$-wavenumbers $k_{z,\longitudinal}$ and $k_{z,\shear}$ are imaginary: specifically, $k_{z,\longitudinal} \approx i |k_\longitudinal| = i\omega/|v_\longitudinal|$ and $k_{z,\shear} \approx i |k_\shear| = i\omega/|v_\shear|$. The hyperbolic trigonometric functions in \cref{eq:xi_z} therefore have imaginary arguments, which makes them equal to circular trigonometric functions with real arguments times $i$. We can then see that the real part of the denominators in \cref{eq:xi_z} goes to zero periodically, specifically with frequency spacing that approaches $\Delta f = |v_\longitudinal|/(2h)$ at high frequencies. The periodic structure of the poles then leads to a decomposition of the PSD into near-resonance terms (where the denominator goes near zero) and off-resonance terms.

For frequencies such that the denominator is nearly zero, $|\text{Im}\left\{\xi_z(k)\right\}|$ therefore becomes large but not infinite due to the nonzero loss angle. Near the maximum, these peaks are Lorentzian functions of $k$ and can be the dominant contribution to the spectrum through the integral in \cref{eq:SMN_Spectrum}.

For high frequencies (such that the $z$-wavenumbers are greater than the cutoff scale $2/w$) and away from the periodic peak structure, the admittance is smooth and the integral is dominated by a low-order Taylor expansion of $\left|\mathrm{Im}\left\{\xi_z(k)\right\}\right|$. As the frequency gets larger, almost all of the frequency space becomes well-approximated by this short Taylor expansion. This term of the expansion captures off-resonant mechanical noise. Notably, the zeroth-order term in this Taylor expansion is linearly proportional to the loss angle, while the Lorentzian functions of $k$ that describe the peaks are not.

Starting from our ansatz~(\cref{eq:xi_z_approx}), we proceed to evaluate each term and assemble the approximation. Firstly, the zeroth-order coefficient $\xi^{(0)}_z$ captures the physics at and around the local minima (`valleys') between the evenly spaced longitudinal eigenmodes with spacing $\Delta \omega= \pi v_\longitudinal/h$. Specifically,
\begin{align}\label{eq:xi0_z}
    \xi^{(0)}_z=-\frac{\cot(k_\longitudinal h)}{\rho v_\longitudinal \omega}.
\end{align}

Second, to evaluate the last term in \cref{eq:xi_z_approx}, we introduce new symbols to shorten the expression for $\xi_z(k)$ to 
\begin{align}
    \xi_z(k) = -\frac{k_{z,\longitudinal} k_\shear^2}{2\mu}\left(\frac{1}{D_{1,\xi}(k)}+\frac{1}{D_{2,\xi}(k)}\right),
\end{align}
where $D_{1,\xi}(k)$ and $D_{2,\xi}(k)$ are the first and second denominators in \cref{eq:xi_z}, respectively. We find the peak wavenumbers $k_p$ by setting $D_{1,\xi}(k_{p,1}) = 0$ and $D_{2,\xi}(k_{p,2}) = 0$ and solving for $k$. To accurately estimate the noise spectrum, we do not need to find all zeros of these equations; due to the Gaussian $e^{-k_p^2w^2/4}$ in \cref{eq:SMN_Spectrum}, there is a cutoff wavenumber $2/w$ such that the contributions of peaks with center frequencies $k_p \gg 2/w$ are negligible. We therefore solve $D_{1,\xi}(k_{p,1}) = 0$ and $D_{2,\xi}(k_{p,2}) = 0$ to leading order ($k^2$) and drop higher-order contributions. We find
\begin{align}\label{eq:kp}
\resizebox{\columnwidth}{!}{$\displaystyle
k_{p,1} = \omega^2\sqrt{\left|\frac{2\cot(k_\longitudinal h/2)}{v_\shear^4D^{(2)}_{1,\xi}}\right|},~~
k_{p,2} = \omega^2\sqrt{\left|\frac{2\tan(k_\longitudinal h/2)}{v_\shear^4D^{(2)}_{2,\xi}}\right|},
$}
\end{align}
where $(')$ denotes a derivative with respect to $k$, and $D^{(2)}_{1,\xi}$ and $D^{(2)}_{2,\xi}$ are the second derivatives of the two denominators of $\xi_z(k)$. By Taylor expanding the denominators of $\xi_z(k)$ as quadratic functions, we easily find their zeros, but this is not necessary. Through algebraic manipulation, we find, to leading order in $k$,
\begin{multline}
    D'_{1,\xi}(k) \approx D^{(2)}_{1,\xi}k = i\Big(8k_\shear^2\cot(k_\longitudinal h/2) - \\\frac{hk_\shear^4}{2k_\longitudinal}\csc^2(k_\longitudinal h/2) - 8k_\longitudinal k_\shear \cot(k_\shear h/2)\Big)k,
\end{multline}
\begin{multline}
    D'_{2,\xi}(k) \approx D^{(2)}_{2,\xi}k = i\Big(-8k_\shear^2\tan(k_\longitudinal h/2) - \\\frac{hk_\shear^4}{2k_\longitudinal}\sec^2(k_\longitudinal h/2) + 8k_\longitudinal k_\shear \tan(k_\shear h/2)\Big)k.
\end{multline}

We now calculate the residue, $\text{Res}\left\{\xi_z(k),k_p+i\Gamma_p\right\}$, to be
\begin{align}
    \text{Res}\left\{\xi_z(k),k_{p,1}+i\Gamma_{p,1}\right\} &\approx -i\frac{k_{\longitudinal} k_\shear^2}{2\mu D'_{1,\xi}(k_{p,1})} \label{eq:xiRes1}, \\
    \text{Res}\left\{\xi_z(k),k_{p,2}+i\Gamma_{p,2}\right\} &\approx -i\frac{k_{\longitudinal} k_\shear^2}{2\mu D'_{2,\xi}(k_{p,2})} \label{eq:xiRes2}.
\end{align}

\begin{figure*}[t]
\includegraphics[width=1\linewidth]{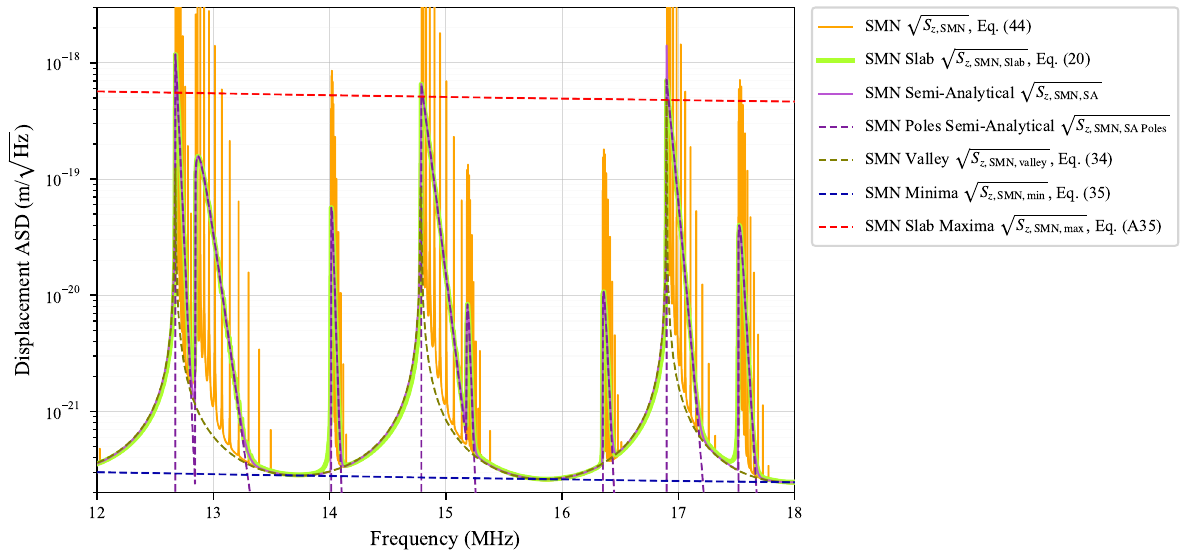}
  \caption{The Modeled Amplitude Spectral Density of the Substrate Mechanical Noise (SMN) for the GQuEST End Mirrors. We compare the cylinder (\cref{eq:SMN_Spectrum_Cyl}) to the infinite slab (\cref{eq:SMN_Spectrum}), just the valley term for the infinite slab (\cref{eq:SMNHFAValley}), just the poles term for the infinite slab (the more accurate version of the second term in \cref{eq:SMNHFA}, as given in~\crefSMNSemi), the sum of these (semi-)analytic parts, and the infinite slab minima and maxima (\cref{eq:SMNMin} and \cref{eq:SMNSlabMax}, respectively). Only 1,000 points were used to generate the curves except \cref{eq:SMN_Spectrum_Cyl}, which has $\Delta f_{\text{sample}} = 3.9$~Hz, the same as~\cref{fig:SMN}. If coarser frequency sampling were done and \cref{eq:SMN_Spectrum_Cyl} were averaged in frequency, the resulting curve would approach that generated by \cref{eq:SMN_Spectrum}. This demonstrates a physical limit where under-sampled power spectra (i.e., spectra with insufficient frequency resolution) that average multiple mirror resonances end up with the apparent spectral density of infinite mirrors. The infinite-mirror spectrum can be approximated analytically; these approximations can be used for computations and give insight into physical reasoning.}
  \label{fig:SMNZoom}
\end{figure*}

Finally, by plugging in the coefficient $\xi^{(0)}_z$ (\cref{eq:xi0_z}), the dominant peak wavenumbers (\cref{eq:kp}), and the residues $\text{Res}\left\{\xi_z(k),k_p+i\Gamma_p\right\}$ (\cref{eq:xiRes1,eq:xiRes2}) into \cref{eq:xi_z_approx}, and evaluating the integral in \cref{eq:SMN_Spectrum}, we obtain our closed-form expression for the noise spectrum in the high-frequency approximation (HFA):
\begin{widetext}
\begin{multline}\label{eq:SMNHFA}
    S_{z,\text{SMN, HFA}}(\omega) = \frac{4k_{\mathrm{B}} T}{\pi \rho |v_\longitudinal| w^2\omega^2}\left|\mathrm{Im}\left\{\cot(k_\longitudinal h)\right\}\right|+\\
    \frac{k_{\mathrm{B}} T}{\omega}\left(\mathrm{Re}\left\{-i\frac{k_{\longitudinal} k_\shear^2}{\mu D'_{1,\xi}(k_{p,1})}\right\}k_{p,1}e^{-k_{p,1}^2w^2/4}+\mathrm{Re}\left\{-i\frac{k_{\longitudinal} k_\shear^2}{\mu D'_{2,\xi}(k_{p,2})}\right\}k_{p,2}e^{-k_{p,2}^2w^2/4}\right).
\end{multline}
\end{widetext}
This approximation is plotted to compare with a finite mirror spectrum and the exact integral form in \cref{fig:SMNZoom}.

There are a few limitations to this approximate formula for the SMN spectrum. The first term is not valid for $k_p \ll 2/w$ as the Taylor series for $\text{Im}\{\xi_z(k)\}$ does not converge under this condition. In addition, the poles $k_p$ and the residues were found by solving $D_{1,\xi}(k_{p,1}) = 0$ and $D_{2,\xi}(k_{p,2}) = 0$ to leading order ($k^2$) and dropping higher-order contributions. This can cause this high-frequency approximation \cref{eq:SMNHFA} to not accurately resemble the spectrum from \cref{eq:SMN_Spectrum}. We leave an exact closed-form spectrum to future work. We note that the Taylor series term combined with solutions from a numerical root finding algorithm for the denominators of $\xi_z(k)$ and the analytic forms of the derivatives of the denominators of $\xi_z(k)$ reproduces \cref{eq:SMN_Spectrum} very well in less computational time than evaluating \cref{eq:SMN_Spectrum} directly, see \crefSMNSemi.

At frequencies between the mechanical resonances of the infinite slab, we can drop the second term in \cref{eq:SMNHFA} as it is only large for $k_p < 2/w$. Under this assumption, we can also simplify the first term and write
\begin{align}\label{eq:SMNHFAValley}
    S_{z,\text{SMN, HFA}}(\omega) = \frac{2k_{\mathrm{B}} T h\phi}{\pi M_0 w^2\omega\sin^2\left(\omega h/|v_\longitudinal|\right)}.
\end{align}

We can see that between the resonances, the first term in \cref{eq:SMNHFA} is the largest term at almost all frequencies and thus produces the periodic U-shapes in the spectrum. The last term in \cref{eq:SMNHFA} comes from the poles of $\xi_z(k)$ and dominates for frequencies at and just above the eigenfrequencies of the infinite slab (where $k_p \lesssim 2/w$).

Lastly, we are interested in the noise level at the local minima of the spectrum. As the local minima occur between resonances, we can evaluate this noise level using \cref{eq:SMNHFAValley}. Quantitatively, the noise is minimized for frequencies $\omega = \pi (m_\longitudinal + 1/2)|v_\longitudinal|/h, \, m_\longitudinal \in \mathbb{N}$, such that $\sin^2(\omega h/|v_\longitudinal|)=1\,$ for some $\, m_\longitudinal$; this is equivalent to approximating $\sin(\cdot)\approx1$ in \cref{eq:SMNHFAValley} at all frequencies. We get the resulting noise spectrum that interpolates between local minima
\begin{align}\label{eq:SMNMin}
    S_{z,\text{SMN min}}(\omega) = \frac{2k_{\mathrm{B}} T h \phi}{\pi M_0 w^2 \omega}.
\end{align}
This expression indicates that thin, stiff mirrors with a low loss angle and a large beam size can be used to minimize substrate mechanical noise in narrow frequency bands between resonances. However, note that \cref{eq:SMNMin} still assumes the radial (lateral) extent of the mirror is infinite.

\subsubsection{Cylindrical Mirror}\label{subsubsec:SMN_cylinder}
In the preceding section, we modeled the mirror as an infinite slab of finite thickness. However, interferometer mirrors are typically short cylinders, and this finite lateral scale introduces additional features and relevant frequency scales in the noise spectrum.

The mechanical noise calculation for a cylinder, with radius $a$, proceeds analogously to that for an infinite slab, i.e., we can still calculate $\widehat{\delta z}$ using an integral over $r$ (see \cref{eq: delta z r integral}, replacing $\infty$ with $a$). However, we rewrite $u_z(z,r,\omega)$ and $I(r)$ not using a continuous Hankel transformation but as a series of discrete modes that (almost) obey the boundary conditions at the `barrel' of the mirror ($r = a$, i.e., on the curved lateral surface of the cylinder). We follow previous literature~\cite{BonduPLA98ThermalNoiseb, LiuPRD00ThermoelasticNoiseb} and ignore the normal stress on the barrel of the cylinder, $\sigma_{rr}(r=a)$, for simplicity. This approximation is reasonable because $\sigma_{rr}(r=a)$ is small when the mirror is thin, and the beam size is small when compared to the radius.\footnote{Previous literature approximates $\sigma_{rr}$ as linear in $z$ to find an approximate correction term, $\Delta \textbf{u}$, that when summed to $\textbf{u}$ gives a solution that obeys all boundary conditions. Because the elastic wavelength is smaller than the mirror thickness, no such linear approximation would be valid, and more accurate correction terms become difficult to concoct. Somiya and Yamamoto~\cite{SomiyaPRD09CoatingThermal} note the correction is important in the regime $h\ll a$, but this is only below the first mechanical eigenfrequency, i.e., the quasistatic limit. \ifdcc We leave this as an approximation (since it only shifts the mode structure) and will provide an exact solution using~\cite{HutchinsonJoAM80VibrationsSolid} in the future. \else  We ultimately let the experimental data verify our choice to ignore $\sigma_{rr}$ and will provide an exact solution using~\cite{HutchinsonJoAM80VibrationsSolid} in the future. \fi}

The choice of $\sigma_{rz}(r=a) = 0$ motivates the ansatz that 
\begin{equation}
\begin{split}
    u_z(z,r,\omega) &= \sum_{\ell=1}^\infty J_0(k_\ell r) \bar{u}_{z,\ell}(z)e^{i\omega t} \\
    u_r(z,r,\omega) &= \sum_{\ell=1}^\infty J_1(k_\ell r) \bar{u}_{r,\ell}(z)e^{i\omega t},
\end{split}
\end{equation}
where $J_m$ are the Bessel functions of the first kind and $k_\ell$ is a solution to $J_1(k_\ell a) = 0$. In other words, $k_\ell = \zeta_\ell/a$, where $\zeta_\ell$ is the $\ell$th zero of $J_1(x)$. To check that we have met our boundary condition, we compute $\sigma_{rz}$ as
\begin{align}
    \sigma_{rz} \propto \sum_{\ell=1}^\infty J_1(k_\ell r)Z_\ell(z),
\end{align}
where we have used the fact that the derivative of the zeroth Bessel function is proportional to the first Bessel function, and $Z_\ell$ is some function of $z$. By construction of $k_\ell$, $\sigma_{rz} = 0$ at $r = a$.

Because we are solving the Helmholtz equation on the cylinder, Bessel functions are the natural basis for the solution. We therefore perform Dini expansions~\cite{BonduPLA98ThermalNoiseb,LiuPRD00ThermoelasticNoiseb} on our quantities to leverage this basis. The Dini expansion of the Gaussian beam weight factor is
\begin{align}
    I(r) = \sum_{\ell=1}^\infty I_\ell J_0(k_\ell r),
\end{align}
where the coefficient $I_\ell$ is
\begin{align}
    I_\ell &= \frac{2}{a^2 J_0^2(\zeta_\ell)}\int_0^a I(r) J_0 (k_\ell r)r\, dr\\ &\approx \frac{1}{\pi a^2 J_0^2(\zeta_\ell)}e^{-k_\ell^2w^2/8}.
\end{align}
For the last equality in the equation for $I_\ell$, we have invoked the approximation that the beam size is smaller than the mirror radius $w\ll a$, which is typically the case in interferometers. We then apply the orthogonality of the Bessel functions on the cylinder to write
\begin{align}
    \widehat{\delta z}(\omega) = \sum_{\ell=1}^\infty \bar{u}_{z,\ell}(z=h/2) e^{-k_\ell^2 w^2/8}.
\end{align}
In a similar spirit to the method used for the infinite slab, we find $\bar{u}_{z,\ell}(z)$ (and $\bar{u}_{r,\ell}(z)$) that satisfy the elastic wave equation and the stress boundary conditions at $z=h/2$ and $z=-h/2$. Specifically, $\sigma_{zz, \ell}(z=h/2) = I_\ell$, $\sigma_{zr, \ell}(z=h/2) = \sigma_{zz, \ell}(z=-h/2) = \sigma_{zr, \ell}(z=-h/2) = 0$. The calculation for finding $\bar{u}_{z,\ell}(z=h/2)$ is \textit{identical} to the continuous case for finding $\tilde{u}_z(z=h/2,k)$, where we exchange the continuous variable $k$ for discrete $k_\ell$:
\begin{align}
    \bar{u}_{z,\ell}(z = h/2) = F_0I_\ell \xi_{z,\ell},
\end{align}
where the receptance density for the cylinder is
\begin{widetext}
\begin{multline}\label{eq:xi_zl_definition}
        \xi_{z,\ell} = -\frac{k_{z,\longitudinal,\ell} k_\shear^2}{2\mu} \bigg(\frac{1}{(2k_\ell^2-k_\shear^2)^2\coth(k_{z,\longitudinal,\ell} h/2) - 4k_\ell^2k_{z,\longitudinal,\ell}k_{z,\shear,\ell}\coth(k_{z,\shear,\ell} h/2)} +\\
        \frac{1}{(2k_\ell^2-k_\shear^2)^2\tanh(k_{z,\longitudinal,\ell} h/2) - 4k_\ell^2k_{z,\longitudinal,\ell}k_{z,\shear,\ell}\tanh(k_{z,\shear,\ell} h/2)}\bigg).
\end{multline}
\end{widetext}
This yields a similar-looking spectrum (cf. the infinite slab)
\begin{align}\label{eq:SMN_Spectrum_Cyl}
    S_{z,\text{SMN}}(\omega) = \frac{4k_{\mathrm{B}} T}{\pi a^2 \omega}\sum_{\ell=1}^\infty|\mathrm{Im}\left\{\xi_{z,\ell}\right\}|\frac{e^{-k_\ell^2w^2/4}}{J_0^2(\zeta_\ell)}.
\end{align}

In the limit as the radius $a$ approaches infinity, this spectrum tends towards the results of the infinite slab. For the finite-radius cylinder, the high-frequency spectrum is well approximated by \cref{eq:SMNHFAValley} for frequencies such that the function in \cref{eq:xi_zl_definition} does \textit{not} have any poles with $k_p \approx k_\ell \lesssim 2/w$. In fact, the cylinder's spectrum has a correction factor to the infinite slab's spectrum of $1-w^2/a^2$ to leading order in $w/a$ in this frequency regime. For frequencies such that the function in \cref{eq:xi_zl_definition} \textit{does} have a pole with $k_p \approx k_\ell \lesssim 2/w$ (i.e., an eigenfrequency of the cylinder with high coupling to the beam), there will be a large peak in the spectrum.

Due to the equipartition theorem, the energy in each mechanical eigenmode is $k_{\mathrm{B}}T/2$. Therefore, a measurement of this energy could serve as an absolute calibration reference. By measuring the noise power in a mechanical mode peak, one could in principle infer the sensitivity of the interferometer. However, due to limitations in modeling the exact eigenmodes of the mirror and experimental imperfections such as beam location on the mirror, the exact frequencies and magnitudes of these peaks in a measured spectrum will be uncertain. The peaks of individual eigenmodes are, therefore, a poor choice for calibrating the interferometer sensitivity despite their large magnitude. Instead, by averaging the cylinder's spectrum over a wide frequency window to average out the peaks, one recovers the infinite slab result, which can then serve as a calibration reference. Furthermore, since the part of the averaged frequency spectrum with the peaks does not depend on the loss angle, while the rest of the spectrum does, this technique could serve as an independent measurement of the loss angle.

For GQuEST, we find that our choice of mirror radius has minimal effect at frequencies above approximately 10~MHz. At these high frequencies, the eigenmodes of the cylinder with large coupling to the beam are close in frequency to the longitudinal eigenmodes. Therefore, the fraction of frequencies away from these eigenmodes, where \cref{eq:SMNHFAValley} describes the SMN spectrum of the cylinder well, becomes quite large. We plot the GQuEST substrate mechanical noise amplitude spectral density (ASD) in \cref{fig:SMN} with parameters from \cref{app:Tables}.

\subsection{Mechanical Noise in the Mirror Coatings}\label{sec:CMN}
We now turn our attention to the mechanical noise from the inclusion of a thin coating layer on the front surface of a mirror (i.e., coating mechanical noise, CMN). Although this coating is much thinner than the substrate, the coating has been measured to have a loss angle over 100 times larger than that of substrates (fused silica, silicon, sapphire, etc. have very low loss angles comparatively)~\cite{GrasPRD17AudiobandCoating,PennPLA06FrequencySurface, RodriguezSR19DirectDetectiona,UchiyamaPLA99MechanicalQuality}. Previous work~\cite{HarryCQG02ThermalNoise, SomiyaPRD09CoatingThermal, GurkovskyPLA10ThermalNoise,HongPRD13BrownianThermala} has examined coating mechanical noise below the first mechanical eigenfrequency of the mirror. We find that this quasistatic approximation makes inaccurate predictions above the first mechanical eigenfrequency, even at frequencies where the substrate modes are not resonant.

For our model, we assume for simplicity that the elastic wavelength is much larger than the coating thickness $h_{\coating}$. This assumption starts to break down above approximately 50~MHz. The thickest coating layer (the half-wavelength-thick layer made of low-index
material) is 534~nm thick for 1550~nm light; the elastic wavelength is much larger than this scale, below 1~GHz, so the coating can be treated as uniform with regard to solving the wave equation. Here, we ignore the photoelastic effect for the coating, as the beam penetration depth is small, and this effect is assumed to be subdominant. We also treat the coating as spatially isotropic for simplicity.

We start our modeling with Levin's insight~\cite{LevinPRD98InternalThermal} for calculating the noise spectrum using the power dissipated instead of the admittance (see \cref{eq:LevinFDT}). We use this method because it handles different materials more easily than finding the susceptibility as the contributions from the different loss angles and material properties are much less apparent. We can write the dissipated power as
\begin{align}
    W_{\text{diss}}(\omega) = \omega U_\mathrm{\max}(\omega) \phi,
\end{align}
where $U_\mathrm{\max}$ is the maximum stored elastic strain energy during a cycle of a specific mechanical mode. For multiple materials, one sums over them to find the combined $U_\mathrm{\max}\phi$ product. Since isotropic materials actually have two independent loss angles, we express $W_{\text{diss}}(\omega)$ as
\begin{align}
    W_{\text{diss}}(\omega) = \omega \left(U_{\bulk} \phi_{\bulk} + U_{\shear} \phi_{\shear}\right),
\end{align}
where $U_{\bulk}$ and $U_{\shear}$ are the bulk\footnote{Of course, the B subscript in the Boltzmann constant $k_{\mathrm{B}}$ does not stand for bulk.} and shear elastic strain energies computed by \cref{eq:EnergyDensityBulk,eq:EnergyDensityShear}, respectively, and $\phi_{\bulk}$ and $\phi_{\shear}$ are the loss angles associated with bulk and shear moduli~\cite{HongPRD13BrownianThermala}. We follow Hong~\cite{HongPRD13BrownianThermala} and choose to calculate the power dissipated with these two independent loss angles. The bulk loss angle describes dissipation associated with deformation modes that change the material’s volume under uniform surface forces, while the shear loss angle describes dissipation associated with opposing forces that distort the material’s shape without changing its volume. Other linear combinations of loss angles could be chosen. For the remainder of the section, we work to express the stored elastic strain energy in terms of quantities we have derived in \cref{subsec:MechNoiseSub,app:SMN Details}. Because we calculate the noise without the quasistatic approximation, we cannot follow~\cite{HongPRD13BrownianThermala} in their calculation of the stored elastic strain energy.

We express the energies as volume integrals of the energy densities, specifically
\begin{align}\label{eq:EnergyDensityBulk}
    U_{\bulk} = \frac{1}{2}\int K_0\left|\Theta\right|^2dV
\end{align}
and
\begin{align}\label{eq:EnergyDensityShear}
U_{\shear} = \int \mu_0\Xi_{ij}\Xi_{ij}^*\,dV,
\end{align}
where $K=\lambda + 2\mu/3$ is the bulk modulus, $\Theta$ is the expansion scalar, $\Xi_{ij}$ is the shear tensor (note that in the above integral we are summing the components indexed with $i$ and $j$ under Einstein sum notation), and $^*$ indicates the complex conjugate. The absolute values are needed as the integrand is otherwise complex due to the loss angle. We express the expansion scalar as
\begin{align}
    \Theta = \epsilon_{ii} = \epsilon_{zz} + \epsilon_{rr} + \epsilon_{\varphi\varphi}
\end{align}
and the shear tensor as
\begin{align}
    \Xi_{ij} = \frac{1}{2}(\epsilon_{ij}+\epsilon_{ji}) - \frac{1}{3}\delta_{ij}\Theta
\end{align}
where $\delta_{ij}$ is the Kronecker delta and the strain, $\epsilon_{ij}$, is defined as
\begin{align}
   \epsilon_{ij} = \frac{1}{2}\left(\frac{\partial u_i}{\partial x_j} + \frac{\partial u_j}{\partial x_i}\right).
\end{align}
By axial symmetry, $\epsilon_{r\varphi} = \epsilon_{z\varphi} = 0$, but $\epsilon_{\varphi\varphi} \neq 0$. We can write the nonzero strains in axisymmetric cylindrical coordinates as
\begin{align}
\epsilon_{zz} = \frac{\partial u_z}{\partial z}, \,
\epsilon_{rr} = \frac{\partial u_r}{\partial r}, \,
\epsilon_{\varphi\varphi} = \frac{u_r}{r}, \\
\intertext{and}
   \epsilon_{rz} = \frac{1}{2}\left(\frac{\partial u_r}{\partial z} + \frac{\partial u_z}{\partial r}\right).
\end{align}

Our target is to express the energy densities in terms of easily computed quantities, specifically the surface pressure $\sigma_{zz}$ and the radial displacement $u_r$, which we solved for in \cref{subsec:MechNoiseSub,app:SMN Details}. We can avoid expressing the energy densities in terms of $u_z$ and favor $\sigma_{zz}$ instead of $u_z$ because $\sigma_{zz}$ has a simple form in the coating, and it has nice boundary conditions at the coating-substrate interface; $u_z$ and $\epsilon_{zz}$ are more complicated.

We note the relationship between stress and strain as
\begin{align}
    \sigma_{ij} = \lambda\Theta\delta_{ij} + 2\mu\epsilon_{ij}.
\end{align}
Because the coating is thinner than the elastic wavelength and stress perpendicular to a boundary is continuous across it, we approximate that the stress is constant in the coating and therefore the strain is constant in each coating material. We write the strain in the coating as
\begin{align}
    \epsilon_{zz} = \frac{\sigma_{zz}-\lambda(\epsilon_{rr}+\epsilon_{\varphi\varphi})}{\lambda + 2\mu};~\epsilon_{rz} = \frac{\sigma_{rz}}{2\mu}=0.
\end{align}
By using this relation, we can write our expression for $\Theta$ simply by writing it in terms of just $\epsilon_{rr}+\epsilon_{\varphi\varphi}$ and $\sigma_{zz}$:
\begin{align}
    \Theta = \frac{1}{\lambda+2\mu}\left(\sigma_{zz}+2\mu(\epsilon_{rr}+\epsilon_{\varphi\varphi})\right).
\end{align}
Similarly, we can simplify our expression for $ \Xi_{ij}\Xi^*_{ij}$. By starting with
\begin{align}
    \Xi_{ij}\Xi^*_{ij} = \frac{1}{2}|\epsilon_{rr}-\epsilon_{\varphi\varphi}|^2 + \frac{1}{6}\left|(\epsilon_{rr}+\epsilon_{\varphi\varphi})-2\epsilon_{zz}\right|^2,
\end{align}
we get
\begin{align}
    \Xi_{ij}\Xi^*_{ij} = \frac{1}{2}|\epsilon_{rr}-\epsilon_{\varphi\varphi}|^2 + \frac{\left|2\sigma_{zz}-(3\lambda+2\mu)(\epsilon_{rr}+\epsilon_{\varphi\varphi})\right|^2}{6|\lambda+2\mu|^2}.
\end{align}

Now that we can express the integrands in terms of $\sigma_{zz}$, $\epsilon_{rr}+\epsilon_{\varphi\varphi}$, and $\epsilon_{rr}-\epsilon_{\varphi\varphi}$, we compute these quantities. $\sigma_{zz}$ is simply the applied stress on the mirror from the fictitious drive. We will find $\epsilon_{rr}+\epsilon_{\varphi\varphi}$ and $\epsilon_{rr}-\epsilon_{\varphi\varphi}$ in the coating by computing them in the substrate and assuming that the coating is sufficiently thin and that the coating and substrate materials have similar enough material properties that this is a good approximation in solving the elastic wave equation. Note that at a boundary between different materials, the in-plane normal strains, $\epsilon_{rr}$ and $\epsilon_{\varphi\varphi}$, are equal on either side of a constant-$z$ boundary~\cite{HarryCQG02ThermalNoise}. We thus write
\begin{align}\label{eq:Theta_c}
    |\Theta_\coating|^2 = \frac{1}{|\lambda_\coating+2\mu_\coating|^2}\left|\sigma_{\substrate,zz}+2\mu_\coating(\epsilon_{\substrate,rr}+\epsilon_{\substrate,\varphi\varphi})\right|^2,
\end{align}
where we now label variables with the subscripts $\coating$ for coating and $\substrate$ for substrate. $\Xi_{ij}\Xi^*_{ij}$ can be written similarly. Invoking again the approximation that the coating is thin to assert the strain is constant within each material throughout the coating, we convert the volume elements $dV$ in \cref{eq:EnergyDensityBulk,eq:EnergyDensityShear} to a sum of the different materials in the coating, $dV = \sum_j h_{\coating,j} d^2r$, where $\sum_j h_{\coating,j} = h_{\coating}$, the former being the thickness of the $j$th material. We find 
\begin{widetext}
\begin{align}\label{eq:BulkEnergyInCoating}
    U_{\coating,\bulk} = \sum_j\frac{|3\lambda_{\coating,j}+2\mu_{\coating,j}|h_{\coating,j}}{6|\lambda_{\coating,j}+2\mu_{\coating,j}|^2}\int \left|\sigma_{\substrate,zz}+2\mu_{\coating,j}(\epsilon_{\substrate,rr}+\epsilon_{\substrate,\varphi\varphi})\right|^2 \,d^2r
\end{align}
and
\begin{align}\label{eq:ShearEnergyInCoating}
        U_{\coating,\shear} = \sum_j\frac{|\mu_{\coating,j}| h_{\coating,j}}{2}\int \Big(|\epsilon_{\substrate,rr}-\epsilon_{\substrate,\varphi\varphi}|^2 
        + \frac{\left|2\sigma_{\substrate,zz}-(3\lambda_{\coating,j}+2\mu_{\coating,j})(\epsilon_{\substrate,rr}+\epsilon_{\substrate,\varphi\varphi})\right|^2}{3|\lambda_{\coating,j}+2\mu_{\coating,j}|^2}\Big)\,d^2r.
\end{align}
\end{widetext}

Similarly to the substrate noise in the infinite slab found in \cref{subsec:MechNoiseSub}, we transform the integrands of \cref{eq:BulkEnergyInCoating,eq:ShearEnergyInCoating} using the Hankel transform and Parseval's theorem. We can write
\begin{align}\label{eq:epsilon_to_u_sr}
\tilde{\epsilon}_{\substrate,rr}+\tilde{\epsilon}_{\substrate,\varphi\varphi} = k \tilde{u}_{\substrate,r}(k);~
    \tilde{\epsilon}_{\substrate,rr}-\tilde{\epsilon}_{\substrate,\varphi\varphi} = -k \tilde{u}_{\substrate,r}(k)
\end{align}
and
\begin{align}\label{eq:u_sr}
    \tilde{u}_{\substrate,r}(z = h/2,k) = F_0\tilde{I}(k) \xi_{\substrate,r}(k).
\end{align}

We find $\xi_{\substrate,r}(k)$ in the exact same manner as $\xi_{\substrate,z}(k)$ for the substrate calculation in \cref{subsec:MechNoiseSub,app:SMN Details} by forcing $\bm{u}$ to obey the wave equation inside the infinite slab with the relevant stresses at $z=\pm h/2$. Please note that $\xi_{\substrate,r}(k)$ and $\xi_{\substrate,z}(k)$ have different forms. We ascertain $\xi_{\substrate,r}(k)$ to be
\begin{widetext}
\begin{multline}\label{eq:xi_coating_definition}
    \xi_{\substrate,r}(k) = -\frac{k}{2\mu_\substrate} \bigg(\frac{(2k^2-k_{\substrate,\shear}^2)\coth(k_{\substrate,z,\longitudinal} h/2)-2k_{\substrate,z,\longitudinal}k_{\substrate,z,\shear}\coth(k_{\substrate,z,\shear} h/2)}{(2k^2-k_{\substrate,\shear}^2)^2\coth(k_{\substrate,z,\longitudinal} h/2) - 4k^2k_{\substrate,z,\longitudinal}k_{\substrate,z,\shear}\coth(k_{\substrate,z,\shear} h/2)} +\\
    \frac{(2k^2-k_{\substrate,\shear}^2)\tanh(k_{\substrate,z,\longitudinal} h/2)-2k_{\substrate,z,\longitudinal}k_{\substrate,z,\shear}\tanh(k_{\substrate,z,\shear} h/2)}{(2k^2-k_{\substrate,\shear}^2)^2\tanh(k_{\substrate,z,\longitudinal} h/2) - 4k^2k_{\substrate,z,\longitudinal}k_{\substrate,z,\shear}\tanh(k_{\substrate,z,\shear} h/2)}\bigg).
\end{multline}

Finally, we are ready to find the spectrum for the coating mechanical noise. We plug \cref{eq:u_sr} into \cref{eq:epsilon_to_u_sr} and these equations into \cref{eq:BulkEnergyInCoating,eq:ShearEnergyInCoating}. We also know $\tilde{\sigma}_{\substrate,zz} = F_0\tilde{I}(k)$ at the front surface and therefore within the coating. With the energies $U_{\coating,\bulk}$ and $U_{\coating,\shear}$ fully defined in terms of known quantities, we get the coating mechanical power spectral density by substituting the dissipated power into \cref{eq:LevinFDT}. We find the spectrum for the coating mechanical noise, in terms of $\xi_{\substrate,r}(k)$ from \cref{eq:xi_coating_definition}, to be
\begin{multline}\label{eq:CMN_Spectrum}
    S_{z,\text{CMN}}(\omega) = \frac{2k_{\mathrm{B}} T}{3\pi\omega}\sum_j\frac{h_{\coating,j}}{|\lambda_{\coating,j}+2\mu_{\coating,j}|^2  }\bigg[|3\lambda_{\coating,j}+2\mu_{\coating,j}|\phi_{\coating,j,\bulk}\int_0^\infty \left|1+2\mu_{\coating,j} k\xi_{\substrate,r}(k)\right|^2 ke^{-k^2w^2/4}\,dk \\
    + 4|\mu_{\coating,j}|\phi_{\coating,j,\shear}\int_0^\infty \left(1-\mathrm{Re}\left\{(3\lambda_{\coating,j}+2\mu_{\coating,j})k\xi_{\substrate,r}(k)\right\}+\left|(3(\lambda_{\coating,j}+\mu_{\coating,j})^2+\mu_{\coating,j}^2)k^2\xi_{\substrate,r}(k)^2\right|\right) ke^{-k^2w^2/4}\,dk\bigg],
\end{multline}
\end{widetext}
where $\phi_{\coating,j,\bulk}$ and $\phi_{\coating,j,\shear}$ are the bulk and shear loss angles for the individual coating layers. For a cylindrical mirror instead of an infinite slab, we simply replace the $k$-space integral in \cref{eq:CMN_Spectrum} with a sum over $k = \zeta_\ell/a$, as with the substrate model.

To compare to previous results, we take the quasistatic ($\omega \rightarrow 0$), infinite-$h$ limit, where $h$ is the mirror's thickness, giving us:
\begin{align}
    \xi_{\substrate,r,\text{quasistatic}}(k) = \frac{1}{2k(\lambda_\substrate+\mu_\substrate)} = \frac{1-2\nu_\substrate}{2\mu_\substrate k}.
\end{align}

In the quasistatic, infinite-$h$ regime that has been previously modeled in~\cite{HongPRD13BrownianThermala}, we agree exactly with their Eq. (94) under our assumption that the field does not penetrate into the coating, which sets their $\epsilon_j(z) = 0$.\footnote{The agreement between our spectrum and the literature is not immediately apparent. We point out that their Eqs. (62), (63), (67), (68), and (69) appear to be too large by exactly a factor of two compared to the actual result presented later in~\cite{HongPRD13BrownianThermala}. Eq. (1) in~\cite{YamPRD15MultimaterialCoatingsa} provides an oft-cited result summarizing the result in~\cite{HongPRD13BrownianThermala}. However, their Eq. (1) contains a different error that~\cite{TaitPRL20DemonstrationMultimaterial} corrects.}

\subsubsection{Analytic Coating Mechanical Noise Spectrum at High Frequencies}

We approximate the CMN spectrum in the same limits as in \cref{subsubsec:SMN_Analytic}. In these limits, away from the mirror mechanical resonances (frequencies such that $\xi_{\substrate,r}(k)$ does not have Lorentzian peaks for $k\ll2/w$), $\xi_{\substrate,r}(k)$ can be well approximated by a linear Taylor expansion, specifically
\begin{align}
    \xi_{\substrate,r}(k) \approx \xi^{(1)}_{\substrate,r} k
\end{align}
making
\begin{multline}
    \xi^{(1)}_{\substrate,r} = \frac{1}{\rho_\substrate\omega^2} \bigg(1 - \frac{k_{\substrate,\longitudinal}}{k_{\substrate,\shear}}\Big(\tan(k_{\substrate,\longitudinal} h/2)\cot(k_{\substrate,\shear} h/2) + \\\cot(k_{\substrate,\longitudinal} h/2)\tan(k_{\substrate,\shear} h/2)\Big)\bigg).
\end{multline}
We define the dimensionless parameter $q_j$ that is roughly the lateral-to-vertical strain-energy ratio in the high-frequency limit as
\begin{align}\label{eq:q_j}
    q_j = \frac{16|\mu_{\coating,j}|\mathrm{Re}\left\{\xi^{(1)}_{\substrate,r}\right\}}{w^2} \approx \frac{16|\mu_{\coating,j}|}{\rho_\substrate w^2\omega^2},
\end{align}
where the last approximation roughly holds away from the mechanical resonances of the mirror. With these approximations, we can evaluate \cref{eq:CMN_Spectrum}, writing the high-frequency spectrum as
\begin{widetext}
\begin{multline}\label{eq:CMN_Spectrum_Analytic}
    S_{z,\text{CMN, HFA}}(\omega) = \frac{4k_{\mathrm{B}} T}{3\pi w^2\omega}\sum_j\frac{h_{\coating,j}}{|\lambda_{\coating,j}+2\mu_{\coating,j}|^2} \Bigg[|3\lambda_{\coating,j}+2\mu_{\coating,j}|\phi_{\coating,j,\bulk}+4|\mu_{\coating,j}|\phi_{\coating,j,\shear} +|3\lambda_{\coating,j}+2\mu_{\coating,j}|(\phi_{\coating,j,\bulk}-\phi_{\coating,j,\shear})q_j +\\ \frac{1}{2}\left(|3\lambda_{\coating,j}+2\mu_{\coating,j}|\phi_{\coating,j,\bulk}+\left|\frac{3(\lambda_{\coating,j}+\mu_{\coating,j})^2}{\mu_{\coating,j}}+\mu_{\coating,j}\right|\phi_{\coating,j,\shear}\right)q_j^2\Bigg]
\end{multline}
\end{widetext}

By approximating $\phi_{\coating,j,\bulk}=\phi_{\coating,j,\shear}=\phi_{\coating,j}$ since the separate loss angles are not accurately measured in the MHz band, we get
\begin{align}\label{eq:CMN_Spectrum_Analytic_Simple}
\resizebox{\columnwidth}{!}{$\displaystyle
    S_{z,\text{CMN, HFA}}(\omega) = \frac{4k_{\mathrm{B}} T}{\pi w^2 \omega}\sum_j \frac{h_{\coating,j}\phi_{\coating,j}}{|M_{\coating,j}|}\left(1+\frac{q_j^2}{2(1-2|\nu_{\coating,j}|)}\right). $}
\end{align}
For commonly used mirror substrate materials and geometries and at large frequencies away from the mirror eigenfrequencies, $q_j\ll1$, so we will ignore the $q_j^2$ term in \cref{eq:CMN_Spectrum_Analytic_Simple}, leaving
\begin{align}\label{eq:CMN_Spectrum_Analytic_Simple_no_q}
    S_{z,\text{CMN, HFA}}(\omega) \approx \frac{4k_{\mathrm{B}} T}{\pi w^2 \omega}\sum_j \frac{h_{\coating,j}\phi_{\coating,j}}{|M_{\coating,j}|}.
\end{align}
We define $|M_\coating|$ as the harmonic mean of the coating-layer P-wave moduli magnitudes since it is in the denominator (otherwise, there is an inconsistency when considering other average properties like loss angle). This gives us an effective high-frequency coating loss angle, P-wave modulus, and PSD, respectively, as
\begin{align}\label{eq:CMNphi}
    \phi_\coating=\frac{|M_\coating|}{h_{\coating}}\sum_j \frac{h_{\coating,j}}{|M_{\coating,j}|}\phi_{\coating,j};~|M_\coating|^{-1} = \frac{1}{h_\coating}\sum_j \frac{h_{\coating,j}}{|M_{\coating,j}|}
\end{align}
and
\begin{align}\label{eq:CMNMin}
    S_{z,\text{CMN min}}(\omega) = \frac{4k_{\mathrm{B}} T h_{\coating} \phi_\coating}{\pi |M_\coating| w^2 \omega}.
\end{align}
We note that this is equal to \cref{eq:CMN_Spectrum_Analytic_Simple_no_q}. The remarkably simple structure of the coating mechanical noise in \cref{eq:CMNMin}, which applies to high frequencies away from the mechanical resonances that couple strongly to the laser beam, does not depend on any substrate terms. This is unlike the formulae that appear in~\cite{HongPRD13BrownianThermala,YamPRD15MultimaterialCoatingsa}. The $\xi_{\substrate,r}(k)$ term in \cref{eq:CMN_Spectrum}, the only term in \cref{eq:CMN_Spectrum} that depends on the substrate, contributes meaningfully to the overall spectrum amplitude in the quasistatic regime. However, at high frequencies, the substrate dynamics away from resonance become negligible, and we are left with just the term from the surface pressure. In fact, the spectrum given in \cref{eq:CMNMin} is identical to the quasistatic spectrum given in~\cite{HongPRD13BrownianThermala} if we assume an infinitely stiff substrate.

By using the Levin formalism~\cite{LevinPRD98InternalThermal}, we can see mathematically that the coating noise and substrate noise add incoherently. Each layer (including the substrate here) contributes linearly to the total power dissipated $W_{\text{diss}}$, which is directly proportional to the PSD. We use the thin-coating approximation and assert the coating's contribution to the substrate's strain is minimal. We see in the next section how this assumption leads to an inaccurate noise model at the mirror eigenfrequencies.

\begin{figure*}[t]
\includegraphics[width=1\linewidth]{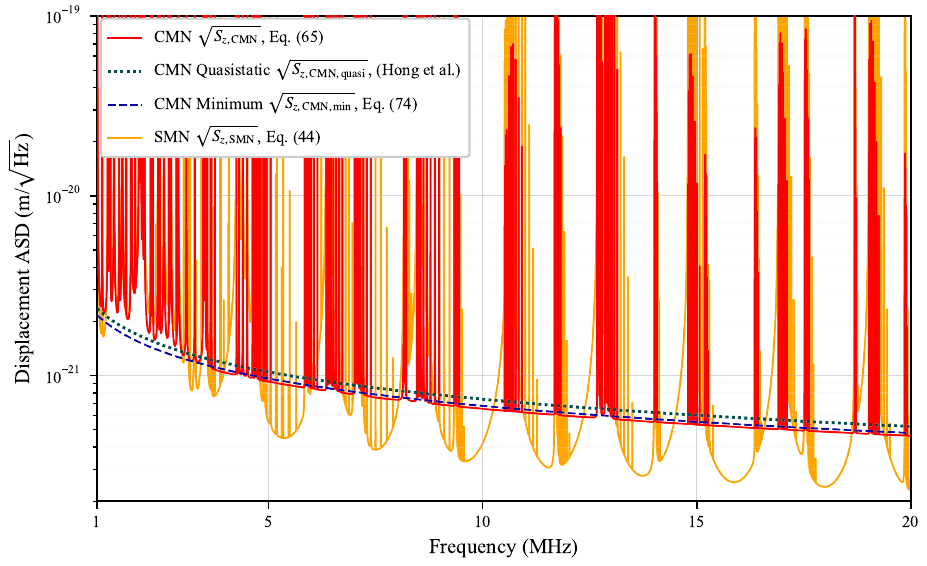}
  \caption{The Modeled Amplitude Spectral Density of the Coating Mechanical Noise (CMN) for the GQuEST End Mirrors. The red curve plots the coating mechanical noise for a cylindrical end mirror, which is the cylinder version of the CMN for an infinite slab defined in \cref{eq:CMN_Spectrum}. The dotted green curve plots the quasistatic, infinite-$h$ limit from the literature~\cite{HongPRD13BrownianThermala}. The dashed blue curve plots the noise floor defined in \cref{eq:CMNMin}, which does not have any mirror resonances. The SMN ASD is shown for reference; note that the CMN resonances happen at the same frequencies as SMN resonances because the substrate's mechanics drive the coating's mechanics. Not all SMN resonances coincide with particularly large CMN resonances since some substrate modes do not excite the coating. We use $\Delta f_{\text{sample}} = 3.9$~Hz to generate this plot.}
  \label{fig:CMN}
\end{figure*}

\subsubsection{Analytic Coating Mechanical Noise Spectrum at Mirror Eigenfrequencies}

At the mirror eigenfrequencies, the coating mechanical noise depends on the substrate's loss angle. If the substrate's loss angle were equal to zero, then the coating mechanical noise would appear to diverge (due to the presence of $\xi_{\substrate,r}(k)$ in \cref{eq:CMN_Spectrum}) at the mirror's mechanical eigenfrequencies, which is an undesirable result. Ultimately, we have relied on the assumption that the coating and substrate materials are sufficiently similar that we can use the substrate's $\xi_r(k)$ in the coating. On resonance, the loss angles of the materials become important for the behavior of $\xi_r(k)$; if the loss angles are many orders of magnitude different in the substrate and coating, then the similar-material approximation fails.

Instead of a full treatment, we could use a combined loss angle for the entire mirror given by
\begin{align}
    \phi_{\mathrm{m}} = \frac{U_\substrate\phi_\substrate + U_{\coating,\bulk} \phi_{\coating,\bulk} + U_{\coating,\shear} \phi_{\coating,\shear}}{U_\substrate+U_{\coating,\bulk}+U_{\coating,\shear}}.
\end{align}
This weighted loss angle appears in the literature~\cite{HarryCQG02ThermalNoise,HongPRD13BrownianThermala}, although not specifically for calculating the noise at a mirror resonance but instead as a heuristic for the total mechanical noise at a specific mode. By using $\phi_{\mathrm{m}}$ instead of $\phi_\substrate$ in $\xi_{\substrate,r}(k)$, the noise for eigenfrequencies that are not the longitudinal eigenfrequencies ($\omega \neq \pi m_\longitudinal |v_\longitudinal|/h$) is more properly handled. For the longitudinal eigenfrequencies, $\phi_{\mathrm{m}}\approx \phi_\substrate$, so the CMN still diverges as $\phi_\substrate \rightarrow0$. To solve this problem, we believe the full elastic wave equations must be solved outside of the perturbative limit we have used for Levin's method of computing energy dissipation rather than admittance. We would also need to sum the coating and substrate mechanical noises coherently to take into account the fully coupled elastic wave equation. We leave this for future work. We plot the GQuEST coating mechanical noise ASD in \cref{fig:CMN}.

\subsection{Mechanical Noise Model Limitations}
We summarize here some assumptions we have made in our mechanical noise model. We treat the substrates and coatings as isotropic, even though crystalline coatings and substrate materials such as crystalline silicon and sapphire are anisotropic. For crystalline silicon, taking the least-stiff $\langle100\rangle$ P-wave modulus instead of the most-stiff $\langle111\rangle$ P-wave modulus increases the noise floor by less than 30\%.

We ignore any curvature of the optics. This will change the mode structure slightly. We approximate the coating thickness, $h_\coating$, as smaller than the inverse elastic wavenumber;\footnote{The inverse elastic wavenumber is a more appropriate length scale than the wavelength.} this approximation is valid until approximately 50~MHz. We expect the noise to decrease above 50~MHz because the strain in the part of the coating closer to the substrate will be lower than the strain at the top layers. The approximation that the individual coating layers are smaller than the inverse elastic wavenumber is valid into the GHz band. We ignore the photoelastic effect in the coating as Hong~\cite{HongPRD13BrownianThermala} found this effect to be small, and we see no reason to doubt this in our frequency regime. We also rely on the coating's thinness and the similar material properties of the coating and substrate to avoid solving the fully coupled elastic wave equation for the coating and substrate. As previously discussed, we believe this is a good approximation except at and around the eigenfrequencies of the mirror.

We ignore the boundary condition $\sigma_{rr}(r=a) = 0$ for the cylindrical mirror and cannot easily apply a ``Saint-Venant correction"~\cite{BonduPLA98ThermalNoiseb, LiuPRD00ThermoelasticNoiseb} when not in the quasistatic regime. To estimate the effect of $\sigma_{rr}(r=a) \neq 0$, we could calculate the power injected from the residual stress at the boundary compared to the power injected from the laser beam.\ifdcc We leave this as an approximation, awaiting future work using~\cite{HutchinsonJoAM80VibrationsSolid}, and do not address it here, since this error only slightly shifts the mode structure. \else We leave this as an approximation, awaiting future work using~\cite{HutchinsonJoAM80VibrationsSolid}, and examine experimental data in \cref{sec:Comparison_with_Experimental_Data}.\fi

The high-frequency approximation for \cref{eq:SMN_Spectrum} uses pole approximations for ease of calculation and explanation as described in \cref{subsubsec:SMN_Analytic}. We leave an analytic expression for \cref{eq:SMN_Spectrum} at high frequencies to future work.

\section{Substrate Thermoelastic Noise}\label{sec:STENoise}
With our modeling of mechanical noise, we can now readily solve for another noise source, substrate thermoelastic (STE) noise. In this noise source, random thermal fluctuations cause the mirror substrate to expand and contract, changing the physical path length of the laser beam when it reflects off the surface. Thermal fluctuations also change the optical path length via the thermorefractive coefficient, $\beta = \partial n/\partial T$~\cite{BraginskyPLA04CornerReflectors,BenthemPRD09ThermorefractiveThermochemical}. Since the light does not penetrate into the end mirror's substrate, we ignore this effect here. We treat beamsplitter thermoelastic noise (BSTE) in \crefBSTE~but find no additional treatment is needed for beamsplitter thermorefractive noise into the MHz band~\cite{BenthemPRD09ThermorefractiveThermochemical}.

Braginsky calculates the substrate thermoelastic noise spectrum (in one method among many in their paper) using a framework of heat driving mechanics as the noise source~\cite{BraginskyPLA99ThermodynamicalFluctuationsa}. However, using the FDT and Levin's insight, we can start exactly as Levin prescribes, using a fictitious pressure on the front surface of the mirror and calculating the power dissipated. Rather than calculating the power dissipated with some loss angle (due to effects like structural damping), this source requires calculating the power dissipated from thermoelastic damping. We use the formula for the power dissipated due to a temperature gradient from Landau and Lifshitz~\cite{Landau09TheoryElasticity},
\begin{align}\label{eq:ThermalPowerDissipated}
    W_{\text{diss}} = \left\langle \int \frac{\kappa}{T} (\vec{\nabla} \delta T)^2 dV \right \rangle,
\end{align}
where $\kappa$ is the thermal conductivity, $T$ is the physical temperature of the mirror, and $\delta T$ is a scalar field in time and space that represents the fictitious change in temperature due to the fictitious injection on the mirror surface; it is distinct from the actual temperature $T$. The volume integral is over all of space, although in practice, we can simplify it to an integral over the mirror's volume. $\langle ... \rangle$ denotes a time average and will yield a factor of $1/2$.

The thermal field is generated via coupling between expansion and temperature with the (linear) coefficient of thermal expansion $\alpha = \frac{1}{L}\frac{\partial L}{\partial T}$. The temperature field can be expressed in terms of material parameters and the expansion scalar 
\begin{align}\label{eq:TemperatureField}
    \delta T = \frac{-\alpha Y T}{C_V (1-2\nu)}\Theta = \frac{-3\alpha K}{C_V}T\Theta,
\end{align}
where $C_V$ is the {\textbf{volumetric}} heat capacity.  This simple expression requires being in the adiabatic limit, when the thermal diffusion length, $r_\thermal$, is smaller than the beam spot size, the mirror dimensions, and the elastic wavelength. This approximation can break down above 1~GHz for commonly used mirror substrate materials. 
The temperature field and expansion scalar in the adiabatic regime are related via a Grüneisen parameter $3\alpha K/C_V$. The thermal diffusion length, $r_\thermal$, is the key length scale for thermo-optic (thermoelastic and thermorefractive) noise and is given by
\begin{align}
    r_\thermal = \sqrt{\frac{\kappa}{C_V \omega}}.
\end{align}
We substitute $\delta T$ from \cref{eq:TemperatureField} into \cref{eq:ThermalPowerDissipated} and get (first shown in~\cite{LiuPRD00ThermoelasticNoiseb})
\begin{align}
    W_{\text{diss}} = \frac{\kappa T}{2}\left|\frac{\alpha Y}{C_V (1-2\nu)}\right|^2\int|\vec{\nabla} \Theta|^2 dV.
\end{align}

In this work, the expansion scalar $\Theta$ contains the full physics of the elastic wave equation without the quasistatic approximation. We therefore extend previous STE work beyond the quasistatic limit.

In this paper, we do not calculate the elastic fields generated from thermal excitations; see the last term in~\cite{FejerPRD04ThermoelasticDissipation} Eq. (C2). While in principle the thermal and elastic fields are coupled, we can calculate the thermoelastic noise by only considering the thermal fields from elastic excitations. We can neglect the thermally generated elastic fields as the thermoelastic loss angle, $\phi \lesssim Y_0 \alpha^2 T/C_V$~\cite{LifshitzPRB00ThermoelasticDamping,FejerPRD04ThermoelasticDissipation}, is very small compared to unity. This loss angle determines the round-trip coupling from the wave equation to the heat equation back to the wave equation. By not solving the fully coupled PDEs, we only introduce error on the order of the small thermoelastic loss angle.

To calculate the substrate thermoelastic noise, we express the volume integral with axially symmetric Hankel-transformed cylindrical coordinates, specifically
\begin{align}
    W_{\text{diss}} = \frac{\kappa T}{2}\left|\frac{\alpha Y}{C_V (1-2\nu)}\right|^2 2\pi\int_{-h/2}^{h/2}\int_0^\infty|\vec{\nabla} \tilde{\Theta}|^2 k\,dk\,dz.
\end{align}
We seek to express the gradient of the expansion scalar in terms of quantities we have already calculated (in the main text and \cref{app:SMN Details}). In calculating $\tilde{\Theta} = \vec{\nabla} \cdot \tilde{\bm{u}}$, we note $\tilde{\bm{u}} = \nabla \tilde{\Phi} + \nabla \times \tilde{\bm{\Psi}}$. Since $\tilde{\bm{\Psi}}$ is twice differentiable, $\vec{\nabla} \cdot \left(\nabla \times \tilde{\bm{\Psi}}\right) = 0$. We then write $\tilde{\Theta} = \vec{\nabla}^2\tilde{\Phi}$. Since $\tilde{\Phi}$ obeys the longitudinal wave equation (see \cref{eq:PotentialWaveEquations}), with a short algebraic manipulation we can write $\tilde{\Theta} = -k_{\longitudinal}^2\tilde{\Phi}$. With this, we express $|\vec{\nabla} \tilde{\Theta}|^2$ as
\begin{align}
    |\vec{\nabla} \tilde{\Theta}|^2 = |k_{\longitudinal}|^4 \left(\left|\frac{\partial\tilde{\Phi}}{\partial z}\right|^2 + k^2\left|\tilde{\Phi}\right|^2\right).
\end{align}

We have already solved for $\tilde{\Phi}$ in \cref{app:SMN Details}, in particular \cref{eq:PhiPsi,eq:aleph,eq:beth}. After calculating $|\vec{\nabla} \tilde{\Theta}|^2$ in terms of $A(k)$ and $B(k)$ and performing the $z$ integral by direct computation, we obtain the power dissipated as a $k$-space integral. By defining $\bar{A}(k) = A(k)/(F_0 \tilde{I}(k))$ and $\bar{B}(k) = B(k)/(F_0 \tilde{I}(k))$, we can write the substrate thermoelastic noise spectrum (using Levin's FDT \cref{eq:LevinFDT}) as
\begin{widetext}
\begin{multline}\label{eq:STE}
    S_{z,\text{STE}}(\omega) = \frac{k_{\mathrm{B}} T^2 \kappa \rho^2 (1+\nu_0)^2 \alpha^2 \omega^2 }{\pi C_V^2 (1-\nu_0)^2} \int_0^\infty\bigg(\left(|\bar{A}(k)|^2 + |\bar{B}(k)|^2 \right)\frac{k^2+|k_{z,\longitudinal}|^2}{\mathrm{Re}\{k_{z,\longitudinal}\}}\sinh(\mathrm{Re}\{k_{z,\longitudinal}\}h) +\\ \left(|\bar{A}(k)|^2 - |\bar{B}(k)|^2 \right)\frac{k^2-|k_{z,\longitudinal}|^2}{\mathrm{Im}\{k_{z,\longitudinal}\}}\sin(\mathrm{Im}\{k_{z,\longitudinal}\}h)\bigg)ke^{-k^2w^2/4}\,dk
\end{multline}
\end{widetext}

In the quasistatic (small-$\omega$), infinite-$h$ limit, we find that the second term in the integrand of \cref{eq:STE} vanishes and that the resulting spectrum matches the literature result (Eq. (12) in~\cite{BraginskyPLA99ThermodynamicalFluctuationsa} and Eq. (18) in \cite{LiuPRD00ThermoelasticNoiseb}). For a noise spectrum in the regime such that $r_\thermal\ll w$ is no longer a valid approximation, see~\cite{CerdonioPRD01ThermoelasticEffects}.

We approximate the STE spectrum in the same limits as in \cref{subsubsec:SMN_Analytic}. In these limits, away from the mirror mechanical resonances, the second term in the integrand of \cref{eq:STE} is negligible, and we find that the high-frequency approximation for the STE noise is 
\begin{align}\label{eq:STE_HFA}
    S_{z,\text{STE, HFA}}(\omega) = \frac{2k_{\mathrm{B}} T^2 \kappa (1+\nu_0)^2\alpha^2 h}{\pi C_V^2 (1-\nu_0)^2 |v_\longitudinal|^2 w^2 \sin^2(\omega h/|v_\longitudinal|)}.
\end{align}
This reaches a minimum value for $\omega = \pi(m_\longitudinal+1/2)|v_\longitudinal|/h$, specifically
\begin{align}\label{eq:STE_Min}
    S_{z,\text{STE, min}}(\omega) = \frac{2k_{\mathrm{B}} T^2 \kappa (1+\nu_0)^2\alpha^2 h}{\pi C_V^2 (1-\nu_0)^2 |v_\longitudinal|^2 w^2}.
\end{align}

We note that this flat PSD matches the predictions from~\cite{LifshitzPRB00ThermoelasticDamping}, specifically an effective loss angle that is linear in frequency below the inverse thermal relaxation time (typically in the many GHz to THz regime for commonly used optical substrates at room temperature). Above this frequency, which roughly corresponds to the thermal diffusion length being equal to the inverse elastic wavenumber, the STE noise minima will decrease.

Similarly to the mechanical noise, we could approximate the integrand of \cref{eq:STE} to find a more accurate high-frequency approximation. However, as the STE noise is subdominant to mechanical noise for commonly used interferometer substrate materials like fused silica and silicon, we choose to end the modeling here. 

We note that thermoelastic loss can be thought of as contributing to the (frequency-dependent) mechanical loss angle rather than as a separate noise source. We choose to consider thermoelastic noise separately from mechanical noise. Mechanical noise is typically dominated by structural damping~\cite{HarryAOA06ThermalNoise}, which is why these noises are treated independently. There is a risk of double-counting the noise if thermoelastic loss were the dominant loss mechanism and were used as the loss angle for mechanical noise.

\begin{figure*}[t]
\includegraphics[width=1\linewidth]{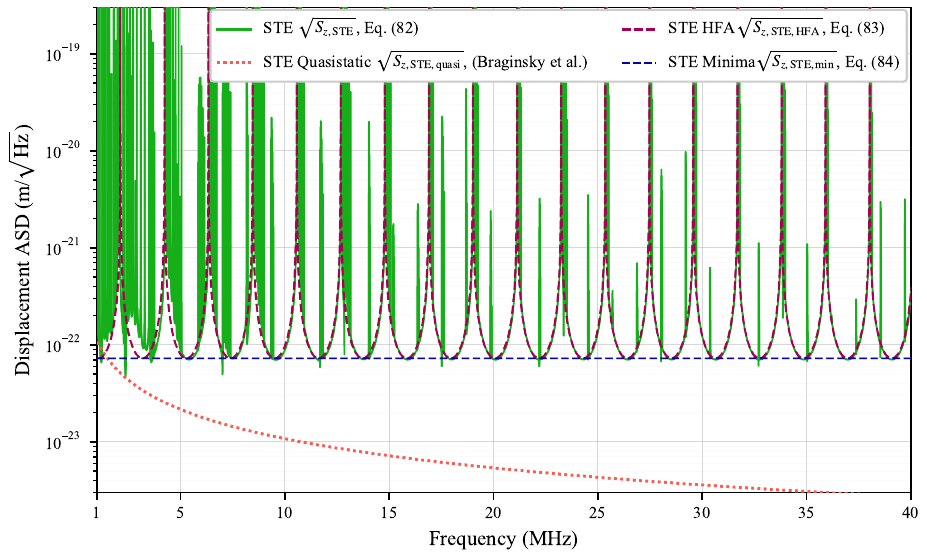}
  \caption{The Modeled Amplitude Spectral Density of the Substrate Thermoelastic Noise (STE) for the GQuEST End Mirrors. The STE curve is given by the cylindrical version of \cref{eq:STE}. The substrate thermoelastic noise high-frequency approximation is given by \cref{eq:STE_HFA}. The substrate thermoelastic noise minima are given by \cref{eq:STE_Min}. The quasistatic, infinite-half-space approximation from the literature~\cite{BraginskyPLA99ThermodynamicalFluctuationsa, LiuPRD00ThermoelasticNoiseb} (both references give identical spectra) is also given. A sample width of $\Delta f_{\text{sample}} = 3.9$~Hz was used.}
  \label{fig:STE}
\end{figure*}

We plot the GQuEST substrate thermoelastic noise ASD in \cref{fig:STE}.

\section{Coating Thermo-Optic Noise}\label{sec:CTONoise}

\begin{figure}[t]
\includegraphics[width=\linewidth]{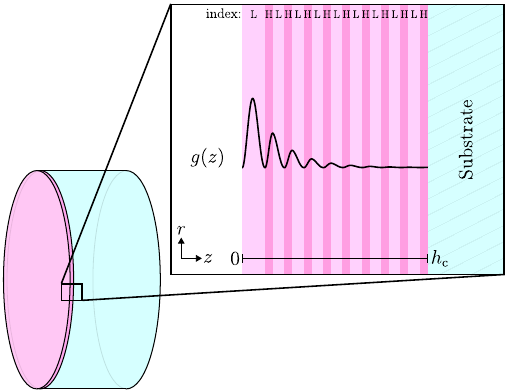}
  \caption{An example of a Bragg coating with alternating high-index and low-index layers. The first layer on the left is a low-index material cap whose purpose is to minimize the electric field on the front face of the optic. This minimizes the effects of surface defects on the light. Within the coating is a modified version of the intensity function, $g(z)$, presented in \cref{eq:g(z)}, which was modified in the figure to accurately stretch and contract as the index of refraction varies across layers. The index of the layer is labeled at the top with an $\mathrm{H}$ denoting a high-index and an $\mathrm{L}$ denoting a low-index material. This example coating's thickness is exaggerated in the 3-dimensional drawing on the left, and the number of layers is reduced for simplicity. Please note that for this section, the front surface of the mirror is defined as $z=0$ as shown in this figure; this figure uses the convention of \cref{sec:CTONoise}, where the zero of $z$ is redefined to be on the face of the coating, breaking from the convention of previous sections. The $r$-axis denotes the radial direction, measured outward from the origin/center.}
  \label{fig:bragg_layers}
\end{figure}

\subsection{Background}
We now move to a different source of classical noise in Michelson interferometers, coating thermo-optic (CTO) noise. Thermal fluctuations couple to the phase of the interferometer's light by affecting the optical path length through thermal expansion and thermorefraction, i.e., an index of refraction that is temperature dependent. Previous analysis of coating thermo-optic noise indicates that it would be a dominant noise source in the GQuEST experiment~\cite{EvansPRD08ThermoopticNoise,VermeulenPRX25PhotonCountingInterferometry}. More precise modeling presented in this paper indicates that it is subdominant to mechanical noise in the coating.

Specifically, we will consider thermal fluctuations of a thin (smaller than the elastic wavelength) coating on a thick (much larger than the thermal diffusion length) substrate with different material properties from each other. The coupling of temperature to thermal expansion is given by the coefficient of thermal expansion, $\alpha$. The coupling of temperature to the index of refraction is given by the thermorefractive coefficient, $\beta = \partial n/\partial T$.

Thermal expansion couples to the phase of the light in the interferometer arm by changing the length that the light travels. Positive thermal expansion in the end mirror decreases the length that the light travels from the beamsplitter to the end mirror, decreasing the phase accumulated by the light.

Thermorefraction couples to the phase of the light in an interferometer arm by changing the index of refraction of the light when traveling through the outermost part of the coating. An increase in the refractive index of the coating increases the phase accumulated by the light.

The key length scale for this noise is the thermal diffusion length, $r_\thermal(f)$, i.e., the characteristic distance over which a temperature oscillation at frequency $f$ diffuses. This length scale establishes what other length scales need to be considered. If something has a length scale much smaller than the thermal diffusion length, for example, the different coating layers at low frequencies, then their material properties can be averaged, and less careful consideration of their structure is needed. If something has a length scale much larger than the thermal diffusion length, like the beam spot size at high frequencies, then noise from fluctuations will be reduced because they will average out. The two noise processes we are considering, thermoelastic noise and thermorefractive noise, occur on different length scales and can therefore be considered as independent (and thus uncorrelated) noises when the thermal diffusion length is shorter than their length scales, namely $h_\coating$ for coating thermoelastic (CTE) noise and the beam penetration depth $\bar{\lambda}$ for coating thermorefractive (CTR) noise. Because the thermal diffusion length is no longer the smallest length scale in the problem, we cannot use the adiabatic approximation as we did in \cref{sec:STENoise}.

There are many length scales that affect the behavior of the coating thermo-optic noise spectrum. Roughly in order of smallest to largest, they are the inverse wavenumber of the laser beam standing wave in the coating, $1/(2k_{\text{laser}})$, which is of the same order as the individual coating layer thickness; the $1/e$ beam intensity penetration depth, $\bar{\lambda}$; the coating thickness, $h_{\coating}$; the Gaussian beam spot radius ($1/e^2$ in intensity), $w$; the mirror thickness, $h$; and the mirror radius, $a$. In solving the heat equation, we will ignore the mirror's finite dimensions, treating them as infinite where convenient and appropriate. For thermorefractive noise, this choice is fine except for very low frequencies when the thermal diffusion length is on the order of or larger than the mirror parameters. See~\cite{SomiyaPRD09CoatingThermal} for an in-depth discussion of how the mirror thickness and radius affect the noise spectrum at low frequencies, and see \cref{app:Tables} for values of these length scales. As seen in \cref{sec:STENoise}, the mirror's finite dimensions have a large impact on the thermoelastic noise at high frequencies as the quasistatic approximation fails. We ignore some aspects of the dielectric coating stack and treat the coating as a material with average properties for ease of calculation. This will miss out on some physics, especially the coating thermoelastic noise at frequencies such that the thermal diffusion length is of a similar scale to the individual coating layer thicknesses.

Previous work considered the regime $\bar{\lambda} \ll r_\thermal$~\cite{BraginskyPLA00ThermorefractiveNoisea,EvansPRD08ThermoopticNoise,SomiyaPRD09CoatingThermal}. This work relaxes the assumption that $\bar{\lambda} \ll r_\thermal$ and allows for $r_\thermal \ll 1/(2k_{\text{laser}})$. For simplicity, we will assume $r_\thermal \ll w$ because of the vastly different frequency scales to which $\bar{\lambda}$ and $w$ correspond.

Unlike earlier in the paper (\cref{sec:MechNoise,sec:STENoise,fig:Mirror1}) in which the front surface was defined at $z=h/2$ to utilize symmetry, the front surface in this section is defined as $z = 0$ and the back surface at $z=h$. We do not require the specific form of any elastic quantities beyond the coating thickness, so we make this more natural coordinate choice for this problem.

To solve for the CTO noise, we will use the FDT (\cref{eq: fdt}), Levin's insight (\cref{eq:LevinFDT}), and the power dissipated from a thermal gradient (\cref{eq:ThermalPowerDissipated}). We will find the temperature field and then its gradient by solving the heat equation with driving terms related to the noise source.

\subsubsection{Setting up the Thermoelastic Power Injection}
To find the thermoelastic component of CTO noise, we cannot use the adiabatic approximation because the thermal diffusion length is no longer the smallest length scale in the problem; the inverse wavenumber of the laser beam standing wave is smaller than $r_\thermal$ until the frequency is around 10-30~MHz (depending on the laser beam wavelength). We instead find the fictitious thermoelastic power injected by solving the elastic wave equation with the usual periodic pressure on the front surface (see \cref{eq:periodic_drive}) and noting the fictitious power injected per unit volume $P_j/\delta V$ (analogous to the fictitious force in Levin's formalism) is~\cite{LifshitzPRB00ThermoelasticDamping,FejerPRD04ThermoelasticDissipation,SomiyaPRD09CoatingThermal}
\begin{align}\label{eq:Power_from_Expansion}
    \frac{P_j}{\delta V} = \omega \frac{\alpha_j |Y_j| T}{1-2|\nu_j|}\Theta_j.
\end{align}
Here, $j=c$ and $j=s$ correspond to the coating (itself composed of two materials) and substrate, respectively. This equation appears in the literature (in the time domain, i.e., with a time derivative instead of $\omega$; see~\cite{FejerPRD04ThermoelasticDissipation,LifshitzPRB00ThermoelasticDamping}).\footnote{Under the adiabatic approximation, this equation and the heat equation, \cref{eq:Heat_Equation}, reduce immediately to \cref{eq:TemperatureField}.} Because thermal expansion in both the coating and substrate couples to the laser's phase, we have a power injection term in the coating and substrate and must solve the heat equation in the coating and substrate.

Unlike previous work~\cite{EvansPRD08ThermoopticNoise}, the power injection will have a nontrivial frequency dependence based on the mechanics of the mirror because the expansion scalar contains these dynamics. In fact, we have already solved for the expansion scalar in the coating in \cref{sec:CMN} in $k$-space as
\begin{align}
    \tilde{\Theta}_\coating = \frac{1+2\mu_c k \xi_{\substrate,r}(k)}{\lambda_\coating + 2\mu_\coating}F_0\tilde{I}(k)\sin(\omega t).
\end{align}
We then $k$-space transform the power injection (\cref{eq:Power_from_Expansion}), plug the expansion scalar into it, and re-express the elastic moduli. This yields
\begin{align}\label{eq:CTE_Power_injection}
    \frac{\tilde{P}_\coating}{\delta V} = \omega \tilde{I}(k)TF_0\sin(\omega t)\frac{1+|\nu_\coating|}{1-|\nu_\coating|}(1+2\mu_\coating k\xi_{\substrate,r}(k))\alpha_\coating.
\end{align}
In order to easily carry around this term for calculations, we define a new dimensionless function in $k$-space,
\begin{align}
    \varrho_j(k) = \frac{1+|\nu_j|}{1-|\nu_j|}(1+2\mu_j k\xi_{\substrate,r}(k)).
\end{align}
Note that the substrate subscript on $\xi_r$ is correct as $\xi_r$ is identical in the substrate and coating near the front surface. We emphasize that the dynamics of $\xi_r$ are given by solving the wave equation in the substrate alone.

By transforming $\tilde{I}(k)$ back into $I(r)$, we recover most of the coating thermoelastic power injection formula from~\cite{EvansPRD08ThermoopticNoise}. We note that in the quasistatic, infinite-$h$ limit, our $\varrho_{j,\text{quasistatic}}(k) \alpha_j$ is equivalent to their $\bar{\alpha}_j$ (see their Eq. (A1)).

The additional factor of $\omega$ in \cref{eq:CTE_Power_injection} compared to~\cite{EvansPRD08ThermoopticNoise} also appears in~\cite{SomiyaPRD09CoatingThermal} and comes from this being a power injection, not energy. We thus get a factor of $\omega$ from taking a time derivative. This deviation from~\cite{EvansPRD08ThermoopticNoise} allows us to use a consistent definition of \cref{eq:LevinFDT} throughout this paper (contrast with~\cite{EvansPRD08ThermoopticNoise} Eq. (9)). Since we are interested in the magnitude squared of the temperature field, we ignore the factor of $i$ present in the drive term of~\cite{SomiyaPRD09CoatingThermal}.

One should note that we have not subtracted off the substrate contribution to the power injection like in~\cite{EvansPRD08ThermoopticNoise}. We will see how this term arises naturally from the continuity of temperature at $z=h_{\coating}$. We are left with
\begin{align}
    \frac{\tilde{P}_\coating}{\delta V} = \omega \tilde{I}(k)TF_0\sin(\omega t)\varrho_\coating(k)\alpha_\coating.
\end{align}
The temperature profile $\delta T$ affects the expansion scalar, and one would, in theory, have to solve both together as coupled PDEs, but the coupling is governed by the thermoelastic loss angle, $\phi \lesssim Y_0 \alpha^2 T/C_V$~\cite{LifshitzPRB00ThermoelasticDamping,FejerPRD04ThermoelasticDissipation}, which in practice is very small compared to unity.

We lastly need to set up the substrate power injection. The direct impact of the substrate thermoelastic noise was calculated in \cref{sec:STENoise}. However, we cannot ignore the substrate thermoelastic coefficient in this section, as it would give us inaccurate boundary conditions at the coating-substrate interface. By Hankel-transforming \cref{eq:Power_from_Expansion} and noting $\tilde{\Theta} = -k_{\longitudinal}^2\tilde{\Phi}$, we can write
\begin{equation}\label{eq:Substrate_Power_injection}
\resizebox{\columnwidth}{!}{$\displaystyle
    \frac{\tilde{P}_\substrate}{\delta V} = -\omega k_{\longitudinal}^2 \frac{\alpha_\substrate |Y_\substrate| T}{1-2|\nu_\substrate|}\left(A(k)\cosh(k_{z,\longitudinal} z) + B(k)\sinh(k_{z,\longitudinal} z)\right).$}
\end{equation}

\subsubsection{Setting up the Thermorefractive Power Injection}
So far, we have just introduced half of the picture for coating thermo-optic power injection. We now set up the thermorefractive term. Since there are no elastic mechanics that correspond to the index of refraction's temperature sensitivity $\beta$ (we ignore all coating photoelastic effects in this paper), the thermorefractive power injection is
\begin{align}
    \frac{\tilde{P}_\coating}{\delta V} = -\omega \tilde{I}(k) TF_0 \sin(\omega t)  g(z) \bar{\beta}\lambda_{\text{laser}},
\end{align}
where $g(z)$ is the laser's relative intensity as a function of the depth $z$ (what we call the ``Bragg profile" and is specified in \cref{eq:g(z)} but we keep it general here), $\bar{\beta}$ is the effective thermorefractive coefficient (see Eq. (B20) in~\cite{EvansPRD08ThermoopticNoise}), and $\lambda_{\text{laser}}$ is the laser wavelength in vacuum. Since the beam does not penetrate into the substrate (we are only modeling high-reflectivity mirrors in this section, not the beamsplitter), the thermorefractive power injection in the substrate is zero.

The laser's relative intensity, $g(z)$, is proportional to the beam intensity within the coating. It is a generalization of the delta function in~\cite{EvansPRD08ThermoopticNoise} and has units of inverse length. We multiply the thermorefractive term by $g(z)$ because fluctuations in the index of refraction only couple to the phase of the light proportionally to its intensity. The thermoelastic term, however, does not get this factor because any thermal expansion changes the location of the mirror's front surface and therefore the phase of the light.

The negative sign of this power injection compared to the thermoelastic power injection is important for frequencies such that the thermal diffusion length is not smaller than the coating thickness. The minus sign allows for coherent cancellation of low-frequency thermo-optic noise as discussed above. 

\subsubsection{Setting up the Heat Equation}

We now need to find the thermal gradient, $\vec{\nabla} \delta T$, arising from the fictitious power injection. To do this, we use the $k$-space-transformed heat equation with the source terms related to our noise process. Because we are considering frequencies such that the thermal diffusion length is much smaller than the transverse beam size $w$, we will ignore heat flow (and thus spatial derivatives) in axes transverse to the $z$-axis ($r$, the radial axis transverse to $z$, and $\varphi$, the azimuthal axis). We write the heat equation~\cite{EvansPRD08ThermoopticNoise} as
\begin{align}\label{eq:Heat_Equation}
    C_V\frac{\partial \widetilde{\delta T}}{\partial t} = \kappa\frac{\partial^2 \widetilde{\delta T}}{\partial z^2} + \frac{\tilde{P}}{\delta V},
\end{align}
where $\widetilde{\delta T}$ is the Hankel-transformed temperature field. We will solve this equation in both the coating and the substrate, with material parameters taking their local values. The power injected per unit volume for the substrate has already been given as \cref{eq:Substrate_Power_injection}. The coating power injected per unit volume is the sum of the CTE and CTR power injection terms, specifically
\begin{align}\label{eq:Power_injection}
    \frac{\tilde{P}_\coating}{\delta V} = \omega \tilde{I}(k) TF_0 \sin(\omega t)\left(\varrho_\coating(k)\alpha_\coating - g(z) \bar{\beta}\lambda_{\text{laser}} \right).
\end{align}

\Cref{eq:Power_injection} (and similarly \cref{eq:Substrate_Power_injection}) can be broken up into 4 parts, each addressing one of Levin's requirements for the fictitious injection. $\tilde{I}(k)$ is the laser beam spatial profile (in $k$-space), $T F_0$ is the amplitude (a factor of temperature is present because this is a noise process from temperature fluctuations), $\sin(\omega t)$ adds the periodicity at the (angular) frequency of interest, and finally the term in parentheses is the coupling between the thermal fluctuations and the length fluctuations.

Following the standard method for solving differential equations with time-periodic boundary conditions, we remove the time dependence and make the following separation of variables
\begin{align}
    \widetilde{\delta T}(k,z,t) = \frac{TF_0}{C_\coating} \tilde{I}(k)\mathrm{Re}\{e^{i\omega t} \theta(k,z)\},
\end{align}
where $C_\coating$ is the average volumetric heat capacity in the coating and $\mathrm{Re}\{\theta(k,z)\}$ is the real part of a complex-valued function, $\theta(k,z)$, which itself has units of $\mathrm{m}^{-2}\mathrm{K}^{-1}$ (and nothing to do with an angle). Please note the different (but similar) definitions for $\theta(k,z)$ here and in~\cite{EvansPRD08ThermoopticNoise, SomiyaPRD09CoatingThermal,FejerPRD04ThermoelasticDissipation}. After performing some algebraic manipulation on \cref{eq:Heat_Equation}, we then get the following ODEs in the coating and substrate, respectively:
\begin{align}\label{CoatingODE}
    \theta_\coating(k,z) - \frac{1}{\gamma_\coating^2} \theta_\coating''(k,z) = -\varrho_\coating(k) \alpha_\coating + g(z) \bar{\beta}\lambda_{\text{laser}}
\end{align}
\begin{multline}\label{SubstrateODE}
    \theta_\substrate(k,z) - \frac{1}{\gamma_\substrate^2} \theta_\substrate''(k,z) = \\
    \frac{C_\coating}{C_\substrate}\frac{k_{\longitudinal}^2|Y_\substrate|}{1-2|\nu_\substrate|}\left(\bar{A}(k)\cosh(k_{z,\longitudinal} z) + \bar{B}(k)\sinh(k_{z,\longitudinal} z)\right) \alpha_\substrate,
\end{multline}
where $C_\substrate $ is the volumetric heat capacity in the substrate, and we have defined $'$ as a spatial derivative in $z$ and
\begin{align}
    \gamma_j \equiv \sqrt{\frac{i\omega C_j}{\kappa_j}} = \frac{\sqrt{i}}{r_{\thermal,j}}.
\end{align}

We are now in a position to evaluate \cref{eq:ThermalPowerDissipated}. Using the separation of variables for $\widetilde{\delta T}$, we can express the power dissipated as
\begin{align}\label{eq:WdissTOwithTheta}
    W_{\text{diss}} = \frac{TF_0^2}{4\pi C_\coating^2}\int_0^\infty k e^{-k^2w^2/4}\int_0^\infty \kappa \left|\frac{\partial \theta(k,z)}{\partial z}\right|^2 dz\,dk.
\end{align}

Because the thermal diffusion length is much smaller than the mirror thickness, we take the $z$-integral upper bound to be infinity. We can then substitute the equation for the power dissipated into the fluctuation-dissipation theorem, \cref{eq:LevinFDT}, to get our coating and substrate thermo-optic (TO) PSD.
\begin{align}\label{eq:SLCTOwithTheta}
    S_{z,\,\text{TO}} = \frac{2k_{\mathrm{B}} T^2}{\pi \omega^2C_\coating^2}\int_0^\infty ke^{-k^2w^2/4}\int_0^\infty \kappa \left|\frac{\partial \theta(k,z)}{\partial z}\right|^2 dz\,\,dk.
\end{align}
If $\varrho_j(k)$ does not depend on $k$ (such as in the quasistatic limit), then the $k$-integral becomes easy, and we get
\begin{align}
    S_{z,\,\text{TO}} = \frac{4k_{\mathrm{B}} T^2}{\pi w^2 \omega^2C_\coating^2}\int_0^\infty \kappa \left|\frac{\partial \theta(z)}{\partial z}\right|^2 dz.
\end{align}
The substrate's contribution to the PSD will be easily subtracted off when applying boundary conditions, and we will be left with the CTO PSD.

To summarize what we have done in this section, we have used the FDT and Levin's method to require only the power dissipated from a fictitious force related to CTO noise to find the PSD of CTO noise. To find the power dissipated, we must find the temperature from this fictitious injection, take its spatial gradient, and integrate over the mirror's volume. To find the temperature, we solve the heat equation with the fictitious power injection. To simplify the heat equation, we only consider variations in the $z$-axis and time. We then perform a separation of variables. Thus, we have reduced this problem to a second-order linear ordinary differential equation, albeit with some complications due to the complex values in the equation. The fact that the drive term has $k$ dependence in $\varrho_j(k)$ does not affect the solving of the differential equation, as $\varrho_j(k)$ is constant in $z$ both in and near the coating; we therefore drop the $k$ argument from $\theta(k,z)$ and simply write $\theta(z)$ for clarity.

\subsection{Solving the Differential Equation}

To extend previous models to the regime $r_\thermal \sim 1/(2k_{\text{laser}}) \ll \bar{\lambda}$, the Bragg profile $g(z)$ cannot be approximated as a delta function like in~\cite{EvansPRD08ThermoopticNoise} or a decaying exponential like in~\cite{SomiyaPRD09CoatingThermal}. Instead, we assign 
\begin{equation}\label{eq:g(z)}
    g(z) = \frac{N}{\bar{\lambda}}e^{-z/\bar{\lambda}}\sin^2(k_{\text{laser}}z).
\end{equation}
The Bragg profile has this form because the Bragg coating creates a standing wave that reflects a certain amount of power at each interface between coating layers. A standing wave term of this form is also chosen to model \textit{substrate} thermorefractive noise~\cite{BenthemPRD09ThermorefractiveThermochemical}. $\sin^2(k_{\text{laser}}z)$ is chosen instead of $\cos^2(k_{\text{laser}}z)$ because the electric field is often engineered to be zero at the front face of a Bragg coating to reduce damage from dust or scratches on the surface interacting with a field maximum. The results are quite similar in any case. $\overline{\lambda}$ and $k_{\text{laser}}$ are given by
\begin{align}
    \overline{\lambda} &= \frac{\lambda_{\text{laser}}}{8 \ln(n_{\text{H}}/n_{\text{L}})}\left( \frac{1}{n_{\text{L}}} + \frac{1}{n_{\text{H}}} \right);
    &
    k_{\text{laser}} = \frac{2\pi n_{\text{avg}}}{\lambda_{\text{laser}}},
\end{align}
where $n_{\text{H}}$ and $n_{\text{L}}$ are the indices of refraction for the high-index and low-index coating layers, and $n_{\text{avg}}$ is the harmonic mean of the coating indices of refraction. $N$ is a normalization factor approximately equal to 2, defined by normalizing the integral of $g(z)$ to 1 to match the behavior of other Bragg profiles. Its exact form is
\begin{equation}
    N = \frac{2}{1-\frac{1}{1+(2k_{\text{laser}}\bar{\lambda})^2}}.
\end{equation}
We can rewrite $g(z)$ into a form such that \cref{CoatingODE} can be easily solved,
\begin{equation}
    g(z) = \frac{N}{\bar{\lambda}}e^{-z/\bar{\lambda}}\frac{1}{2}\left(1- \cos(2k_{\text{laser}}z)\right).
\end{equation}
The particular solution to \cref{CoatingODE} is
\begin{multline}
    \theta_{\coating,\particular}(z) = Ee^{-z/\bar{\lambda}} + G e^{-z/\bar{\lambda}}\cos(2k_{\text{laser}}z)+\\
    He^{-z/\bar{\lambda}}\sin(2k_{\text{laser}}z) -\varrho_\coating(k)\alpha_\coating.
\end{multline}
The coefficients $E$, $G$, and $H$ are given by  
\begin{align}
    E = -\frac{N\bar{\beta}\lambda_{\text{laser}}}{2} \frac{\gamma_\coating^2\bar{\lambda}}{1-\gamma_\coating^2\bar{\lambda}^2},
\end{align}
\begin{align}
    G = -\frac{N\bar{\beta}\lambda_{\text{laser}}}{2\bar{\lambda}} \left(\Gamma + \frac{\Lambda^2}{\Gamma}\right)^{-1},
\end{align}
and
\begin{align}
    H = -\frac{N\bar{\beta}\lambda_{\text{laser}}}{2\bar{\lambda}} \left(\Lambda + \frac{\Gamma^2}{\Lambda}\right)^{-1},
\end{align}
where $\Gamma$ and $\Lambda$ are given by
\begin{align}
    \Gamma &= 1-\frac{1}{\gamma_\coating^2\bar{\lambda}^2}+\frac{4k_{\text{laser}}^2}{\gamma_\coating^2};
    &
    \Lambda &= \frac{4k_{\text{laser}}}{\gamma_\coating^2\bar{\lambda}}.
\end{align}
For the substrate, we get
\begin{multline}
    \theta_{\substrate,\particular}(z) = \frac{1}{1-k_{z,\longitudinal}^2/\gamma_\substrate^2} \times\\ \frac{C_\coating}{C_\substrate}\frac{k_{\longitudinal}^2|Y_\substrate|}{1-2|\nu_\substrate|}\left(\bar{A}(k)\cosh(k_{z,\longitudinal} z) + \bar{B}(k)\sinh(k_{z,\longitudinal} z)\right) \alpha_\substrate.
\end{multline}
Note that the particular solutions never diverge since $1-\gamma_\coating^2\bar{\lambda}^2 \neq 0$ and $1-k_{z,\longitudinal}^2/\gamma_\substrate^2 \neq 0$ and since $\gamma_\coating^2\bar{\lambda}^2$ and $k_{z,\longitudinal}^2/\gamma_\substrate^2$ are purely imaginary and thus never equal to 1 for all frequencies.

We now look more closely at $\theta_{\substrate,\particular}(z)$. In particular, $|k_{z,\longitudinal}^2/\gamma_\substrate^2| \ll 1$ when the thermal diffusion length is smaller than both the beam spot size, which we have already assumed, and the elastic wavelength, which is easily the case below 1~GHz for commonly used substrate materials like fused silica and crystalline silicon at room temperature. We then recover the adiabatic approximation as expected. When performing our integral to calculate the noise, we ignore the particular solution in the substrate, as it has already been addressed in \cref{sec:STENoise}. For the small thermal diffusion lengths we are considering, we do not need to worry about the noise adding coherently in the coating and substrate. We cannot ignore the particular solution in the substrate for the boundary conditions. We thus express the substrate term near the coating as 
\begin{align}
    \frac{\tilde{P}_\substrate}{\delta V} = \omega \tilde{I}(k)TF_0\sin(\omega t)\varrho_\substrate(k)\alpha_\substrate
\end{align}
and get $\theta_{\substrate,\particular}(z)$ near the coating as 
\begin{align}
    \theta_{\substrate,\particular}(z) = -\frac{C_\coating}{C_\substrate}\varrho_\substrate(k)\alpha_\substrate.
\end{align}
Note that this does not have any $z$-dependence, as we are using the approximation that the elastic wavelength is much larger than the coating thickness. 

The homogeneous solutions to \cref{CoatingODE,SubstrateODE} can be expressed as 
\begin{align}
    \theta_{\coating,\homogeneous}(z) = \mathcal{A} \cosh(\gamma_\coating z) + \mathcal{B} \sinh(\gamma_\coating z)
\end{align}
and
\begin{align}
    \theta_{\substrate,\homogeneous}(z) = \mathcal{C}e^{\gamma_\substrate z} + \mathcal{D}e^{-\gamma_\substrate z},
\end{align}
respectively. Here, $\mathcal{A}$, $\mathcal{B}$, $\mathcal{C}$, and $\mathcal{D}$ are constants which will be solved for using four boundary conditions: no heat flow (i.e., $d\theta /dz = 0$) at $z = 0$ or $z = \infty$, continuity of temperature between the coating and the substrate (at $z = h_{\coating}$), and conservation of energy at this same interface.

Requiring no heat flow at $z = \infty$ sets $\mathcal{C} = 0$ (thankfully, so we can avoid future confusion with the volumetric heat capacity). Putting this together, the full solution is
\begin{multline}
    \theta_\coating(z) = 
    \mathcal{A}\cosh(\gamma_\coating z) + \mathcal{B}\sinh(\gamma_\coating z) - \varrho_\coating(k) \alpha_\coating \\
    +Ee^{-z/\bar{\lambda}} + G e^{-z/\bar{\lambda}}\cos(2k_{\text{laser}}z)+He^{-z/\bar{\lambda}}\sin(2k_{\text{laser}}z)
\end{multline}
and
\begin{align}
    \theta_\substrate(z) = \mathcal{D}e^{-\gamma_\substrate z}-\frac{C_\coating}{C_\substrate}\varrho_\substrate(k)\alpha_\substrate.
\end{align}

We solve for $\mathcal{B}$ by applying the no-heat-flow boundary condition at $z=0$, which yields
\begin{align}
    \mathcal{B} = \frac{1}{\gamma_\coating\bar{\lambda}}(E + G) - \frac{2k_{\text{laser}}}{\gamma_\coating}H.
\end{align}
The last two coefficients, $\mathcal{A}$ and $\mathcal{D}$, can be solved by applying continuity of temperature and conservation of energy. The latter is equivalent to imposing continuity of heat flow at the interface. Mathematically,
\begin{align}
    \theta_\coating(h_{\coating}) = \theta_\substrate(h_{\coating});
    &~~~~~
    \kappa_\coating\theta_\coating'(z)|_{z = h_{\coating}} = \kappa_\substrate\theta_\substrate'(z)|_{z = h_{\coating}}.
\end{align}
We find that 
\begin{equation}
\resizebox{\columnwidth}{!}{$\displaystyle
\begin{aligned}
    \mathcal{A} &= \Bigg(\varrho_\coating(k) \alpha_\coating - \frac{C_\coating}{C_\substrate}\varrho_\substrate(k) \alpha_\substrate - \mathcal{B}(\sinh(\gamma_\coating h_{\coating})+R\cosh(\gamma_\coating h_{\coating})) \\
    &\quad +e^{-h_{\coating}/\bar{\lambda}}(E + G\cos(2k_{\text{laser}}h_{\coating})+H\sin(2k_{\text{laser}}h_{\coating}))\Big(\frac{R}{\bar{\lambda}\gamma_\coating}-1\Big) \\
    &\quad -\frac{2k_{\text{laser}}R}{\gamma_\coating}e^{-h_{\coating}/\bar{\lambda}}(H\cos(2k_{\text{laser}}h_{\coating})-G\sin(2k_{\text{laser}}h_{\coating})) \Bigg) \Bigg/ \\
    &\quad \Big(\cosh(\gamma_\coating h_{\coating})+R\sinh(\gamma_\coating h_{\coating})\Big)
\end{aligned}
$}
\end{equation}
and
\begin{multline}
    \mathcal{D} = 
    e^{\gamma_\substrate h_{\coating}}\Big(- \varrho_\coating(k) \alpha_\coating + \frac{C_\coating}{C_\substrate}\varrho_\substrate(k) \alpha_\substrate \\
    +\mathcal{A}\cosh(\gamma_\coating h_{\coating}) + \mathcal{B}\sinh(\gamma_\coating h_{\coating}) \\
    + e^{-h_{\coating}/\bar{\lambda}}(E + G\cos(2k_{\text{laser}}h_{\coating})+H\sin(2k_{\text{laser}}h_{\coating}))\Big).
\end{multline}
$R$ has the same definition as in other papers~\cite{FejerPRD04ThermoelasticDissipation,EvansPRD08ThermoopticNoise,SomiyaPRD09CoatingThermal}:
\begin{align}
    R = \sqrt{\frac{\kappa_\coating C_\coating}{\kappa_\substrate C_\substrate}}.
\end{align}

Finally, we have recovered the subtraction of the substrate contribution to the thermoelastic component. In the quasistatic limit, $\varrho_\coating(k) \alpha_\coating - C_\coating\varrho_\substrate(k) \alpha_\substrate/C_\substrate = \Delta \bar{\alpha}$ from~\cite{EvansPRD08ThermoopticNoise} (their Eq. (18)).

\subsection{Results}
\begin{figure*}[t]
\includegraphics[width=1\linewidth]{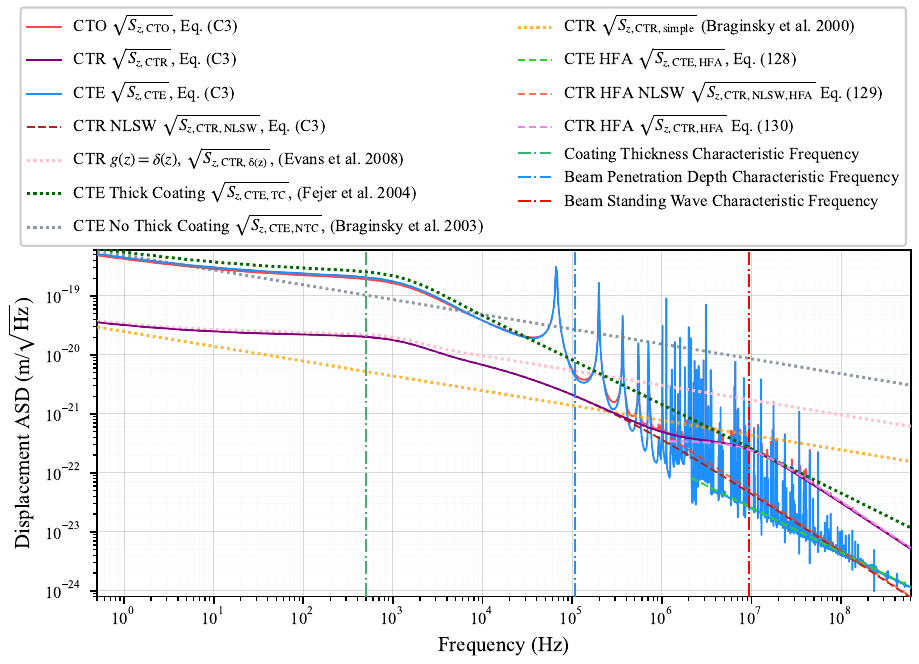}
  \caption{The Modeled Amplitude Spectral Density of the Coating Thermo-optic (CTO) Noise and Its Constituents for the GQuEST End Mirrors. Please note the comparisons to different models and limits. The red curve is from \cref{eq:CTO_Full}, and the purple CTR and blue CTE curves are also from \cref{eq:CTO_Full} but setting the CTE and CTR terms to zero, respectively. The CTO curve is behind the CTE and CTR curves for most frequencies. The No Laser Standing Wave (NLSW) term is from \cref{eq:CTO_Full} but ignoring terms that come from the laser standing wave (i.e., setting $G=H=0$ and re-normalizing $E$). This NLSW ASD is approximately equal to the full CTR ASD for frequencies such that the individual coating layer thickness is much smaller than the thermal diffusion length. We compare these to four historical references. The CTR noise model with $g(z)=\delta(z)$ is from~\cite{EvansPRD08ThermoopticNoise} and incorporates physics of a thick coating layer. The Thick Coating CTE ASD is from~\cite{FejerPRD04ThermoelasticDissipation} and the No Thick Coating CTE ASD is from~\cite{Braginsky03ThermodynamicalFluctuations}; the former model includes physics from the thermal diffusion length becoming comparable to or smaller than the coating thickness. For these two CTE models, we do not include the finite-sized mirror correction from~\cite{Braginsky03ThermodynamicalFluctuations}. The simple, thin coating CTR ASD is from~\cite{BraginskyPLA00ThermorefractiveNoisea}. We then give three high-frequency approximations: in order, they are the CTE noise floor minima (\cref{eq:CTEHFA}), the CTR noise with no standing wave (\cref{eq:CTRHFANLSW}), and CTR noise with a standing wave (\cref{eq:CTRHFA}). Substrate thermo-optic noise is not included in this model, even though it would be added coherently to CTO noise at low frequencies. Also note the characteristic frequencies that give the approximate frequency scale of the different terms. 500 geometrically distributed frequency points are used below 1~MHz, and $4\cdot10^5$ linearly distributed frequency points are used above 1~MHz, yielding $\Delta f_{\text{sample}} = 1.5$ kHz, which means the CTE spectrum is undersampled. See~\cref{fig:GQuEST} for an adequately sampled spectrum.}
  \label{fig:CTO}
\end{figure*}

Now that $\theta(z)$ is fully solved, we can calculate the full spectrum. See the mathematical details in \cref{app:CTO Details}. We first compare with previous literature results. For $\bar{\lambda}\ll r_{\thermal,\coating}$ and in the quasistatic, infinite-$h$ limit, we find full agreement with the spectrum of~\cite{EvansPRD08ThermoopticNoise}. Relaxing $\bar{\lambda}\ll r_{\thermal,\coating}$ but staying in the quasistatic limit (i.e., an infinite-half-space mirror) to compare results with Fig. 3 of~\cite{BallmerPRD15PhotothermalTransfer}, we find very good qualitative agreement between our new analytic model and that of~\cite{BallmerPRD15PhotothermalTransfer}. We make slightly different assumptions about the injection term, and we average our material properties, so we do not expect full agreement.

To easily deal with the thermoelastic noise, we choose to report the noise at high frequencies (the same limits as in \cref{subsubsec:SMN_Analytic}) and away from the eigenfrequencies of the infinite slab. At these ``valley" frequencies,
\begin{align}
    \varrho_{j, \text{valley}}(k) \approx \frac{1+|\nu_j|}{1-|\nu_j|}(1+2\mu_j\xi^{(1)}_{\substrate,r}k^2).
\end{align}
Performing the $k$-integral in \cref{eq:SLCTOwithTheta} and normalizing by $\int_0^\infty k e^{-k^2w^2/4}dk = 2/w^2$ gives
\begin{multline}
    \frac{w^2}{2}\int_0^\infty |\varrho_{j, \text{valley}}(k)| k e^{-k^2w^2/4}\,dk \approx \\
    \frac{1+|\nu_j|}{1-|\nu_j|}\left(1+\frac{8|\mu_j|\mathrm{Re}\{\xi^{(1)}_{\substrate,r}\}}{w^2}\right) = \frac{1+|\nu_j|}{1-|\nu_j|}\left(1+\frac{q_j}{2}\right),
\end{multline}
where $q_j$ is defined in \cref{eq:q_j}. At high frequencies, for commonly used substrate and coating materials and beam sizes, and away from the mirror mechanical eigenfrequencies, $q_j \approx 0$. We thus write
\begin{align}
    \varrho_{j, \text{min}}(k) \approx \frac{1+|\nu_j|}{1-|\nu_j|}.
\end{align}
We then define
\begin{align}
    \bar{\alpha}_{\coating, \text{min}} &= \sum_{j\in \text{coating}}\frac{h_{\coating,j}}{h_{\coating}}\frac{1+|\nu_j|}{1-|\nu_j|}\alpha_j,\\
    \bar{\alpha}_{\substrate, \text{min}} &= \frac{1+|\nu_\substrate|}{1-|\nu_\substrate|} \alpha_\substrate,\\
    \Delta \bar{\alpha}_{\text{min}} &= \bar{\alpha}_{\coating, \text{min}} - \frac{C_\coating}{C_\substrate}\bar{\alpha}_{\substrate, \text{min}},
\end{align}
so we see the coating thermoelastic noise is slightly smaller in the valleys than the quasistatic prediction.

This paper's focus is on frequencies such that the CTE and CTR noises have separated, so we report them separately for clarity. For completeness, we will report each spectrum with its temperature fluctuation spectrum pre-factor (see~\cite{BraginskyPLA00ThermorefractiveNoisea,LevinPLA08FluctuationDissipation,EvansPRD08ThermoopticNoise}) by itself and then the spectrum in simplified form.

For CTE noise, the spectrum at the minima is approximately
\begin{equation}\label{eq:CTEHFA}
\begin{split}
    S_{z,\text{CTE,HFA}}(\omega) &=
    \frac{
    2\sqrt{2}k_{\mathrm{B}}T^2}{\pi w^2 \sqrt{\kappa_\substrate C_\substrate \omega}
    } \frac{r_{\thermal,\coating}^2}{R(1+R)}
    \left(\Delta \bar{\alpha}_{\text{min}}\right)^2 \\
    &= \frac{
    2\sqrt{2}k_{\mathrm{B}}T^2}{\pi w^2 (1+R)}\sqrt{\frac{\kappa_\coating}{C_\coating^3\omega^3}}
    \left(\Delta \bar{\alpha}_{\text{min}}\right)^2.
\end{split}
\end{equation}

We present the CTR noise in several limiting cases. If the standing-wave contribution (the $\sin^2(k_{\text{laser}}z)$ term) were ignored, then we would get the following noise spectrum for $r_{\thermal,\coating} \ll \bar{\lambda}$,
\begin{equation}\label{eq:CTRHFANLSW}
\begin{split}
    S_{z,\text{CTR,NLSW,HFA}}(\omega) &= \frac{2\sqrt{2}k_{\mathrm{B}}T^2}{\pi w^2 \sqrt{\kappa_\substrate C_\substrate \omega}} \frac{r_{\thermal,\coating}^3}{\sqrt{2}R\bar{\lambda}^3}(\bar{\beta}\lambda_{\text{laser}})^2 \\
    &=\frac{2k_{\mathrm{B}}T^2\kappa_\coating}{\pi w^2 C_\coating^2 \bar{\lambda}^3 \omega^2}(\bar{\beta}\lambda_{\text{laser}})^2.
\end{split}
\end{equation}
This thermorefractive term was published without calculations in~\cite{VermeulenPRX25PhotonCountingInterferometry}. If we do not ignore the standing-wave contribution to the noise, then we would get the following noise spectrum for $r_{\thermal,\coating} \lesssim 1/(2k_{\text{laser}})$,
\begin{multline}\label{eq:CTRHFA}
    S_{z,\text{CTR,HFA}}(\omega) = \\ \frac{k_{\mathrm{B}}T^2\kappa_\coating N^2k_{\text{laser}}^2}{\pi w^2 C_\coating^2 \bar{\lambda}\omega^2}\frac{1}{1+(2k_{\text{laser}}r_{\thermal,\coating})^4}(\bar{\beta}\lambda_{\text{laser}})^2.
\end{multline}
The denominator of the last fraction also appears in the equation for high-frequency \emph{substrate} thermorefractive noise~\cite{BenthemPRD09ThermorefractiveThermochemical}. This equation can be further simplified by applying more specific frequency regimes. The frequency at which $r_{\thermal,\coating} \sim1/(2k_{\text{laser}})$ is approximately 10-30~MHz in $\mathrm{Ta}_2\mathrm{O}_5$-$\mathrm{SiO}_2$ coatings for near IR laser wavelengths, which is a particularly useful regime for experiments like GQuEST~\cite{VermeulenPRX25PhotonCountingInterferometry}. For $r_{\thermal,\coating} \approx 1/(2k_{\text{laser}})$, we find that
\begin{equation}\label{eq:CTRHFAMid}
\resizebox{\columnwidth}{!}{$\displaystyle
\begin{aligned}
    S_{z,\text{CTR,HFA,mid}}(\omega) &= \frac{2\sqrt{2}k_{\mathrm{B}}T^2}{\pi w^2 \sqrt{\kappa_\substrate C_\substrate \omega}}\frac{N^2}{64\sqrt{2}Rk_{\text{laser}}^2\bar{\lambda}r_{\thermal,\coating}}(\bar{\beta}\lambda_{\text{laser}})^2 \\
    &=\frac{k_{\mathrm{B}}T^2N^2}{32\pi w^2 \kappa_\coating k_{\text{laser}}^2\bar{\lambda}}(\bar{\beta}\lambda_{\text{laser}})^2.
\end{aligned}$}
\end{equation}
At even higher frequencies, i.e., for $r_{\thermal,\coating} \ll 1/(2k_{\text{laser}})$,
\begin{equation}\label{eq:CTRHFAHigh}
\begin{split}
    S_{z,\text{CTR,HFA,high}}(\omega) &=
    \frac{2\sqrt{2}k_{\mathrm{B}}T^2}{\pi w^2 \sqrt{\kappa_\substrate C_\substrate \omega}} \frac{N^2k_{\text{laser}}^2r_{\thermal,\coating}^3}{2\sqrt{2}R\bar{\lambda}}(\bar{\beta}\lambda_{\text{laser}})^2 \\
    &=
    \frac{k_{\mathrm{B}}T^2\kappa_\coating N^2k_{\text{laser}}^2}{\pi w^2 C_\coating^2 \bar{\lambda}\omega^2}(\bar{\beta}\lambda_{\text{laser}})^2.
\end{split}
\end{equation}

We plot the exact form and the approximations for the GQuEST coating thermo-optic noise ASD in \cref{fig:CTO}.

\subsection{Coating Thermo-Optic Noise Model Limitations}
A main limitation of this model is the simplification of the multi-layer stack into a single coating layer. Fejer~\cite{FejerPRD04ThermoelasticDissipation} predicts and roughly calculates ``dissipation peaks" (``Debye peaks"~\cite{LifshitzPRB00ThermoelasticDamping}) when the thermal diffusion length is equal to the coating thickness and individual coating layer thickness. For the former effect, the noise spectrum does not rise as a function of frequency, but its decay does plateau. We assume a similar plateau will occur for the latter effect.

We also do not consider how changes in each coating layer's length and index of refraction affect the phase of the transmitted light at the boundary between layers. This will change the complex reflectivity and will likely also increase the noise. The averaging process is also approximate, as it treats the layers as one material in some instances and individual layers in others.

This model also does not extend down below $\sim 4$~Hz where the (substrate) thermal diffusion length is on the order of the beam spot size, the mirror radius, and/or the mirror thickness. See~\cite{SomiyaPRD09CoatingThermal} for more on this low-frequency range. The model also does not coherently sum STE and CTO noise at low frequencies, which is relevant when the thermal diffusion length is larger than the coating thickness, i.e., frequencies below approximately 1~kHz.

\section{\ifdcc Other Noise Sources and Projections for GQuEST\else Comparison with Experimental Data\fi}\label{sec:Comparison_with_Experimental_Data}
\ifdcc In this section, we apply our models to GQuEST, an experiment under construction. \else To verify our models, we examine the data from a previous experiment, the Fermilab Holometer, which comprised a pair of 39-m-arm-length Michelson interferometers, designed to search for spacetime fluctuations~\cite{ChouCQG17HolometerInstrument}. We then model the classical noise for GQuEST, an experiment under construction. \fi The noise model for GQuEST is particularly important as GQuEST's science sensitivity depends on how low the classical noise can be engineered. For each experiment, we list the relevant experimental parameters and a global list of material parameters in \cref{app:Tables}.

\subsection{Other Noise Sources}\label{sec:Tidbits}
In a Michelson interferometer with a single beamsplitter and no arm cavities, the noise PSD from a single interaction with the beamsplitter should be multiplied by a transfer function $X(\omega)=2\cos^2(\omega L_{\text{arm}}/c)$ where $L_{\text{arm}}$ is the arm length and $c$ here is the speed of light. The factor of 2 comes from two interactions with the beamsplitter. The intuition behind the $\cos^2(\cdot)$ part of the transfer function is that the noise at frequency $\omega \approx \pi c/(2L_{\text{arm}})$ that is imparted by the beamsplitter upon reflection/transmission is out of phase with the noise imparted by the second reflection from/transmission through the beamsplitter for $t = \pi/\omega$. It takes time $t = 2L_{\text{arm}}/c$ to travel down and back through the arms, so noise around frequencies $\omega =  \pi c/(2L_{\text{arm}})$ will be canceled out. If two beamsplitters are used, say in a traveling-wave interferometer to avoid the standing-wave contribution of substrate thermorefractive  noise~\cite{BenthemPRD09ThermorefractiveThermochemical,VermeulenPRX25PhotonCountingInterferometry}, then there is no $\cos^2(\cdot)$ term as each beamsplitter's thermal noise is independent\footnote{An exception to thermal noise is seismic noise.}. Due to the interferometer DC offset and contrast defect, the transfer function $X(\omega)$ will not be perfectly zero at its nulls.

Substrate thermorefractive noise (STR)~\cite{BenthemPRD09ThermorefractiveThermochemical, HeinertPRD11ThermorefractiveNoise}, laser phase noise (LPN)~\cite{VermeulenPRX25PhotonCountingInterferometry,McCuller15TestingModel}, and substrate charge carrier refractive noise (SCC)~\cite{SiegelPRD23RevisitingThermal} have been adequately modeled into the MHz band, and we will include them in our analysis as they are not negligible compared to the noise sources modeled in this work. Thermochemical noise~\cite{BenthemPRD09ThermorefractiveThermochemical}, photo-thermal noise~\cite{BraginskyPLA99ThermodynamicalFluctuationsa, BallmerPRD15PhotothermalTransfer}, residual gas noise~\cite{Weiss89ScatteringResidual, Whitcomb97MeasurementOptical}, quantum radiation pressure noise~\cite{DanilishinLRR12QuantumMeasurement,CavesPRD81QuantummechanicalNoise}, control noise (no controller used has a MHz bandwidth), seismic and Newtonian noise (seismic and Newtonian fluctuations have a low-frequency cutoff)~\cite{SaulsonPRD84TerrestrialGravitationala,TrozzoG22SeismicNewtonian}, scattered stray light~\cite{VinetPRD96ScatteredLight}, and noise from electromagnetic fields (this should be extremely small for fused silica substrates and still quite small for silicon) have been adequately modeled into the MHz band and are negligible compared to other noise sources so we do not model them in this paper. Interferometers with DC readout, in which the same laser beam supplies the signal and noise sidebands along with the local oscillator, are insensitive to readout air diffusion noise~\cite{VermeulenPRX25PhotonCountingInterferometry} by common mode rejection, so we ignore it in our analysis here, as the Holometer used DC readout (instead of balanced homodyne)~\cite{ChouCQG17HolometerInstrument} and we restrict our analysis of GQuEST to thermal noise. We do not include thermorefringent noise~\cite{KryhinPRD23ThermorefringentNoise} under our isotropic assumption but estimate it to be subdominant to other noise sources in GQuEST. Noise from the power-recycling optics also causes noise at the interferometer output. However, this noise is suppressed by the common-mode rejection of the interferometer, which was modeled to be subdominant and thus not included.

\subsection{Holometer}\label{sec:Holometer}
We compare our models to measurement data from one of the two interferometers of the Holometer~\cite{ChouCQG17HolometerInstrument}. The Holometer interferometers used 12.7-mm-thick fused silica optics. The end mirrors had a radius of 25.4 mm, and the beamsplitter had a radius of 38.1 mm. These optics used $\mathrm{Ta}_2\mathrm{O}_5$-$\mathrm{SiO}_2$ Bragg coatings for 1064~nm laser light (see \cref{tab:Holometer} for details). We use a dataset obtained through averaging the cross-correlations of two photodiodes at the interferometer output~\cite{ChouCQG17HolometerInstrument,MartynovPRA17QuantumCorrelationa}. This technique enables quantum shot noise to be averaged away over time, revealing the interferometer classical noise. This cross-spectral density (CSD) data has appeared in Figure 6.25 and Figure 6.26 of~\cite{McCuller15TestingModel}, and the same dataset has been published in~\cite{ChouCQG17HolometerInstrument} (but only up to 10 MHz there).

\begin{figure*}[t]
\includegraphics[width=1\linewidth]{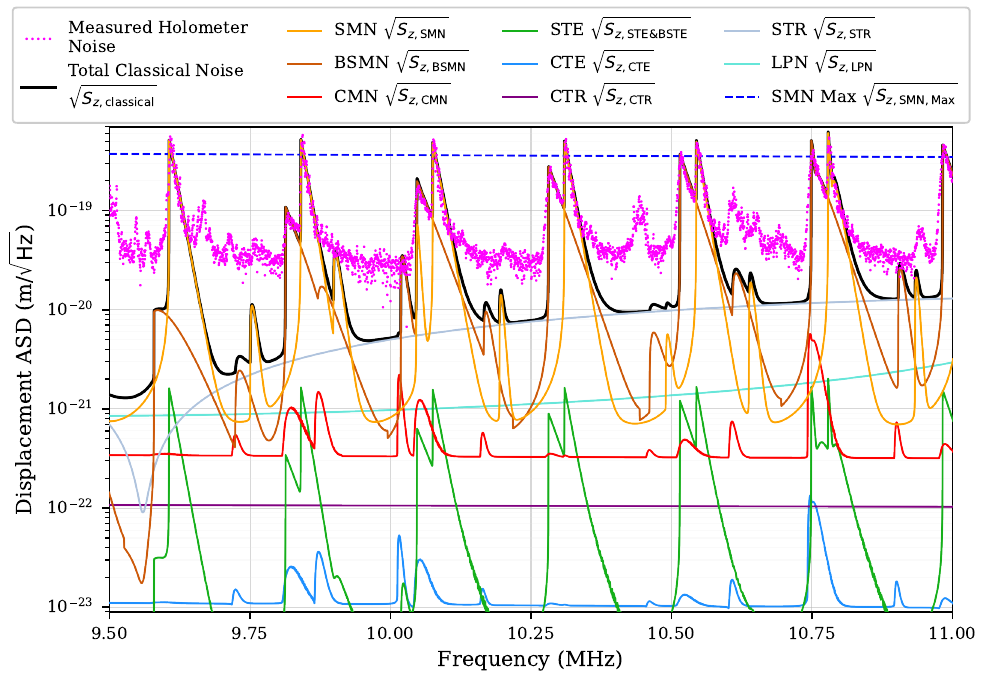}
  \caption{The Amplitude Spectral Density for the Holometer. The parameters are from \cref{tab:Holometer} and the data is from~\cite{McCuller15TestingModel}. Note that the BSTE noise has been combined with the STE noise curve. The frequency bin width is 381.5~Hz and the integration time is 145 hours. The models use this bin width as $\Delta f_{\text{sample}}$. Because of the large data bin width compared to the Lorentzian peak bandwidth, an infinite slab model is used instead of a finite cylinder since an average in frequency of a spectrum produced by the cylinder model yields the spectrum produced by the infinite slab model by the equipartition theorem. The SMN Maximum curve (\cref{eq:SMNSlabMax}) is included as a reference.}
  \label{fig:Holometer}
\end{figure*}

We compare our models to the data from this experiment in \cref{fig:Holometer}. During this data run, the single interferometer shot noise was around the level of $\sqrt{S_{z,\text{shot}}} = 2\cdot10^{-18}~\text{m}/\sqrt{\text{Hz}}$, the frequency bin width ($\Delta f_{\text{bin}}$) is 381.5~Hz, and the measurement time ($\tau$) was 145 hours, over which the uncorrelated shot noise was averaged down. The averaged down ASD, $\sqrt{S_{z,\text{avg}}}$ is given by
\begin{align}
    \sqrt{S_{z,\text{avg}}} = \mathcal{N}^{-1/4}\sqrt{S_{z,\text{shot}}},
\end{align}
where $\mathcal{N}=\Delta f_{\text{bin}} \tau$ is the number of spectra. A floor of correlated noise was found at $3\cdot 10^{-20}~\text{m}/\sqrt{\text{Hz}}$ (of unknown source, see below), while $\sqrt{S_{z,\text{avg}}} < 2\cdot 10^{-20}~\text{m}/\sqrt{\text{Hz}}$ meaning further averaging would not improve sensitivity.

The primary features that correspond between model and data are maxima in the ASD shaped like an inverse sawtooth wave (with the sharp up-ramp at lower frequencies and then a more gradual down-ramp at higher frequencies); these features are caused by the substrate mechanical noise and beamsplitter mechanical noise (BSMN). This is a confirmation of these noise models in \cref{subsec:MechNoiseSub} and \crefBSMN. The only free parameters in the model that were fit to match the data in this plot are the exact thicknesses of the mirrors, which set the location and spacing in frequency of the maxima (the fit value is well within the error margin for the measured thickness, i.e., it is within 50 $\mu$m of the nominal thickness). The beam size sets the rate of fall of the sawtooth, as it arises from the Gaussian term $\tilde{I}(k)^2$ present in \cref{eq:rcptnc_to_admtnc} and subsequent expressions. 

Based on the SMN model for the cylinder (see~\cref{eq:SMN_Spectrum_Cyl} and the discussion after it), one could initially expect to see Lorentzian peaks in the data at the cylinder mechanical eigenfrequencies. These peaks do not appear in the data in~\cref{fig:Holometer} because the frequency bin width of 381.5~Hz is much larger than the bandwidth of the mechanical mode (approximately $\phi_\substrate \cdot f_\text{mode} = 10$~Hz), and therefore the data points represent CSDs that are averages over a narrow noise peak and a floor. One can show that averaging the noise spectrum for the cylinder (\cref{eq:SMN_Spectrum_Cyl}) over a large frequency bin width recovers the noise spectrum for the infinite slab (\cref{eq:SMN_Spectrum}), i.e., the infinite slab model captures the \emph{average} spectral density from the thermal occupation of each mechanical mode of the optics. This recovery of the slab model is a consequence of the equipartition theorem, as each mode always contains $k_{\mathrm{B}}T/2$ of thermal energy, concentrated near its eigenfrequency and independent of the loss angle. The noise model chosen for comparison is therefore one that treats the Holometer optics as infinite slabs, rather than finite diameter cylinders with discrete eigenmodes. Note the excellent agreement between the data and the model at the mechanical noise peaks in~\cref{fig:Holometer}.

There is excess measured noise in the minima between the mechanical noise peaks (e.g., at 9.95~MHz in~\cref{fig:Holometer}) compared to the modeled noise. In order to estimate the substrate loss angle and eliminate SMN and BSMN as causes of the excess measured noise, we turned to another dataset (detailed below) that indicates $\phi_\substrate < 10^{-5}$, and thus we believe the Holometer mirror substrates have a loss angle that is consistent with the known substrate loss angle of fused silica of $\phi_\substrate \sim 10^{-6}$~\cite{PennPLA06FrequencySurface} in their high-frequency modes. We therefore use this literature reference value for~\cref{fig:Holometer}. This upper bound on the substrate loss angle of $\phi_\substrate < 10^{-5}$ rules out SMN and BSMN as the cause of the noise floor. The heights of the observed BSMN peaks are modulated according to the beamsplitter transfer function $X(\omega)$, while the observed noise floor is not (there is no decrease in the noise floor near 9.6~MHz), which excludes noises from the beamsplitter, like BSMN, BSTE (see \crefBSTE), and STR~\cite{BenthemPRD09ThermorefractiveThermochemical}. This noise floor is flat from 5~MHz to 12~MHz (see~\cite{ChouCQG17HolometerInstrument,McCuller15TestingModel}), which rules out coating noises as their slopes are not flat in that frequency range (despite looking flat in~\cref{fig:Holometer}). Furthermore, the coating loss angle, $\phi_c$, would have to be above 1 for CMN to match the measured amplitude, which is highly unlikely. The excess noise in the valleys is unlikely to be laser noise~\cite{McCuller15TestingModel} (although the peak at 10.44~MHz is expected to be laser noise from higher-order optical modes), and other modeled noise sources like STE noise are expected to be very subdominant. We speculate that the correlated noise floor may be due to electronic crosstalk between cables; however, the magnitude of the noise floor then implies that the two channels are isolated by 40 dB (the ratio of shot noise to the correlated noise floor), which is unexpectedly poor performance for the cables and ADCs used in the experiment. We note that the circulating laser beam in the interferometer was modeled to be the fundamental Gaussian mode. A full treatment of the laser beam with higher-order optical modes will likely predict the laser noise peak at 10.44~MHz and may explain the noise above the noise floor at 10.6~MHz.

We would like to use the plotted high-frequency cross-correlation data to measure the substrate loss angle, $\phi_\substrate$, in the MHz band. This loss angle has direct implications for the feasibility of photon-counting interferometry (see~\cref{eq:SMNMin} and~\cite{VermeulenPRX25PhotonCountingInterferometry}). While GQuEST is designed to use crystalline silicon substrates, not fused silica like the Holometer, the observed loss angle matching the material loss angle would indicate that this low material loss angle can be achieved without suspending the optics. Clamps raise the effective loss angle in the quasistatic limit, but, due to impedance mismatches, we hypothesize they can be used for MHz-band experiments without increasing the effective loss angle. However, we are unable to infer the substrate loss angle in this data set. The loss angle does not influence the shape of the sawtooth peaks in the infinite-slab model, see \cref{eq:SMNHFA}. The height of the valleys between these peaks of the SMN spectrum does depend on the loss angle, see \cref{eq:SMNHFAValley}. However, this noise valley is not observed as an unknown noise source dominates there, and hence we cannot use the data to infer the loss angle. 

Although the 381.5~Hz bin-width dataset does not allow us to extract the substrate loss angle, a 23.84~Hz bin width dataset taken over 5 minutes (not shown) does allow us to bound it. For this 5-minute dataset, the spectrum is not statistically resolved, as the shot noise level was $2\cdot10^{-18}~\text{m}/\sqrt{\text{Hz}}$ and was not fully averaged down to resolve the correlated noise floor. We are able to identify peaks (at frequency $f_{\text{peak}}$) in the MHz band which are well above the statistical noise. We found a frequency bin representing a local maximum where its neighboring frequency bins were half as large in magnitude, due to spectral leakage from windowing. This spectral leakage is exactly the feature that would be produced by a noise peak with a width smaller than the data bin width (i.e., leaving the peak unresolved in frequency) after application of a Hann window. Assuming this feature is a substrate mechanical noise peak, we claim $\phi_\substrate < \Delta f_{\text{bin}}/f_{\text{peak}} \approx 10^{-5}$  but cannot bound the substrate loss angle further. We note a failed attempt to extract the substrate loss angle from a 23.84~Hz bin width dataset taken over 6 (non-consecutive) hours (also not shown). This data is composed of the same 5-minute datasets with which we did observe Lorentzian peaks, and therefore the 6-hour dataset has a lower statistical noise level. Nevertheless, we do \emph{not} see Lorentzian peaks in this data set. Here, we suspect that on timescales larger than 5 minutes, the mirror eigenmodes that couple to the beam are changing, and therefore there are no high-Q peaks that appear in the averaged dataset. The measured frequencies of maxima may change due to the beam position on the optic drifting, causing different modes to couple to the beam, or due to changes in the optic's temperature, which changes the eigenfrequency of the modes. The latter explanation seems likely, as the mirror temperature only needs to change 0.01~K in order to change the eigenfrequency of a mode around 10~MHz by 10~Hz~\cite{IdeJG37VelocitySound,SpinnerJACS56ElasticModuli}.

One goal attempted in \cite{ChouCQG17HolometerInstrument} was to utilize the peaks in the thermal noise spectrum to confirm the Holometer instrument calibration, which was limited by the unknown beam position and modal coupling factors (see their section 6.5). \Cref{fig:Holometer} shows that the analytically determined maximum-infinite-slab noise is consistent with the Holometer data. This indicates that, while the PSD from individual mechanical modes may depend on too many unknown parameters, the aggregation of the noise from many mechanical modes of the substrate produces a predictable average spectral density which can be used for calibration.

\subsection{GQuEST}\label{sec:GQuEST}
\begin{figure*}[t]
\includegraphics[width=1\linewidth]{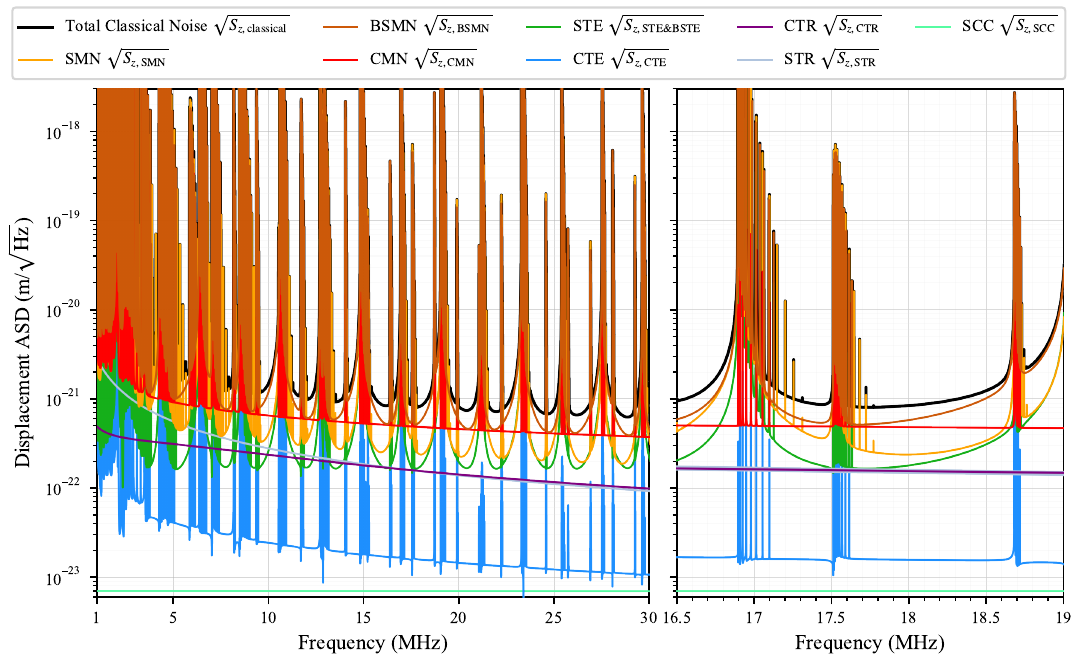}
  \caption{The Modeled Amplitude Spectral Density of Thermal Noise for the GQuEST Experiment. The parameters are from \cref{tab:GQuEST}. Note that the BSTE noise has been combined with the STE noise curve. We use the STR noise from~\cite{BenthemPRD09ThermorefractiveThermochemical} (drawn thicker in the right panel for visibility behind the CTR curve) and the SCC noise from~\cite{SiegelPRD23RevisitingThermal}. We use $\Delta f_{\text{sample}} = 5$~Hz.}
  \label{fig:GQuEST}
\end{figure*}

We model the under-construction GQuEST experiment using parameters from~\cite{VermeulenPRX25PhotonCountingInterferometry}. GQuEST has been specifically engineered to have low classical noise in frequency bands centered around 18~MHz; see \cref{fig:GQuEST}. The amplitude of beamsplitter mechanical noise (see \crefBSMN) is approximately equal to that of the coating mechanical noise at these frequencies, while substrate mechanical noise is slightly subdominant. GQuEST uses a traveling-wave configuration to avoid the standing wave in the beamsplitter; the presence of the standing wave causes a large increase in STR noise~\cite{BenthemPRD09ThermorefractiveThermochemical}. The total projected noise at 18~MHz is $8\cdot10^{-22}~\text{m}/\sqrt{\text{Hz}}$. We do not include laser phase noise or readout air diffusion noise in our analysis of GQuEST, but they are projected to be subdominant to thermal noise~\cite{VermeulenPRX25PhotonCountingInterferometry}.

\cref{subsubsec:SMN_cylinder} converts the infinite-slab model to a cylindrical-mirror model, showing that the integrals convert to sums. We apply this conversion to other noise sources so that the plots can be presented for finite-radius mirrors.

We correct our previous model of the beamsplitter (substrate) thermo-optic noise~\cite{VermeulenPRX25PhotonCountingInterferometry} here. The fiducial GQuEST thermo-optic noise should not be subject to the beamsplitter transfer function, as we have two beamsplitters to avoid the standing-wave contribution to the substrate thermorefractive noise. The thermoelastic noise in the beamsplitter substrate should be added coherently with the thermorefractive noise, as both originate from the same thermal fluctuations. An increase in temperature reduces the phase of the reflected light and increases the phase through thermal expansion and thermal refraction of the transmitted light, so these effects add coherently. We have also updated a few material properties, so the noise model plots in this work might be slightly different from those in~\cite{VermeulenPRX25PhotonCountingInterferometry}, even if the underlying physics has remained the same.

\section{Summary \& Outlook}\label{sec:conclusion}
Detecting weak signals with a classical noise background is only feasible if the classical noise background is understood. We calculate classical thermal noises in optics up to and including the MHz band, whereas they were previously only calculated up to and including the kHz band. \ifdcc We find agreement with the established low-frequency models for all of the examined noises. \else We find the dominant classical noise for the Holometer is substrate mechanical noise, not the coating mechanical noise that dominates experiments like LIGO. \fi For GQuEST, which was designed to have frequency bands with low substrate mechanical noise, coating mechanical noise and substrate mechanical noise (including beamsplitter contributions) are the dominant noise sources. The updated total noise at 18~MHz is $8\cdot10^{-22}~\text{m}/\sqrt{\text{Hz}}$; the total classical noise estimated by applying the models of previous literature~\cite{EvansPRD08ThermoopticNoise,HongPRD13BrownianThermala, BonduPLA98ThermalNoiseb} is $1.4\cdot10^{-21}~\text{m}/\sqrt{\text{Hz}}$. At frequencies near the longitudinal eigenfrequencies of the mirror, the updated noise models predict significantly more noise than previous models that use the quasistatic approximation.

As both GQuEST and lower-frequency experiments like LIGO are limited by coating mechanical noise, it would be valuable to make mirror coatings with low mechanical loss angle in the MHz band as well as the 100~Hz band. Therefore, a broadband reduction in the coating loss angle would improve the search for both gravitational waves and quantum-gravity signals. There has already been much ongoing work to lower coating mechanical noise, for example~\cite{AbernathyPLA18BulkShear,GrasPRD18DirectMeasurement,VajentePRD20MethodExperimental,GranataCQG20AmorphousOptical,TaitPRL20DemonstrationMultimaterial,AmatoPRD21OpticalMechanical,NunezAOA23AmorphousDielectric,McGheePRL23TitaniaMixed,ChabbraPRD25MeasurementThermooptic}.

The modeling in this paper primarily focuses on the MHz band due to its applicability to GQuEST. In future work, we will extend these models into the kHz band of interferometers like LIGO, Cosmic Explorer, Einstein Telescope, and GEO600. This work approximates the eigenmodes of the cylindrical mirrors, both for ease of calculation and because single eigenmodes are difficult to measure in the MHz band. In the kHz band, the exact eigenmodes can be modeled and resolved. Cosmic Explorer will have mechanical eigenmodes in its planned sensitivity band, so the models developed here, combined with the work of Hutchinson~\cite{HutchinsonJoAM80VibrationsSolid}, will be particularly relevant to its sensitivity. Furthermore, the success of proposed techniques to avoid the quantum shot noise background~\cite{Payne25PhotonCounting} depends on the classical noise, making the classical noise modeling for Cosmic Explorer especially important for these ideas.

\section{Acknowledgments}
D.G. thanks Jacques Ding for correcting minor constant factors from an earlier version of this manuscript. This article was prepared using the resources of the Fermi National Accelerator Laboratory (Fermilab), a U.S. Department of Energy, Office of Science, Office of High Energy Physics, HEP User Facility. Fermilab is managed by Fermi Research Alliance, LLC (FRA), acting under Contract No. DE-AC02-07CH11359. The GQuEST project is funded in part by the Heising-Simons Foundation through Grant No. 2025-5992. This paper has been assigned LIGO DCC number LIGO-P2600194.




\appendix
\renewcommand{\thesubsection}{\Alph{section}.\arabic{subsection}}
\section{SMN Calculation Details}\label{app:SMN Details}
\subsection{Main Calculation Details}
In this section, we present the details of the substrate mechanical noise calculation. We seek to get a closed form for $\tilde{u}_z(z=h/2,k,\omega)$ starting from the elastic wave equation in $z$, $r$, and $t$ coordinates,
\begin{align}
    \rho \frac{\partial^2 \bm{u}}{\partial t^2} = (\lambda + 2\mu)\nabla (\nabla \cdot \bm{u}) - \mu \nabla \times (\nabla \times \bm{u}),
\end{align}
where $\rho$ is the substrate density, $\bm{u}$ is the displacement vector (that specifies the displacement of a point relative to its stationary position), $t$ is time, and $\lambda$ and $\mu$ are the two Lam\'e parameters~\cite{Bedford94IntroductionElastic}. We are in two spatial dimensions ($z$ and $r$), so we will have four boundary conditions for this second-order differential equation. We express the boundary conditions as a stress $\sigma$, with units of pressure, as
\begin{align}
    \sigma_{ij} = \lambda(\nabla\cdot \bm{u})\delta_{ij} + \mu \left(\frac{\partial u_i}{\partial x_j} + \frac{\partial u_j}{\partial x_i}\right), 
\end{align}
where $i, j \in \{z,r\}$ (some orthogonal spatial coordinate generally), $\delta_{ij}$ is the Kronecker delta, and $x_i$ is some spatial coordinate, here $x_i \in \{z,r\}$. Our front surface boundary condition will be that the stress $\sigma_{zz}(z = h/2, r)$ equals the incoming pressure, $F_0I(r)$.\footnote{Note that there is a sign convention with the pressure. We follow the convention of~\cite{LiuPRD00ThermoelasticNoiseb, SomiyaPRD09CoatingThermal} such that the pressure is going away from the surface when solving the time-independent wave equation. Of course, all spectra will agree regardless of convention; the key is consistency.} Our other three boundary conditions are $\sigma_{rz}(z = h/2, r) = \sigma_{zz}(z = -h/2, r) = \sigma_{rz}(z = -h/2, r) = 0$, so-called stress-free boundary conditions. As this problem is an infinite slab, there are no further boundary conditions, as no more surfaces exist. 

Using the Helmholtz decomposition, we can rewrite $\bm{u}$ as
\begin{align}\label{eq: displacement vector}
    \bm{u} = \nabla \Phi + \nabla \times \bm{\Psi}
\end{align}
where $\Phi$ is the scalar potential and $\bm{\Psi}$ is the vector potential. This allows us to write the elastic wave equation as 
\begin{align}
    \nabla\left[\frac{\partial^2 \Phi}{\partial t^2} - \frac{\lambda + 2\mu}{\rho}\nabla^2 \Phi\right] + \nabla \times \left[\frac{\partial^2 \bm{\Psi}}{\partial t^2} - \frac{\mu}{\rho}\nabla^2 \bm{\Psi}\right] = 0.
\end{align}
where $\Phi$ and $\bm{\Psi}$ are functions of $z$, $r$, and $t$. We have multiple wave equations equal to zero (a gauge choice),
\begin{align}
    \frac{\partial^2 \Phi}{\partial t^2} = \frac{\lambda + 2\mu}{\rho}\nabla^2 \Phi;~\frac{\partial^2 \bm{\Psi}}{\partial t^2} = \frac{\mu}{\rho}\nabla^2 \bm{\Psi}.
\end{align}
The first equation describes longitudinal waves, and the second describes shear waves. We write 
\begin{align}
    v_\longitudinal = \sqrt{\frac{\lambda + 2\mu}{\rho}} = \sqrt{\frac{M}{\rho}};~v_\shear = \sqrt{\frac{\mu}{\rho}},
\end{align}
where $v_\longitudinal$ and $v_\shear$ are the longitudinal and shear speeds of sound, respectively, and $M$ is the P-wave modulus. By taking a time-dependent ansatz of $\cos(\omega t)$, we can write

\begin{align}
    \nabla^2 \Phi(z, r) + \frac{\omega^2}{v_\longitudinal^2}\Phi(z,r)  = 0;~\nabla^2 \bm{\Psi}(z,r)+ \frac{\omega^2 }{v_\shear^2}\bm{\Psi}(z,r) = 0.
\end{align}
Similarly to $\bm{u}$, we wish to express $\Phi(z, r)$ and $\bm{\Psi}(z,r)$ as $\tilde{\Phi}(z, k)$ and $\bm{\tilde{\Psi}}(z,k)$ using a Hankel transformation. We then express the Laplacian in transformed cylindrical coordinates as 
\begin{align}
    \nabla^2 = \frac{\partial^2}{\partial z^2} - k^2.
\end{align}
We finally get to the easily solvable differential equations
\begin{align}\label{eq:PotentialWaveEquations}
    \frac{\partial^2\tilde{\Phi}}{\partial z^2} - k_{z,\longitudinal}^2\tilde{\Phi}  = 0;~\frac{\partial^2\bm{\tilde{\Psi}}}{\partial z^2} - k_{z,\shear}^2\bm{\tilde{\Psi}} = 0,
\end{align}

where 
\begin{align}
    k_{z,\longitudinal} = \sqrt{k^2 - k_\longitudinal^2};~k_{z,\shear} = \sqrt{k^2 - k_\shear^2};~k_\longitudinal=\frac{\omega}{v_\longitudinal};~k_\shear=\frac{\omega}{v_\shear}.
\end{align}
These linear ordinary differential equations are solved by
\begin{align}\label{eq:PhiPsi}
    \tilde{\Phi}(z,k) &= A(k)\cosh(k_{z,\longitudinal}z) + B(k)\sinh(k_{z,\longitudinal}z)\\
    \tilde{\Psi}_{\varphi}(z,k) &= C(k)\cosh(k_{z,\shear} z) + D(k)\sinh(k_{z,\shear} z)\\
    \tilde{\Psi}_z(z,k) &= \tilde{\Psi}_r(z,k) = 0.
\end{align}
A convenient gauge choice is made from the symmetry of the problem for $\bm{\Psi}$. For the remainder of this paper, we drop the azimuthal label $\varphi$ and write $\Psi \equiv \Psi_\varphi$. The coefficients $A(k),~B(k),~C(k)$ and $D(k)$, which depend on $k$, are found through our boundary conditions. We can express the spatially transformed boundary conditions on the potentials as
\begin{align}
    \tilde{\sigma}_{zz}(z,k) = \mu (2k^2-k_\shear^2)\tilde{\Phi} + 2\mu k \frac{d\tilde{\Psi}}{dz}
\end{align}
and
\begin{align}
    \tilde{\sigma}_{rz}(z,k) = -2\mu k \frac{d\tilde{\Phi}}{dz} - \mu(2k^2-k_\shear^2)\tilde{\Psi}.
\end{align}

The solutions to the elastic wave equation are either symmetric or anti-symmetric about $z=0$. We can then decompose our boundary conditions into two sets of boundary conditions, one symmetric and one anti-symmetric, and sum the solutions to each to arrive at our actual boundary conditions. For both sets of boundary conditions, $\tilde{\sigma}_{rz}(z = h/2) =\tilde{\sigma}_{rz}(z = -h/2) = 0$ as before. For the symmetric boundary conditions, $\tilde{\sigma}_{zz}(z = h/2) = \tilde{\sigma}_{zz}(z = -h/2) = F_0\tilde{I}(k)/2$. For the anti-symmetric boundary conditions, $\tilde{\sigma}_{zz}(z = h/2) = F_0\tilde{I}(k)/2$ and $\tilde{\sigma}_{zz}(z = -h/2) = -F_0\tilde{I}(k)/2$.

The symmetric solution restricts $\tilde{\Phi}$ to be even and $\tilde{\Psi}$ to be odd because of the derivative in $\tilde{\sigma}_{zz}$. Similarly, the anti-symmetric solution restricts $\tilde{\Phi}$ to be odd and $\tilde{\Psi}$ to be even.

Evaluating \cref{eq:PhiPsi} and their $z$-derivatives at the two surfaces $z=\pm h/2$, and defining
\begin{align}
\Upsilon &\equiv 2k^2-k_\shear^2,
\end{align}
\begin{align}
c_\longitudinal &\equiv \cosh(k_{z,\longitudinal}h/2), \quad s_\longitudinal \equiv \sinh(k_{z,\longitudinal}h/2),
\end{align}
\begin{align}
c_\shear &\equiv \cosh(k_{z,\shear}h/2), \quad s_\shear \equiv \sinh(k_{z,\shear}h/2),
\end{align}
we obtain
\begin{align}
\resizebox{\columnwidth}{!}{$\displaystyle
\tilde{\sigma}_{zz}(\pm h/2,k)
=\mu\Big[\Upsilon(A c_\longitudinal \pm B s_\longitudinal)
      +2k k_{z,\shear}(\pm C s_\shear + D c_\shear)\Big], $}
\end{align}
and
\begin{align}
\resizebox{\columnwidth}{!}{$\displaystyle
\tilde{\sigma}_{rz}(\pm h/2,k)
=-\mu\Big[2k k_{z,\longitudinal}(\pm A s_\longitudinal + B c_\longitudinal)
      +\Upsilon(C c_\shear \pm D s_\shear)\Big]. $}
\end{align}
Relative to the trigonometric convention sometimes used in elasticity texts, our sign choice in the wavenumbers $k_{z,\longitudinal}$ and $k_{z,\shear}$ (see~\cref{eqs:wavenumbers}) leads to hyperbolic rather than circular functions in the through-thickness dependence. The algebra is otherwise identical.

Because the elastic solutions are either symmetric or anti-symmetric about the midplane $z=0$, it is natural to replace the top/bottom surface stresses by combinations with definite parity. This is the step that makes the symmetry decomposition explicit at the level of the boundary-condition equations. We therefore define
\begin{align}
\begin{pmatrix}
\Sigma_{\text{sym}}\\ T_{\text{sym}}\\ \Sigma_{\text{anti}}\\ T_{\text{anti}}
\end{pmatrix}
&=
\frac{1}{2}
\begin{pmatrix}
1 & 0 & 1 & 0\\
0 & -1 & 0 & 1\\
1 & 0 & -1 & 0\\
0 & -1 & 0 & -1
\end{pmatrix}
\begin{pmatrix}
\tilde{\sigma}_{zz}(h/2)\\ \tilde{\sigma}_{rz}(h/2)\\ \tilde{\sigma}_{zz}(-h/2)\\ \tilde{\sigma}_{rz}(-h/2)
\end{pmatrix}.
\end{align}
Substituting the surface stresses above into these combinations gives a linear system for the coefficient functions $A$, $B$, $C$, and $D$. In this basis, the boundary conditions separate into independent symmetric and anti-symmetric sectors, making the parity structure explicit. The resulting block-diagonal system is
\begin{align}
\resizebox{\columnwidth}{!}{$\displaystyle
\begin{pmatrix}
\Sigma_{\text{sym}}\\ T_{\text{sym}}\\ \Sigma_{\text{anti}}\\ T_{\text{anti}}
\end{pmatrix}
=
\mu
\begin{pmatrix}
\Upsilon c_\longitudinal & 2k k_{z,\shear}c_\shear & 0 & 0\\
2k k_{z,\longitudinal}s_\longitudinal & \Upsilon s_\shear & 0 & 0\\
0 & 0 & \Upsilon s_\longitudinal & 2k k_{z,\shear}s_\shear\\
0 & 0 & 2k k_{z,\longitudinal}c_\longitudinal & \Upsilon c_\shear
\end{pmatrix}
\begin{pmatrix}
A\\ D\\ B\\ C
\end{pmatrix}.
$}
\end{align}
Thus, the symmetric sector involves only $(A, D)$, while the anti-symmetric sector involves only $(B, C)$, explicitly isolating the even and odd coefficient functions in \cref{eq:PhiPsi}. Using
\begin{align}
(\Sigma_{\text{sym}},T_{\text{sym}})^T &= \left(F_0\tilde{I}(k)/2,\,0\right)^T
\end{align}
and
\begin{align}
(\Sigma_{\text{anti}},T_{\text{anti}})^T &= \left(F_0\tilde{I}(k)/2,\,0\right)^T,
\end{align}
the two $2\times 2$ systems are
\begin{align}
\mu
\begin{pmatrix}
\Upsilon c_\longitudinal & 2k k_{z,\shear}c_\shear\\
2k k_{z,\longitudinal}s_\longitudinal & \Upsilon s_\shear
\end{pmatrix}
\begin{pmatrix}
A\\ D
\end{pmatrix}
&=
\frac{F_0\tilde{I}(k)}{2}
\begin{pmatrix}
1\\0
\end{pmatrix}
\end{align}
and
\begin{align}
\mu
\begin{pmatrix}
\Upsilon s_\longitudinal & 2k k_{z,\shear}s_\shear\\
2k k_{z,\longitudinal}c_\longitudinal & \Upsilon c_\shear
\end{pmatrix}
\begin{pmatrix}
B\\ C
\end{pmatrix}
&=
\frac{F_0\tilde{I}(k)}{2}
\begin{pmatrix}
1\\0
\end{pmatrix},
\end{align}
from which \cref{eq:aleph,eq:beth,eq:C,eq:D} follow directly.

Solving these boundary-condition systems yields the coefficient functions
\begin{widetext}
\begin{align}\label{eq:aleph}
   A(k) = \frac{2k^2-k_\shear^2}{2\mu\left((2k^2-k_\shear^2)^2\cosh(k_{z,\longitudinal} h/2) - 4k^2k_{z,\longitudinal}k_{z,\shear}\sinh(k_{z,\longitudinal}h/2)\coth(k_{z,\shear} h/2)\right)}F_0\tilde{I}(k)
\end{align}
\begin{align}\label{eq:beth}
   B(k) = \frac{2k^2-k_\shear^2}{2\mu\left((2k^2-k_\shear^2)^2\sinh(k_{z,\longitudinal} h/2) - 4k^2k_{z,\longitudinal}k_{z,\shear}\cosh(k_{z,\longitudinal}h/2)\tanh(k_{z,\shear} h/2)\right)}F_0\tilde{I}(k)
\end{align}
\begin{align}\label{eq:C}
   C(k) = -\frac{2k k_{z,\longitudinal}\cosh(k_{z,\longitudinal}h/2)}{\cosh(k_{z,\shear}h/2)}
   \frac{1}{2\mu\left((2k^2-k_\shear^2)^2\sinh(k_{z,\longitudinal} h/2) - 4k^2k_{z,\longitudinal}k_{z,\shear}\cosh(k_{z,\longitudinal}h/2)\tanh(k_{z,\shear} h/2)\right)}F_0\tilde{I}(k)
\end{align}
\begin{align}\label{eq:D}
   D(k) = -\frac{2k k_{z,\longitudinal}\sinh(k_{z,\longitudinal}h/2)}{\sinh(k_{z,\shear}h/2)}
   \frac{1}{2\mu\left((2k^2-k_\shear^2)^2\cosh(k_{z,\longitudinal} h/2) - 4k^2k_{z,\longitudinal}k_{z,\shear}\sinh(k_{z,\longitudinal}h/2)\coth(k_{z,\shear} h/2)\right)}F_0\tilde{I}(k)
\end{align}
Using these, we can express the displacement $\tilde{\bm{u}}$ as
\begin{align}
   \tilde{u}_z = \frac{\partial \tilde{\Phi}}{\partial z} + k \tilde{\Psi};~
   \tilde{u}_r = -k\tilde{\Phi} -\frac{\partial \tilde{\Psi}}{\partial z}.
\end{align}
We then evaluate $\tilde{u}_z(z,k,\omega)$ and $\tilde{u}_r(z,k,\omega)$ at the front surface of the mirror, $z=h/2$, yielding
\begin{equation}
\resizebox{\columnwidth}{!}{$\displaystyle
   \tilde{u}_z(h/2)= k_{z,\longitudinal}A(k)\sinh(k_{z,\longitudinal}h/2) + k_{z,\longitudinal}B(k)\cosh(k_{z,\longitudinal}h/2) + k C(k)\cosh(k_{z,\shear} h/2) + kD(k)\sinh(k_{z,\shear} h/2)\equiv F_0\tilde{I}(k)\xi_z(k).
$}
\end{equation}
and 
\begin{equation}
\resizebox{\columnwidth}{!}{$\displaystyle
   \tilde{u}_r(h/2) = -kA(k)\cosh(k_{z,\longitudinal}h/2) - kB(k)\sinh(k_{z,\longitudinal}h/2) - k_{z,\shear}C(k)\sinh(k_{z,\shear} h/2) - k_{z,\shear}D(k)\cosh(k_{z,\shear} h/2)\equiv F_0\tilde{I}(k)\xi_r(k).
$}
\end{equation}
\end{widetext}

\subsection{Substrate Mechanical Noise Maximum Height for the Infinite Slab}
In the main text, we calculated the noise minima between the longitudinal eigenfrequencies by setting $\omega = \pi (m_\longitudinal+1/2)|v_\longitudinal|/h$ for natural number $m_\longitudinal$. If, instead, we wish to compute the noise maximum value, which occurs at the longitudinal eigenfrequencies, we set $\omega = \pi m_\longitudinal |v_\longitudinal|/h$. Under the assumption that
\begin{align}
    \frac{h}{\pi m_\longitudinal} \ll w \ll \frac{h}{\pi m_\longitudinal} \phi^{-1/2},
\end{align}
the noise spectrum at the longitudinal modes is
\begin{align}\label{eq:SMNSlabMax}
    S_{z,\text{SMN max}}(\omega) = \frac{k_\mathrm{B} T}{M_0h\omega}.
\end{align}
Note that this does not depend on the beam size $w$. This makes intuitive sense in the eigenmode decomposition framework~\cite{GillespiePRD95ThermallyExcited}. At the longitudinal eigenfrequencies, the entire mirror face moves in unison, so the mean wavefront displacement $\delta z$ does not depend on the beam size. This is described in~\cite{GillespiePRD95ThermallyExcited} as the ``effective mass coefficient" (a parameter related to the inverse of the mode-to-mean-displacement coupling) being constant for the longitudinal eigenmodes.

\subsection{Alternative Method to Calculate the Substrate Mechanical Noise Spectrum for the Infinite Slab}\label{sec:SMNSemi}

In \cref{subsubsec:SMN_Analytic}, we presented a closed-form expression for the substrate mechanical noise spectrum for an infinite slab. While this expression is easy to compute and highlights the key physics for substrate mechanical noise, the approximations used to generate it are not appropriate at all frequencies, and the approximation therefore fails. In this section, we provide an alternative method to evaluating the $k$-space integral in \cref{eq:SMN_Spectrum}. We use the same ansatz for $\text{Im}\{\xi_z(k)\}$ (\cref{eq:xi_z_approx}). Instead of finding the poles $k_p$ using a simplification of the denominators of $\xi_z(k)$, we find them numerically (say using Brent's method). We compute the residues by computing the derivative of the denominators of $\xi_z(k)$ without approximation (in the main text, we approximate the derivatives for simplicity). The integral in \cref{eq:SMN_Spectrum} can then be evaluated analytically, with the numerically determined pole locations $k_p$. To highlight the success of this method, we plot direct, numerical $k$-space integral evaluation of \cref{eq:SMN_Spectrum} alongside this method's evaluation in~\cref{fig:SMNZoom}.

\subsection{Beam Not Centered on Finite Mirror}
The main-text work assumes that the beam is centered on the mirror. Experimentally, this is not always the case, either due to alignment limitations or the need to avoid problematic spots on the mirror, like point absorbers. Because the noise can be dominated by a few Bessel modes, we examine what happens if the beam is displaced from the center of the mirror with some impact parameter $b$. Due to the comparative difficulty, we model the mirror in this instance with a square face instead of a circular face and the beam displaced from the center along a transverse axis (say the $x$-axis). We get the following spectrum for a square-faced mirror with side length $2a$ (to keep $a$ the ``radius"-like quantity),
\begin{widetext}
\begin{equation}\label{eq:DisplacedBeamNoiseSpectrum}
\resizebox{\columnwidth}{!}{$\displaystyle
    S_z(\omega) = \frac{4k_\mathrm{B} T}{a^2 \omega}\left(\sum_{\ell=0}^\infty\sum_{m=0}^\infty\eta^2_\ell\eta^2_m\left|\mathrm{Im}\left\{\xi^{\text{sym}}_{z,\ell,m}\right\}\right|e^{-(k^{\text{sym}}_{\ell,m})^2w^2/4}\cos^2(k^{\text{sym}}_{\ell,0}b)+\sum_{\ell'=0}^\infty\sum_{m=0}^\infty\eta^2_m\left|\mathrm{Im}\left\{\xi^{\text{anti}}_{z,\ell',m}\right\}\right|e^{-(k^{\text{anti}}_{\ell',m})^2w^2/4}\sin^2(k^{\text{anti}}_{\ell',0}b)\right),
$}
\end{equation}
\end{widetext}
where we define 
\begin{align}
    \eta_\ell = 
    \begin{cases} 
      \frac{1}{2} & \ell= 0 \\
      1 & \text{otherwise} \\
    \end{cases}, 
\end{align}
and $\xi^{\text{sym}}_{z,\ell,m}$ and $\xi^{\text{anti}}_{z,\ell',m}$ are the square plate analogs of $\xi_{z,\ell}$ (see~\cref{eq:xi_zl_definition}), with $k_\ell$ replaced with $k^{\text{sym}}_{\ell,m}$ and $k^{\text{anti}}_{\ell',m}$ respectively. $k^{\text{sym}}_{\ell,m}$ and $k^{\text{anti}}_{\ell',m}$ are the square plate analogs of $k_\ell$, where 
\begin{align}
    k^{\text{sym}}_{\ell,m}=\frac{\pi\sqrt{\ell^2+m^2}}{a}; ~k^{\text{anti}}_{\ell',m} = \frac{\pi\sqrt{((2\ell'+1)/2)^2+m^2}}{a}.
\end{align}
Here, $\ell$ and $\ell'$ are dummy variables, but the prime is used to help distinguish the symmetric vs. anti-symmetric terms that arise when the beam is displaced. These boundary conditions come from applying $\sigma_{zx} = \sigma_{zy} = 0$ at the edge of the mirror and ignoring $\sigma_{xx}$ and $\sigma_{yy}$.

Somewhat surprisingly, when computing the noise minimum, there is no dependence on the half-side-length $a$ or the impact parameter $b$, and the noise floor is given by \cref{eq:SMNMin}. This is because $k^{\text{sym}}_{\ell,0} \approx k^{\text{anti}}_{\ell',0} \approx k_\ell$ for the noise floor, and $\cos^2(k_\ell b)+\sin^2(k_\ell b)=1$. More careful analysis with a cylinder may reveal some dependence on the impact parameter, but we leave this for future work.
\section{Mechanical and Thermoelastic Noise in the Beamsplitter}\label{app:BSN}
\subsection{Mechanical and Photoelastic Noise in the Beamsplitter}\label{sec:BSMN}
We turn towards modeling the beamsplitter mechanical noise. We model our beamsplitter as a parallel infinite slab (with infinite radius and no wedge angle) with a 50:50 coating on the front surface (that the input beam hits) and an Anti-Reflective (AR) coating on the back surface (that only the transmitted beam goes through). Since both of these coatings are approximately 10 times thinner than the High-Reflective (HR) coating of the end mirrors, we choose to ignore their effects and focus on noise from the substrate (the substrate has a larger contribution to the total loss angle than the coating when the coating is this thin). Unlike in LIGO, Virgo, and KAGRA, Michelson interferometers that are sensitive in the MHz band have been built and designed without arm cavities~\cite{ChouCQG17HolometerInstrument,PatraPRL25BroadbandLimits, VermeulenPRX25PhotonCountingInterferometry}. Therefore, the beamsplitter's noise is sensed equally to the noise of the end mirrors.  

The mechanical noise of the beamsplitter affects the phase of the reflected light by changing the physical path length. These vibrations also couple to the transmitted light by changing the physical path length through the substrate, which has an index of refraction larger than 1. The mechanical noise also alters the phase of the transmitted light through the photoelastic (PE) effect, in which strain in the optic changes the index of refraction. In an isotropic material, the change in index $\Delta n$ from the strain $\epsilon$ is given by~\cite{HongPRD13BrownianThermala}
\begin{align}
    \Delta n = \beta_{\mathrm{PE},\parallel} \epsilon_{\parallel} + \beta_{\mathrm{PE},\perp} \epsilon_{\perp},
\end{align}
where $\beta_{\mathrm{PE},\parallel}$ and $\beta_{\mathrm{PE},\perp}$ are the ``elastorefractive coefficients" (a name we introduce here to avoid ambiguity with other constants, even though $\beta_{\mathrm{PE}}$ is in the literature, as $\beta$ for example in~\cite{HongPRD13BrownianThermala}) parallel and perpendicular to the beam propagation axis (not the polarization axis), respectively. $\epsilon_{\parallel}$ and $\epsilon_{\perp}$ are the strains parallel and perpendicular to the propagating beam, respectively. Note that the strains are complex due to the mechanical loss angle, and the change in the index of refraction is therefore also complex. However, this does not imply any absorption but instead a phase lag similar to the elastic quantities such as displacement and strain. For linearly polarized light traveling through isotropic materials, like fused silica, or certain anisotropic materials, like (100) crystalline silicon, the change in the index of refraction is first-order insensitive to the shear strain. We therefore continue to approximate (100) crystalline silicon as isotropic. For a beam propagating in the $z$-axis, we write
\begin{align}
    \Delta n = \beta_{\mathrm{PE},\parallel} \epsilon_{zz} + \beta_{\mathrm{PE},\perp} (\epsilon_{rr}+\epsilon_{\varphi\varphi}).
\end{align}
We note $\beta_{\mathrm{PE}}$ is typically negative because a positive strain increases the volume of a material, decreasing its density and therefore its index of refraction. For a linear, isotropic material, we define $\beta_{\mathrm{PE}}$ in terms of the photoelastic tensor coefficients $p_{ij}$~\cite{HongPRD13BrownianThermala},\footnote{Care must be taken when sourcing the elastorefractive coefficients from experimental literature. Due to the Poisson effect, uniaxial stress induces strain in other axes, which mixes the elastorefractive coefficients.}
\begin{align}
    \beta_{\mathrm{PE},\parallel} = -\frac{1}{2}n^3p_{12};~~\beta_{\mathrm{PE},\perp} = -\frac{1}{2}n^3\frac{p_{11}+p_{12}}{2}.
\end{align}
For simplicity, we model our beamsplitter as if the light were propagating perpendicular to its faces (so a \ang{0} angle of incidence instead of \ang{45}). Due to Snell's law, the transmitted light travels a distance greater than the beamsplitter thickness when the angle of incidence is not \ang{0}. In addition, the effective beam spot size on the angled beamsplitter is larger perpendicular to the axis of rotation. These effects will partially cancel and, if we simply follow the beamsplitter corrections in Eq. 2 of~\cite{BenthemPRD09ThermorefractiveThermochemical}, the actual noise PSD will be 40\% smaller for a silicon beamsplitter and 20\% smaller for fused silica, assuming a \ang{45} angle of incidence. 

Light traveling towards and away from the laser source forms a standing wave inside the beamsplitter. We found that this effect contributes less than 1\% to the noise PSD in the valleys between the mechanical mode peaks because its wavelength is much smaller than the elastic wavelength. 

Because the angle of incidence is not \ang{0}, the standing wave forms a striped pattern in the laser intensity inside the beamsplitter. More specifically, for an angle of incidence of $\vartheta$ in the $x$-$z$ plane, the intensity goes as $I(x,y) \propto \cos^2(2\pi x \sin(\vartheta/2)/\lambda_{\text{laser}})$, where $\lambda_{\text{laser}}$ is the laser wavelength in vacuum~\cite{HeinertPRD14ThermalNoise}. Since $\vartheta \approx \ang{45}$, the striped pattern wavelength is on the order of the laser wavelength, which is much smaller than the elastic wavelength below 1 GHz for commonly used mirror materials. Furthermore, Heinert~\cite{HeinertPRD14ThermalNoise} finds a tiny contribution for a large angle of incidence in the substrate mechanical noise (cf. their Eqs. (25) and (26)). We therefore ignore its effects. 

In addition to this, the reflected light's path length change from mechanical noise is a factor of $\sqrt{2}$ smaller than for end mirrors of the same composition due to the \ang{45} angle of incidence. As this angle approaches \ang{90}, i.e., a glancing reflection, the added noise from surface displacement approaches zero. Lastly, we also ignore the fact that the two incident beams on the beamsplitter are not directly opposed because the transmitted beam gets shifted when the angle of incidence is not \ang{0}. This applies a torque to the beamsplitter and may affect the noise; see~\cite{DickmannPRD18ThermalNoise} for a discussion.

From \cref{subsec:MechNoiseSub}, we know the substrate's effect on the reflected light. Thus, our principal task is to calculate the phase difference of the transmitted light through the beamsplitter. We can express the optical path length (OPL) difference of the transmitted light from the mechanical noise of a beamsplitter with thickness $h$ as 
\begin{align}\label{eq:BSMN_OPL}
    \delta z_{\text{OPL}} = (n-1)\bigg(\delta z(z=h/2) - \delta z(z=-h/2) \bigg).
\end{align}
Now writing out $\delta z$, we can write
\begin{equation}
\resizebox{\columnwidth}{!}{$\displaystyle
    \delta z_{\text{OPL}} = (n-1)\int\big(u_z(z=h/2) - u_z(z=-h/2)\big)I(r)d^2r. $}
\end{equation}
Using the fundamental theorem of calculus, we can re-express this difference as
\begin{align}
    \delta z_{\text{OPL}} = (n-1)\int\int_{-h/2}^{h/2}\frac{\partial u_z}{\partial z}dz\,I(r)\,d^2r.
\end{align}
We identify $\partial u_z/\partial z$ as $\epsilon_{zz}$ and write this as one volume integral:
\begin{align}
    \delta z_{\text{OPL}} = (n-1)\int\epsilon_{zz}I(r)\,dV.
\end{align}
Now that we have the volume integral of strain, we can naturally insert the effect of photoelasticity, specifically
\begin{equation}
\resizebox{\columnwidth}{!}{$\displaystyle
    \delta z_{\text{OPL}} = \int\big((n-1+\beta_{\mathrm{PE},\parallel})\epsilon_{zz}+\beta_{\mathrm{PE},\perp} (\epsilon_{rr}+\epsilon_{\varphi\varphi})\big)I(r)dV $}
\end{equation}

We can now easily specify the strains. Following~\cite{Somiya09ThermalNoise}, we apply equal and opposite outward-facing pressures, $F_0I(r)$, on the front and back faces of our beamsplitter. This will cause the stress to be even in $z$.

For simplicity, we will focus on just the $\epsilon_{zz}$ term for the optical path length and ignore the effects from $\beta_{\mathrm{PE},\perp}$. We have modeled its effects; they are small and reported at the end of this section. To proceed with the calculation, we undo the application of the fundamental theorem of calculus and write
\begin{align}
    \delta z_{\text{OPL}} = 2(n-1+\beta_{\mathrm{PE},\parallel})\int u_{z,\text{trans}}(z=h/2)I(r)d^2r,
\end{align}
where $u_{z,\text{trans}}(z=h/2)$ is the solution to the elastic wave equation with symmetric boundary conditions, specifically $\sigma_{zz}(z=\pm h/2) = F_0I(r)$. These boundary conditions arise from sensing changes in the thickness of the optic. Mathematically, we are insensitive to solutions with even $u_z$ in \cref{eq:BSMN_OPL}.

There are two more considerations before we get to our PSD. Changes in the thickness of the beamsplitter affect the path length, not the arm length. Changes in arm length are sensed twice from reflection, so the optical path length changes twice as much as for transmission. Therefore, for the Beamsplitter Mechanical Noise (BSMN) ASD, we need to divide $\delta z_{\text{OPL}}$ by 2 for the correct calibration. The second consideration is that as the beamsplitter expands, the reflected light loses phase and the transmitted light gains phase. This effect is fully coherent, so it is added coherently. Accounting for the beamsplitter's angle of incidence suggests we should only add $1/\sqrt{2}$, but we add a factor of 1 under our assumption that the angle of incidence is 0. We thus get
\begin{multline}
    \delta z_{\text{BS}} = \int \big((n-1+\beta_{\mathrm{PE},\parallel})u_{z,\text{trans}}(z=h/2)+\\u_{z,\text{refl}}(z=h/2)\big)I(r)\,d^2r.
\end{multline}

We have already solved the integral problem in \cref{subsec:MechNoiseSub}, except we only take the symmetric solution for $u_{z,\text{trans}}$ to match the boundary conditions. We keep the symmetric and antisymmetric for $u_{z,\text{refl}}(z=h/2)$, the reflection term. Putting all of this together yields the effective receptance density for the beamsplitter, 
\begin{multline}\label{eq:xi_z_BS}
    \xi_{z,\text{BS}}(k) = -\frac{k_{z,\longitudinal} k_\shear^2}{2\mu} \bigg(\frac{(n+\beta_{\mathrm{PE},\parallel})^2}{D_{1,\xi}(k)} +
    \frac{1}{D_{2,\xi}(k)}\bigg),
\end{multline}
where $D_{1,\xi}(k)$ and $D_{2,\xi}(k)$ are the first and second denominators in \cref{eq:xi_z}, respectively, and
\begin{align}\label{eq:BSMN_Spectrum}
    S_{z,\text{BSMN},\parallel}(\omega) = \frac{2k_{\mathrm{B}} T}{\pi \omega}\int_0^\infty|\mathrm{Im}\left\{\xi_{z,\text{BS}}(k)\right\}|ke^{-k^2w^2/4}\,dk.
\end{align}

\begin{figure*}[t]
\includegraphics[width=1\linewidth]{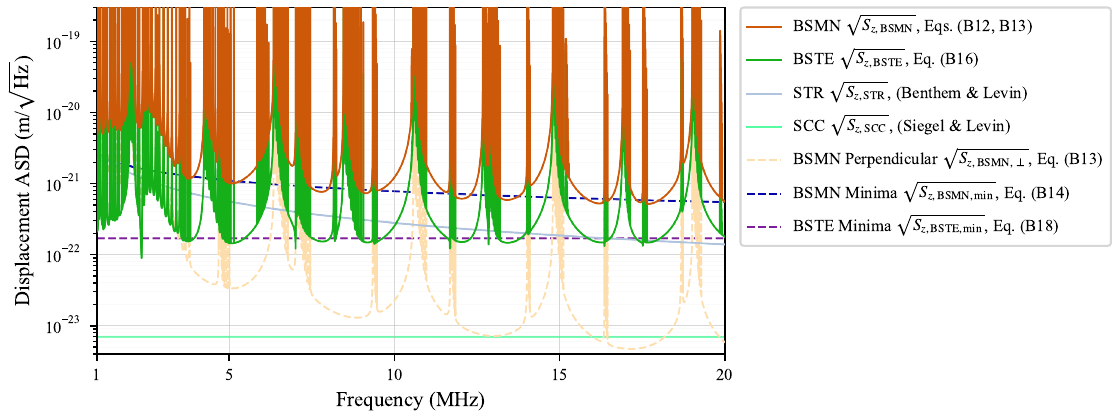}
  \caption{The Modeled Amplitude Spectral Density of the noise for the GQuEST Beamsplitters. The BSMN curve is the sum of \cref{eq:BSMN_Spectrum,eq:BSMN_Spectrum_radial} and the BSTE curve uses \cref{eq:BSTE}. Both are calculated for a cylinder. We use the STR noise from~\cite{BenthemPRD09ThermorefractiveThermochemical} and the SCC noise from~\cite{SiegelPRD23RevisitingThermal}. We also include the plots of \cref{eq:BSMN_Spectrum_radial} by itself and \cref{eq:BSMNMin,eq:BSTE_min} for reference. We use $\Delta f_{\text{sample}} = 3.9$ Hz. To avoid the standing-wave term that greatly increases STR noise~\cite{BenthemPRD09ThermorefractiveThermochemical}, GQuEST is planned to be a traveling-wave interferometer. Therefore two beamsplitters are used and there is no beamsplitter transfer function.}
  \label{fig:BS_Noise}
\end{figure*}

We now briefly turn our attention to the photoelastic noise from the radial strain. As in the coating, we express $\epsilon_{rr}+\epsilon_{\varphi\varphi} = ku_r$ in Hankel-transformed cylindrical coordinates with axial symmetry. Direct integration in the azimuthal and $z$ coordinates gives the noise contribution:
\begin{multline}\label{eq:BSMN_Spectrum_radial}
    S_{z,\text{BSMN},\perp}(\omega) = \frac{2k_{\mathrm{B}}T\beta_{\mathrm{PE},\perp}^2}{\pi \omega}\times \\
    \int_0^\infty\left|\mathrm{Im}\left\{\frac{\nu k_\shear^2}{(1-\nu)\mu k_{z,\longitudinal}D_{1,\xi}(k)}\right\}\right|k^3e^{-k^2w^2/4}\,dk.
\end{multline}
Because $S_{z,\text{BSMN},\parallel}(\omega)$ and $S_{z,\text{BSMN},\perp}(\omega)$ are driven by the same mechanical fluctuations, they should be summed coherently. We plot their contribution for the GQuEST ASD in \cref{fig:BS_Noise}. We finish the section with the evaluation of the PSDs at their noise minima, away from the longitudinal mode resonances. To leading order, the noise minima are
\begin{align}\label{eq:BSMNMin}
    S_{z,\text{BSMN},\parallel,\text{min}}(\omega) &= \frac{k_{\mathrm{B}} Th\phi}{\pi M_0 w^2 \omega}\left(1+|n+\beta_{\mathrm{PE},\parallel}|^2\right)\\
    S_{z,\text{BSMN},\perp,\text{min}}(\omega) &= \frac{4k_{\mathrm{B}} Th\phi \beta_{\mathrm{PE},\perp}^2}{\pi M_0 w^2 \omega}\frac{\nu_0}{1-\nu_0}\frac{|v_\longitudinal|^2}{\omega^2w^2}.
\end{align}

\subsection{Substrate Thermoelastic Noise in the Beamsplitter}\label{sec:BSTE}
Similar to the substrate mechanical noise in the beamsplitter in \crefBSMN, we stick with the simplified model of the beamsplitter. For the transmitted beam, we apply equal and opposite outward-facing pressures on each face of the beamsplitter. We thus set $B(k) = 0$ in the STE spectrum for this contribution. Just like the mechanical noise, the change in the transmitted light's optical path length picks up an additional factor of $n-1+\beta_{\mathrm{PE},\parallel}$. For the reflected arm, the noise is the same as for the end mirror since we still assume the angle of incidence is zero. We thus get the following for the beamsplitter (substrate) thermoelastic (BSTE) spectrum
\begin{widetext}
\begin{multline}\label{eq:BSTE}
    S_{z,\,\text{BSTE}}(\omega) = \frac{k_{\mathrm{B}} T^2 \kappa \rho^2 (1+\nu_0)^2 \alpha^2 \omega^2 }{\pi C_V^2 (1-\nu_0)^2} \int_0^\infty\bigg(\left((n+\beta_{\mathrm{PE},\parallel})^2|\bar{A}(k)|^2 + |\bar{B}(k)|^2 \right)\frac{k^2+|k_{z,\longitudinal}|^2}{\mathrm{Re}\{k_{z,\longitudinal}\}}\sinh(\mathrm{Re}\{k_{z,\longitudinal}\}h) +\\ \left((n+\beta_{\mathrm{PE},\parallel})^2|\bar{A}(k)|^2 - |\bar{B}(k)|^2 \right)\frac{k^2-|k_{z,\longitudinal}|^2}{\mathrm{Im}\{k_{z,\longitudinal}\}}\sin(\mathrm{Im}\{k_{z,\longitudinal}\}h)\bigg)ke^{-k^2w^2/4}\,dk
\end{multline}
\end{widetext}
There is also a component of photoelasticity for strain perpendicular to the propagating beam axis (the $z$-axis), which we choose to ignore for simplicity and because of its small effect on the mechanical noise.

Previous literature~\cite{HeinertPRD14ThermalNoise} has found a rather large increase in the substrate thermoelastic noise when a striped pattern from a standing wave is present on the optic. They give the increase in the quasistatic limit as
\begin{multline}
        S_{z,\text{STE, striped}}(\omega) = S_{z,\text{STE, no stripe}}(\omega)\times\\ \left(1+\frac{2\sqrt{\pi}w\sin(\vartheta/2)}{\lambda_{\text{laser}}}\right).
\end{multline}
Because the beamsplitter substrate thermoelastic noise is small compared to mechanical noise, even with this increase, we choose to ignore this effect and the internal standing wave and leave a full treatment to future work.

We approximate the BSTE spectrum in the same limits as in \cref{subsubsec:SMN_Analytic}. In this limit, away from the mirror mechanical resonances and at high frequencies, the minimum of the BSTE spectrum is
\begin{equation}\label{eq:BSTE_min}
\resizebox{\columnwidth}{!}{$\displaystyle
    S_{z,\,\text{BSTE, min}}(\omega) = \frac{k_{\mathrm{B}} T^2 \kappa (1+\nu_0)^2\alpha^2 h(1+|n+\beta_{\mathrm{PE},\parallel}|^2)}{\pi C_V^2 (1-\nu_0)^2 |v_\longitudinal|^2 w^2}. $}
\end{equation}
Similarly to the previous noises, we could Taylor expand the integrand of \cref{eq:BSTE} for a more accurate high-frequency approximation. However, as we have made significant approximations in \crefBSTE, increased precision would have diminishing returns.

\section{CTO Calculation Details}\label{app:CTO Details}
\subsection{Main Calculation Details}
In this section, we present the details for calculating the coating thermo-optic noise, as no simplified form was derived. Recall that we derived \cref{eq:SLCTOwithTheta} and $\theta(z)$ for all $z$. Not shown here, we analytically differentiate $\theta(z)$, take the squared magnitude, and then integrate. 

We found that our initial implementation of these integrals was unstable at high frequencies. We therefore express these equations as they are implemented in code. Starting with the main text, \cref{eq:SLCTOwithTheta}
\begin{align}
    S_{z,\,\text{TO}} = \frac{2k_{\mathrm{B}} T^2}{\pi \omega^2C_\coating^2}\int_0^\infty ke^{-k^2w^2/4}\int_0^\infty \kappa \left|\frac{\partial \theta(k,z)}{\partial z}\right|^2 \,dz\,dk,
\end{align}

We go from the infinite slab to a cylinder,
\begin{align}
    S_{z,\,\text{TO}} = \frac{4k_{\mathrm{B}} T^2}{\pi a^2\omega^2C_\coating^2}\sum_{\ell=1}^\infty \frac{e^{-k_\ell^2w^2/4}}{J_0^2(\zeta_\ell)}\int_0^\infty \kappa \left|\frac{\partial \theta_\ell(z)}{\partial z}\right|^2 dz,
\end{align}
where $\theta_\ell(z)$ is the generalization of $\theta(k,z)$ from the infinite slab to the cylinder. We then write the PSD by evaluating the integrals
\begin{align}\label{eq:CTO_Full}
    S_{z,\,\text{TO}} = \frac{4k_{\mathrm{B}} T^2}{\pi a^2\omega^2C_\coating^2}\sum_{\ell=1}^\infty \frac{e^{-k_\ell^2w^2/4}}{J_0^2(\zeta_\ell)}\left(\kappa_\substrate I_\substrate + \kappa_\coating\sum_{i=1}^{15} I_i\right)_\ell,
\end{align}
where $I_\substrate$ is from the integration in the substrate and $I_i$ are terms from integration in the coating. Please note there is no relationship between the $I_i$ here and the $I(r)$ for the laser beam profile. The authors apologize for the lack of good and available symbols at this point in the paper. Following~\cite{EvansPRD08ThermoopticNoise,SomiyaPRD09CoatingThermal}, we define a variable (our $\iota$, their $\xi$):
\begin{align}
    \iota = \frac{\sqrt{2}h_\coating}{r_{\thermal,\coating}} = \sqrt{\frac{2\omega C_\coating}{\kappa_\coating}}h_\coating.
\end{align}
$I_\substrate$, $I_1$, $I_2$, and $I_3$ alone replicate the results from~\cite{EvansPRD08ThermoopticNoise} in the limits $\bar{\lambda} \ll r_{\thermal,\coating}$, $1/(2k_{\text{laser}}) \ll r_{\thermal,\coating}$, and the quasistatic limit. In these limits, the remaining integrals are negligible. Note that $I_\substrate$ does not include $\theta_{\substrate,\particular}$, the substrate thermoelastic term. To sum the noises coherently, which is important at low frequencies, one would include it here. These integrals are
\begin{align}
        I_\substrate = \frac{1}{\sqrt{2}r_{\thermal,\substrate}} |\tilde{\mathcal{D}}|^2,
\end{align}
\begin{equation}
\resizebox{\columnwidth}{!}{$\displaystyle
        I_1 = \frac{1}{\sqrt{2}r_{\thermal,\coating}} \frac{|\mathcal{B}(1-R)e^{-\gamma_\coating h_\coating}+\tilde{\mathcal{A}}|^2 (1-e^{-\iota})}{(1+R)^2+2(1-R^2)e^{-\iota}\cos(\iota)+(1-R)^2e^{-2\iota}},$}
\end{equation}
\begin{align}
        I_2 &= \frac{1}{2\sqrt{2}r_{\thermal,\coating}} |(\mathcal{B}+\mathcal{A})(\mathcal{B}-\mathcal{A})^*|(\sin(\iota+\varphi_2)-\sin(\varphi_2)),\\
        I_3 &= \frac{1}{4\sqrt{2}r_{\thermal,\coating}} |\mathcal{B}-\mathcal{A}|^2(1-e^{-\iota}).
\end{align}
We write $\tilde{\mathcal{A}}$ as 
\begin{multline}
        \tilde{\mathcal{A}} = \Bigg[\varrho_{\coating,\ell} \alpha_\coating - \frac{C_\coating}{C_\substrate}\varrho_{\substrate,\ell} \alpha_\substrate \\
        + e^{-h_\coating/\bar{\lambda}}(E + G\cos(2k_{\text{laser}}h_\coating)+H\sin(2k_{\text{laser}}h_\coating))\Big(\frac{R}{\bar{\lambda}\gamma_\coating}-1\Big) \\
        -\frac{2k_{\text{laser}}R}{\gamma_\coating}e^{-h_\coating/\bar{\lambda}}(H\cos(2k_{\text{laser}}h_\coating)-G\sin(2k_{\text{laser}}h_\coating)) \Bigg],
\end{multline}
and we write $\tilde{\mathcal{D}}$ as,
\begin{multline}
    \tilde{\mathcal{D}} = 
    \Big(- \left(\varrho_{\coating,\ell} \alpha_\coating - \frac{C_\coating}{C_\substrate}\varrho_{\substrate,\ell} \alpha_\substrate\right) \\
    +\frac{\tilde{\mathcal{A}}}{1+R\tanh(\gamma_\coating h_\coating)} - \mathcal{B}\frac{2 R e^{-\gamma_\coating h_\coating}}{1+R+(1-R)e^{-2\gamma_\coating h_\coating}} \\
    + e^{-h_\coating/\bar{\lambda}}(E + G\cos(2k_{\text{laser}}h_\coating)+H\sin(2k_{\text{laser}}h_\coating))\Big).
\end{multline}

We also found for numerical stability at high frequencies the need to rewrite $\mathcal{B}+\mathcal{A}$ and $\mathcal{B}-\mathcal{A}$ as
\begin{multline}
        \mathcal{B}+\mathcal{A} = \mathcal{B}(1-R)\frac{1-\tanh(\gamma_\coating h_\coating)}{1+R\tanh(\gamma_\coating h_\coating)} \\
        + \frac{\tilde{\mathcal{A}}}{\cosh(\gamma_\coating h_\coating)+R\sinh(\gamma_\coating h_\coating)},
\end{multline}
\begin{multline}
        \mathcal{B}-\mathcal{A} = \mathcal{B}(1+R)\frac{1+\tanh(\gamma_\coating h_\coating)}  {1+R\tanh(\gamma_\coating h_\coating)} \\
        - \frac{\tilde{\mathcal{A}}}{\cosh(\gamma_\coating h_\coating)+R\sinh(\gamma_\coating h_\coating)},
\end{multline}
and we define $\varphi_2$ as
\begin{align}
        \varphi_2 = \arg((\mathcal{B}+\mathcal{A})(\mathcal{B}-\mathcal{A})^*).
\end{align}
We have extended $\varrho_j(k)$ to the cylinder as $\varrho_{j,\ell}$,
\begin{align}
    \varrho_{j,\ell} = \frac{1+|\nu_j|}{1-|\nu_j|}(1+2\mu_j k_\ell\xi_{\substrate,r,\ell}),
\end{align}
where $\xi_{\substrate,r,\ell}$ is defined as
\begin{widetext}
\begin{multline}
    \xi_{\substrate,r,\ell} = -\frac{k_\ell}{2\mu_\substrate} \bigg(\frac{2k_\ell^2-k_{\substrate,\shear}^2-2k_{\substrate,z,\longitudinal,\ell}k_{\substrate,z,\shear,\ell}\tanh(k_{\substrate,z,\longitudinal,\ell} h/2)\coth(k_{\substrate,z,\shear,\ell} h/2)}{(2k_\ell^2-k_{\substrate,\shear}^2)^2 - 4k_\ell^2k_{\substrate,z,\longitudinal,\ell}k_{\substrate,z,\shear,\ell}\tanh(k_{\substrate,z,\longitudinal,\ell} h/2)\coth(k_{\substrate,z,\shear,\ell} h/2)} +\\
    \frac{2k_\ell^2-k_{\substrate,\shear}^2-2k_{\substrate,z,\longitudinal,\ell}k_{\substrate,z,\shear,\ell}\tanh(k_{\substrate,z,\shear,\ell} h/2)\coth(k_{\substrate,z,\longitudinal,\ell} h/2)}{(2k_\ell^2-k_{\substrate,\shear}^2)^2 - 4k_\ell^2k_{\substrate,z,\longitudinal,\ell}k_{\substrate,z,\shear,\ell}\tanh(k_{\substrate,z,\shear,\ell} h/2)\coth(k_{\substrate,z,\longitudinal,\ell} h/2)}\bigg).
\end{multline}
Here, $I_4$, $I_5$, and $I_6$ do not include the standing wave contribution to the drive term, so using $I_\substrate$ and $I_1$ through $I_6$ and neglecting $I_7$ through $I_{15}$ is valid for the regime $1/(2k_{\text{laser}}) \ll r_{\thermal,\coating}$. Specifically, the integrals are,
\begin{equation}
\resizebox{\textwidth}{!}{$\displaystyle
\begin{aligned}
        I_4 &= -\frac{1}{\bar{\lambda}r_{\thermal,\coating}(\eta_{-}^2+1/(2r_{\thermal,\coating}^2))} \left|(\mathcal{B}+\mathcal{A})E^*\right|\left(e^{-\eta_{-} h_\coating}\left(\frac{1}{\sqrt{2}r_{\thermal,\coating}}\sin\left(\frac{\iota}{2}+\varphi_4\right)-\eta_{-}\cos\left(\frac{\iota}{2}+\varphi_4\right)\right)-\frac{1}{\sqrt{2}r_{\thermal,\coating}}\sin\left(\varphi_4\right)+\eta_{-}\cos\left(\varphi_4\right)\right),
\end{aligned}$}
\end{equation}
\begin{equation}
\resizebox{\textwidth}{!}{$\displaystyle
\begin{aligned}
        I_5 &= -\frac{1}{\bar{\lambda}r_{\thermal,\coating}(\eta_{+}^2+1/(2r_{\thermal,\coating}^2))} \left|(\mathcal{B}-\mathcal{A})E^*\right|\left(e^{-\eta_{+} h_\coating}\left(\frac{1}{\sqrt{2}r_{\thermal,\coating}}\sin\left(\frac{\iota}{2}-\varphi_5\right)-\eta_{+}\cos\left(\frac{\iota}{2}-\varphi_5\right)\right)+\frac{1}{\sqrt{2}r_{\thermal,\coating}}\sin\left(\varphi_5\right)+\eta_{+}\cos\left(\varphi_5\right)\right),
\end{aligned}$}
\end{equation}
\end{widetext}
\begin{align}
        I_6 = \frac{1}{2\bar{\lambda}} |E|^2\left(1-e^{-2h_\coating/\bar{\lambda}}\right),
\end{align}
where we have defined
\begin{align}
        \eta_{+} &= \frac{1}{\bar{\lambda}} + \frac{1}{\sqrt{2}r_{\thermal,\coating}}, \\
        \eta_{-} &= \frac{1}{\bar{\lambda}} - \frac{1}{\sqrt{2}r_{\thermal,\coating}}, \\
        \varphi_4 &= \arg(\sqrt{i}(\mathcal{B}+\mathcal{A})E^*), \\
        \varphi_5 &= \arg(\sqrt{i}(\mathcal{B}-\mathcal{A})E^*).
\end{align}
The remaining integrals are needed when the approximation $1/(2k_{\text{laser}}) \ll r_{\thermal,\coating}$ is not valid:
\begin{widetext}
\begin{multline}
        I_7 = \frac{1}{2r_{\thermal,\coating}} \left|(\mathcal{B}+\mathcal{A})\left(2k_{\text{laser}}H^*-\frac{G^*}{\bar{\lambda}}\right)\right|\Bigg(\frac{e^{-\eta_{-} h_\coating}}{\eta_{-}^2+\chi_{+}^2}\left(\chi_{+}\sin\left(\chi_{+}h_\coating+\varphi_7\right)-\eta_{-}\cos\left(\chi_{+}h_\coating+\varphi_7\right)\right)
        \\
        +\frac{e^{-\eta_{-} h_\coating}}{\eta_{-}^2+\chi_{-}^2}\left(\chi_{-}\sin\left(\chi_{-}h_\coating+\varphi_7\right)-\eta_{-}\cos\left(\chi_{-}h_\coating+\varphi_7\right)\right)-
        \frac{1}{\eta_{-}^2+\chi_{+}^2}\left(\chi_{+}\sin\left(\varphi_7\right)-\eta_{-}\cos\left(\varphi_7\right)\right)
        \\
        -\frac{1}{\eta_{-}^2+\chi_{-}^2}\left(\chi_{-}\sin\left(\varphi_7\right)-\eta_{-}\cos\left(\varphi_7\right)\right)\Bigg),
\end{multline}
\begin{multline}
        I_8 = \frac{1}{2r_{\thermal,\coating}} \left|(\mathcal{B}-\mathcal{A})\left(2k_{\text{laser}}H^*-\frac{G^*}{\bar{\lambda}}\right)\right|\Bigg(\frac{e^{-\eta_{+} h_\coating}}{\eta_{+}^2+\chi_{+}^2}\left(\chi_{+}\sin\left(\chi_{+}h_\coating-\varphi_8\right)-\eta_{+}\cos\left(\chi_{+}h_\coating-\varphi_8\right)\right)
        \\
        +\frac{e^{-\eta_{+} h_\coating}}{\eta_{+}^2+\chi_{-}^2}\left(\chi_{-}\sin\left(\chi_{-}h_\coating-\varphi_8\right)-\eta_{+}\cos\left(\chi_{-}h_\coating-\varphi_8\right)\right)+
        \frac{1}{\eta_{+}^2+\chi_{+}^2}\left(\chi_{+}\sin\left(\varphi_8\right)+\eta_{+}\cos\left(\varphi_8\right)\right)
        \\
        +\frac{1}{\eta_{+}^2+\chi_{-}^2}\left(\chi_{-}\sin\left(\varphi_8\right)+\eta_{+}\cos\left(\varphi_8\right)\right)\Bigg),
\end{multline}
\begin{align}
        I_9 = -\frac{1}{\bar{\lambda}(k_{\text{laser}}^2+1/\bar{\lambda}^2)} \mathrm{Re}\left\{E\left(2k_{\text{laser}}H^*-\frac{G^*}{\bar{\lambda}}\right)\right\}\Bigg(e^{-2h_\coating/\bar{\lambda}}\left(k_{\text{laser}}\sin\left(2 k_{\text{laser}} h_\coating\right)-\frac{1}{\bar{\lambda}}\cos\left(2 k_{\text{laser}} h_\coating\right)\right) + \frac{1}{\bar{\lambda}}\Bigg),
\end{align}
\begin{equation}
\resizebox{\textwidth}{!}{$\displaystyle
        I_{10} = \frac{1}{4} \left|2k_{\text{laser}}H-\frac{G}{\bar{\lambda}}\right|^2\Bigg(\bar{\lambda}\left(1-e^{-2h_\coating/\bar{\lambda}}\right)+\frac{e^{-2h_\coating/\bar{\lambda}}}{4k_{\text{laser}}^2+1/\bar{\lambda}^2} \left(2k_{\text{laser}}\sin\left(4 k_{\text{laser}} h_\coating\right)-\frac{1}{\bar{\lambda}}\cos\left(4 k_{\text{laser}} h_\coating\right)\right)  + \frac{1/\bar{\lambda}}{4k_{\text{laser}}^2+1/\bar{\lambda}^2}\Bigg),$}
\end{equation}
\begin{multline}
        I_{11} = \frac{1}{2r_{\thermal,\coating}} \left|(\mathcal{B}+\mathcal{A})\left(2k_{\text{laser}}G^*+\frac{H^*}{\bar{\lambda}}\right)\right|\Bigg(\frac{e^{-\eta_{-} h_\coating}}{\eta_{-}^2+\chi_{+}^2}\left(\chi_{+}\cos\left(\chi_{+}h_\coating+\varphi_{11}\right)-\eta_{-}\sin\left(\chi_{+}h_\coating+\varphi_{11}\right)\right)
        \\
        -\frac{e^{-\eta_{-} h_\coating}}{\eta_{-}^2+\chi_{-}^2}\left(\chi_{-}\cos\left(\chi_{-}h_\coating+\varphi_{11}\right)-\eta_{-}\sin\left(\chi_{-}h_\coating+\varphi_{11}\right)\right)-
        \frac{1}{\eta_{-}^2+\chi_{+}^2}\left(\chi_{+}\cos\left(\varphi_{11}\right)-\eta_{-}\sin\left(\varphi_{11}\right)\right)
        \\
        +\frac{1}{\eta_{-}^2+\chi_{-}^2}\left(\chi_{-}\cos\left(\varphi_{11}\right)-\eta_{-}\sin\left(\varphi_{11}\right)\right)\Bigg),
\end{multline}
\begin{multline}
        I_{12} = \frac{1}{2r_{\thermal,\coating}} \left|(\mathcal{B}-\mathcal{A})\left(2k_{\text{laser}}G^*+\frac{H^*}{\bar{\lambda}}\right)\right|\Bigg(\frac{e^{-\eta_{+} h_\coating}}{\eta_{+}^2+\chi_{+}^2}\left(\chi_{+}\cos\left(\chi_{+}h_\coating-\varphi_{12}\right)-\eta_{+}\sin\left(\chi_{+}h_\coating-\varphi_{12}\right)\right)
        \\
        -\frac{e^{-\eta_{+} h_\coating}}{\eta_{+}^2+\chi_{-}^2}\left(\chi_{-}\cos\left(\chi_{-}h_\coating-\varphi_{12}\right)-\eta_{+}\sin\left(\chi_{-}h_\coating-\varphi_{12}\right)\right)-
        \frac{1}{\eta_{+}^2+\chi_{+}^2}\left(\chi_{+}\cos\left(\varphi_{12}\right)+\eta_{+}\sin\left(\varphi_{12}\right)\right)
        \\
        +\frac{1}{\eta_{+}^2+\chi_{-}^2}\left(\chi_{-}\cos\left(\varphi_{12}\right)+\eta_{+}\sin\left(\varphi_{12}\right)\right)\Bigg),
\end{multline}
\begin{align}
        I_{13} = -\frac{1}{\bar{\lambda}(k_{\text{laser}}^2+1/\bar{\lambda}^2)} \mathrm{Re}\left\{E\left(2k_{\text{laser}}G^*+\frac{H^*}{\bar{\lambda}}\right)\right\}\Bigg(e^{-2h_\coating/\bar{\lambda}}\left(k_{\text{laser}}\cos\left(2 k_{\text{laser}} h_\coating\right)+\frac{1}{\bar{\lambda}}\sin\left(2 k_{\text{laser}} h_\coating\right)\right) - k_{\text{laser}}\Bigg),
\end{align}
\begin{equation}
\resizebox{\textwidth}{!}{$\displaystyle
        I_{14} = \frac{1}{8k_{\text{laser}}^2+2/\bar{\lambda}^2} \mathrm{Re}\left\{\left(2k_{\text{laser}}H-\frac{G}{\bar{\lambda}}\right)\left(2k_{\text{laser}}G^*+\frac{H^*}{\bar{\lambda}}\right)\right\}\Bigg(e^{-2h_\coating/\bar{\lambda}}\left(2k_{\text{laser}}\cos\left(4 k_{\text{laser}} h_\coating\right)+\frac{1}{\bar{\lambda}}\sin\left(4 k_{\text{laser}} h_\coating\right)\right) - 2k_{\text{laser}}\Bigg),$}
\end{equation}
\begin{equation}
\resizebox{\textwidth}{!}{$\displaystyle
        I_{15} = \frac{1}{4} \left|2k_{\text{laser}}G+\frac{H}{\bar{\lambda}}\right|^2\Bigg(\bar{\lambda}\left(1-e^{-2h_\coating/\bar{\lambda}}\right)-\frac{e^{-2h_\coating/\bar{\lambda}}}{4k_{\text{laser}}^2+1/\bar{\lambda}^2} \left(2k_{\text{laser}}\sin\left(4 k_{\text{laser}} h_\coating\right)-\frac{1}{\bar{\lambda}}\cos\left(4 k_{\text{laser}} h_\coating\right)\right) - \frac{1/\bar{\lambda}}{4k_{\text{laser}}^2+1/\bar{\lambda}^2}\Bigg).$}
\end{equation}
\end{widetext}
We have defined
\begin{align}
        \chi_{+} &= \frac{1}{\sqrt{2}r_{\thermal,\coating}} + 2k_{\text{laser}},\\
        \chi_{-} &= \frac{1}{\sqrt{2}r_{\thermal,\coating}} - 2k_{\text{laser}},\\
        \varphi_7 &= \arg\left(\sqrt{i}(\mathcal{B}+\mathcal{A})\left(2k_{\text{laser}}H^*-\frac{G^*}{\bar{\lambda}}\right)\right),\\
        \varphi_8 &= \arg\left(\sqrt{i}(\mathcal{B}-\mathcal{A})\left(2k_{\text{laser}}H^*-\frac{G^*}{\bar{\lambda}}\right)\right),\\
        \varphi_{11} &= \arg\left(\sqrt{i}(\mathcal{B}+\mathcal{A})\left(2k_{\text{laser}}G^*+\frac{H^*}{\bar{\lambda}}\right)\right),\\
        \varphi_{12} &= \arg\left(\sqrt{i}(\mathcal{B}-\mathcal{A})\left(2k_{\text{laser}}G^*+\frac{H^*}{\bar{\lambda}}\right)\right).
\end{align}

\subsection{Thick Coating Correction}
Similarly to Evans~\cite{EvansPRD08ThermoopticNoise}, we can calculate the correction from the ``thick coating" (compared to the thermal diffusion length) as 
\begin{align}
    \Gamma_{tc} = \frac{S_{z,\,\text{CTO}}}{S_{z,\,\text{CTO, simple}}}.
\end{align}
We define $S_{z,\,\text{CTO, simple}}$ as the spectrum in Eq. (4) from~\cite{EvansPRD08ThermoopticNoise} as
\begin{equation}
      S_{z,\,\text{CTO, simple}}(\omega)
  =
    \frac{
    2\sqrt{2}k_\mathrm{B}T^2
    }{
    \pi w^2 \sqrt{\kappa_\substrate C_\substrate \omega}
    }
    \left(\Delta\bar{\alpha}h_\coating-\bar{\beta} \lambda_{\text{laser}}\right)^2,
\end{equation}
where $\Delta \bar{\alpha}$ is defined in~\cite{EvansPRD08ThermoopticNoise}. Some algebra yields the following equation if $\varrho_j(k)$ does not depend on $k$,
\begin{align}
    \Gamma_{tc} = \frac{\sqrt{2}\kappa_\substrate}{r_{\thermal,\substrate}\omega^2C_\coating^2(\Delta \bar{\alpha} h_\coating - \bar{\beta}\lambda_{\text{laser}})^2}\int \kappa \left|\frac{\partial \theta(z)}{\partial z}\right|^2 dz.
\end{align}
We find the same spectrum as Evans~\cite{EvansPRD08ThermoopticNoise} for $\bar{\lambda} \ll r_{\thermal,\coating}$. Qualitatively, $\Gamma_{tc} \approx 1$ at low frequencies, and at high frequencies the thermoelastic and thermorefractive terms separate. When the thermal diffusion length in the coating is much smaller than $\bar{\lambda}$ and the standing wave in the optical coating is ignored, the thermorefractive term in the thick coating correction is
\begin{align}
    \Gamma_{\overline{\beta}} = \frac{r_{\thermal,\coating}^3}{\sqrt{2}R\bar{\lambda}^3}\frac{(\bar{\beta}\lambda_{\text{laser}})^2}{(\Delta \bar{\alpha} h_\coating - \bar{\beta}\lambda_{\text{laser}})^2}.
\end{align}

\subsection{Average Coating Material Parameters}\label{sec:CTO_Averaging}
Above approximately 2~MHz, the thermal diffusion lengths in $\mathrm{Ta}_2 \mathrm{O}_5$ and $\mathrm{SiO}_2$ are smaller than their respective $\lambda/4$ layer thicknesses for 1064-nm and 1550-nm light. Although the validity is questionable, we still mostly use the averaging calculations from~\cite{FejerPRD04ThermoelasticDissipation}, which are repeated in~\cite{EvansPRD08ThermoopticNoise}, for our entire noise spectrum. We use this existing averaging procedure in part because the properties of the coating materials are fairly similar. Furthermore, the uncertainty in the values of the material parameters for thin coatings is rather large (see, for example,~\cite{WelschIJoT99AbsoluteMeasurement}), making a careful theoretical analysis moot. 

We cannot use previous methods to calculate the average effective thermal expansion coefficient because the methods rely on the quasistatic approximation. Like~\cite{FejerPRD04ThermoelasticDissipation}, we take a simple volume average and write
\begin{align}
    \varrho_\coating(k)\alpha_\coating = \sum_{j\in \text{coating}}\frac{h_{\coating,j}}{h_{\coating}}\varrho_j(k)\alpha_j.
\end{align}

For CTR noise, we could weight the thermal conductivity $\kappa$ and volumetric heat capacity $C_V$ by the fraction of beam penetration that interacts with the layer, but the weighted fractional composition differs from the unweighted fractional composition by only a few percent, so we do not investigate this further.
\section{Implications for Kilometer-Scale Gravitational-Wave Detectors}\label{app:Implications}
A majority of the modeling in this paper is not relevant to the target sensitivity band for existing kilometer-scale gravitational-wave detectors. There may be a slight decrease in the input test mass CTR noise (we do not have a physical motivation for why the end test mass does not have this decrease, but this behavior is also seen in~\cite{BallmerPRD15PhotothermalTransfer}) around 10~kHz. The quasistatic approximation is not appropriate for the sensitivity band of Cosmic Explorer, which will have end mirror mechanical resonances in its frequency band of interest.

Within Cosmic Explorer's sensitivity band and GEO600's very-high-frequency sensitivity band~\cite{JungkindPRD26ProspectsHighfrequencya}, the quasistatic approximation fails. This will change their design curves, so we want to carefully model the noise in the design bands. In order to exactly calculate the modes without approximation (a more important and reasonable goal when the modes are in-band and easily counted compared to the MHz band), we anticipate using the vibrating cylinder solution in~\cite{HutchinsonJoAM80VibrationsSolid} to calculate the mechanical eigenmodes. The STE, CTE, and CTR noises will not separate at low frequencies, so these noises will have to be added coherently.

We also have not examined GEO600's specific configuration, especially the striped pattern on the folding mirror (see~\cite{HeinertPRD14ThermalNoise} for an examination in the quasistatic limit). We expect additional noise when the elastic wavelength is comparable to the striped-pattern wavelength on the folding mirror, which corresponds to around 2~MHz.
\section{Parameters for Noise Curves}\label{app:Tables}
In this section, we compile the parameters used for the noise curves and all of the relevant material properties for high-frequency Michelson interferometers (IFOs).
\ifdcc

\else

\begin{table*}[ht]
  \caption{Holometer Parameters}\label{tab:Holometer}
    \begin{tabular}{lcc}
      \toprule
  PARAMETER & SYMBOL & VALUE \\
    \midrule \midrule
    Laser wavelength in vacuum & $\lambda_{\text{laser}}$ & $1064 \text{ nm}$~\cite{ChouCQG17HolometerInstrument}\\

    Interferometer arm length & $L_{\text{arm}}$ & $39.2 \text{ m}$~\cite{ChouCQG17HolometerInstrument}\\

    Interferometer power recycling cavity length & $L_{\text{PRC}}$ & $39.4 \text{ m}$ (derived from~\cite{McCuller15TestingModel})\\

    Interferometer power recycling cavity finesse & $\mathcal{F}_{\text{PRC}}$ & $3000$ (derived from \cite{ChouCQG17HolometerInstrument})\\


        
    Input laser white phase noise PSD  & $S_{\mathrm{LPN},\mathrm{in}}$ &$(10^{-7}~\mathrm{rad}/\sqrt{\mathrm{Hz}})^2$ \\

    IFO DC offset & $\Lambda_{\text{DC}}$ & 40~ppm~\cite{ChouCQG17HolometerInstrument} \\
    IFO Contrast Defect & $\Lambda_{\text{CD}}$ & 23~ppm~\cite{ChouCQG17HolometerInstrument} \\
    Interferometer arm length difference & $L_{\text{Schnupp}}$ & 0.001~m\\
    Temperature & $T$ & 293~K \\
    Configuration & - & Standing Wave~\cite{ChouCQG17HolometerInstrument} \\
    Interferometer vacuum system pressure & $\rho_{\text{vacuum}}$ & $< 10~\mu$Pa\\
       \midrule
    End mirror $1/e^2~(2\sigma)$ intensity beam radius & $w$ & 5~mm~\cite{ChouCQG17HolometerInstrument}\\
    End mirror radius & $a$ & 25.4~mm~\cite{ChouCQG17HolometerInstrument}\\
    End mirror thickness & $h$ & 12.7~mm~\cite{ChouCQG17HolometerInstrument}\\
    End mirror substrate material & FS & Fused Silica~\cite{ChouCQG17HolometerInstrument}\\
    End mirror coating material & - & $\mathrm{Ta}_2\mathrm{O}_5-\mathrm{SiO}_2$\\
    End mirror coating thickness & $h_\coating$ & 6~$\mathrm{\mu m}$ (derived) \\
    \midrule
    Beamsplitter $1/e^2~(2\sigma)$ intensity beam radius & $w_{\mathrm{BS}}$ & 3.6~mm~\cite{ChouCQG17HolometerInstrument}\\
    Beamsplitter radius & $a_{\mathrm{BS}}$ & 38.1~mm~\cite{ChouCQG17HolometerInstrument}\\
    Beamsplitter thickness & $h$ & 12.7~mm~\cite{ChouCQG17HolometerInstrument}\\
    Beamsplitter substrate material & FS & Fused Silica~\cite{ChouCQG17HolometerInstrument}\\
    \midrule
    Power recycling mirror $1/e^2~(2\sigma)$ intensity beam radius & $w_{\text{PRM}}$ & 3.57~mm~\cite{ChouCQG17HolometerInstrument}\\
    Power recycling mirror radius & $a_{\mathrm{PRM}}$ & 25.4~mm~\cite{ChouCQG17HolometerInstrument}\\
    Power recycling mirror thickness & $h$ & 12.7~mm~\cite{ChouCQG17HolometerInstrument}\\
    Power recycling mirror substrate material & FS & Fused Silica~\cite{ChouCQG17HolometerInstrument}\\
    Power recycling mirror coating material & - & $\mathrm{Ta}_2\mathrm{O}_5-\mathrm{SiO}_2$\\
    Power recycling mirror coating thickness & $h_\coating$ & $3~\mathrm{\mu m}$ (derived) \\
    \bottomrule
    \end{tabular}
\end{table*}

\begin{table*}[ht]
  \caption{GQuEST Parameters (fiducial)}\label{tab:GQuEST}
    \begin{tabular}{lcc}
      \toprule
  PARAMETER & SYMBOL & VALUE \\
    \midrule \midrule
    Laser wavelength in vacuum & $\lambda_{\text{laser}}$ & $1550 \text{ nm}$\\

    Interferometer arm length & $L_{\text{arm}}$ & $7 \text{ m}$\\

    Interferometer power recycling cavity length & $L_{\text{PRC}}$ & $10 \text{ m}$\\

    Interferometer power recycling cavity finesse & $\mathcal{F}_{\text{PRC}}$ & $3000$\\

    Input mode cleaner cavity length & $L_{\text{IMC}}$ & $4.5 \text{ m}$\\

    Input mode cleaner finesse & $\mathcal{F}_{\text{IMC}}$ & $3300$\\
        
    Input laser white phase noise PSD  & $S_{\mathrm{LPN},\mathrm{in}}$ &$(10^{-7}~\mathrm{rad}/\sqrt{\mathrm{Hz}})^2$ \\

    IFO DC offset & $\Lambda_{\text{DC}}$ & 20~ppm \\
    IFO Contrast Defect & $\Lambda_{\text{CD}}$ & 40~ppm\\
    Temperature & $T$ & 294~K \\
    Interferometer arm length difference & $L_{\text{Schnupp}}$ & $<0.01$~m\\
    Configuration & - & Traveling Wave \\
    Interferometer vacuum system pressure & $\rho_{\text{vacuum}}$ & 10 $\mu$Pa\\
       \midrule
    End mirror $1/e^2~(2\sigma)$ intensity beam radius & $w$ & 3~mm\\
    End mirror radius & $a$ & 12.7~mm\\
    End mirror thickness & $h$ & 2~mm\\
    End mirror substrate material & c-Si & Crystalline Silicon\\
    End mirror coating material & - & $\mathrm{Ta}_2\mathrm{O}_5-\mathrm{SiO}_2$\\
    End mirror coating thickness & $h_\coating$ & 10~$\mathrm{\mu m}$ \\
    \midrule
    Beamsplitter $1/e^2~(2\sigma)$ intensity beam radius & $w$ & 3~mm\\
    Beamsplitter radius & $a_{\mathrm{BS}}$ & 19~mm\\
    Beamsplitter thickness & $h$ & 2~mm\\
    Beamsplitter substrate material & c-Si & Crystalline Silicon\\
    \midrule
    Power recycling mirror 1 $1/e^2~(2\sigma)$ intensity beam radius & $w_{\text{PRM}}$ & 0.25~mm\\
    Power recycling mirror 2 $1/e^2~(2\sigma)$ intensity beam radius & $w_{\text{PR2}}$ & 2~mm\\
    Power recycling mirror 3 $1/e^2~(2\sigma)$ intensity beam radius & $w_{\text{PR3}}$ & 3.2~mm\\
    Power recycling optics radius & $a_{\mathrm{PRM}}$ & 12.7~mm\\
    Power recycling optics thickness & $h_{\text{PRM}}$ & 6~mm\\
    Power recycling optics substrate material & FS & Fused Silica\\
    Power recycling optics coating material & - & $\mathrm{Ta}_2\mathrm{O}_5-\mathrm{SiO}_2$\\
    Power recycling mirror 1 coating thickness & $h_\coating$ & 5~$\mathrm{\mu m}$ \\
    Power recycling mirrors 2 and 3 coating thickness & $h_\coating$ & 10~$\mathrm{\mu m}$ \\
    \bottomrule
    \end{tabular}
\end{table*}

\begin{table*}[ht]
  \caption{Material Parameters at room temperature}\label{tab:Materials}
    \begin{tabular}{lcc}
      \toprule
  PARAMETER & SYMBOL & VALUE \\
    \midrule \midrule
    FS Young's modulus & $E$ & 72~GPa~\cite{FejerPRD04ThermoelasticDissipation} \\
    FS Poisson ratio & $\nu$ & 0.17~\cite{FejerPRD04ThermoelasticDissipation}  \\
    FS Density & $\rho$ & 2200~$\mathrm{kg}\,\mathrm{m}^{-3}$~\cite{FejerPRD04ThermoelasticDissipation}\\
    FS Loss angle in substrate & $\phi_\substrate$ & $10^{-6}$~\cite{PennPLA06FrequencySurface}\\
    FS Loss angle in coating & $\phi$ & $5\cdot10^{-5}$~\cite{PrincipePRD15MaterialLoss}\\
    \ifdcc \else FS Index of refraction at $\lambda_{\text{laser}} = 1064$ nm & $n$ & 1.466~\cite{ChabbraPRD25MeasurementThermooptic} \\ \fi
    FS Index of refraction at $\lambda_{\text{laser}} = 1550$ nm & $n$ & 1.45~\cite{GranataCQG20AmorphousOptical}  \\
    FS Elastorefractive (photoelastic) coefficient, $\parallel$ & $\beta_{\mathrm{PE},\parallel}$ &  -0.42 (derived from~\cite{BiegelsenJAP76OpticalFrequency})\\
    FS Elastorefractive (photoelastic) coefficient, $\perp$ & $\beta_{\mathrm{PE},\perp}$ &  -0.30 (derived from~\cite{BiegelsenJAP76OpticalFrequency})\\
    FS Thermal conductivity & $\kappa$ & 1.38~W$\,\mathrm{m}^{-1}\,\mathrm{K}^{-1}$~\cite{FejerPRD04ThermoelasticDissipation} \\
    FS Volumetric heat capacity & $C_V$ & $1.64 \cdot 10^{6}$~J$\,\mathrm{m}^{-3}\,\mathrm{K}^{-1}$~\cite{FejerPRD04ThermoelasticDissipation} \\
    FS Coefficient of thermal expansion & $\alpha$ & $0.51 \cdot10^{-6}$~$\mathrm{K}^{-1}$~\cite{FejerPRD04ThermoelasticDissipation} \\
    FS Thermorefractive coefficient $\partial n/\partial T$ & $\beta$ & $8 \cdot 10^{-6}$~$\mathrm{K}^{-1}$~\cite{EvansPRD08ThermoopticNoise} \\
    FS Effective coefficient of thermal expansion at the coating interface (min) & $\bar{\alpha}_{\substrate, \text{min}}$ & $ 0.7\cdot10^{-6}~\mathrm{K}^{-1}$ \\
    
    \midrule
    c-Si Young's modulus ((100) crystal orientation) & $E_\substrate$ & 130~GPa~\cite{HopcroftJMS10WhatYoungs} \\
    c-Si Poisson ratio ((100) crystal orientation) & $\nu_\substrate$ & 0.28~\cite{HopcroftJMS10WhatYoungs}  \\
    c-Si Density & $\rho_\substrate$ & 2329~$\mathrm{kg}\,\mathrm{m}^{-3}$~\cite{HeninsJRNBSSA64PrecisionDensity}\\
    c-Si Loss angle & $\phi_\substrate$ & $10^{-6}$~\cite{RodriguezSR19DirectDetectiona}\\
    c-Si Index of refraction at $\lambda_{\text{laser}} = 1550$ nm & $n_\substrate$ & $3.48$~\cite{FreyOTA06TemperaturedependentRefractive} \\
    c-Si Elastorefractive (photoelastic) coefficient, $\parallel$ & $\beta_{\mathrm{PE},\parallel}$ &  -0.36 (derived from~\cite{BiegelsenPRL74PhotoelasticTensor})\\
    c-Si Elastorefractive (photoelastic) coefficient, $\perp$ & $\beta_{\mathrm{PE},\perp}$ &  +0.81 (derived from~\cite{BiegelsenPRL74PhotoelasticTensor})\\
    c-Si Thermal conductivity & $\kappa_\substrate$ & 156~W$\,\mathrm{m}^{-1}\,\mathrm{K}^{-1}$~\cite{GlassbrennerPR64ThermalConductivity} \\
    c-Si Volumetric heat capacity & $C_\substrate$ & $1.65 \cdot 10^{6}$~J$\,\mathrm{m}^{-3}\,\mathrm{K}^{-1}$~\cite{ShanksPR63ThermalConductivity} \\
    c-Si Coefficient of thermal expansion & $\alpha_\substrate$ & $2.5 \cdot10^{-6}$~$\mathrm{K}^{-1}$~\cite{OkadaJAP84PreciseDetermination} \\
    c-Si Thermorefractive coefficient $\partial n/\partial T$ at $\lambda_{\text{laser}} = 1550$ nm & $\beta_\substrate$ & $1.8 \cdot 10^{-4}$~$\mathrm{K}^{-1}$~\cite{FreyOTA06TemperaturedependentRefractive, KommaAPL12ThermoopticCoefficient} \\
    c-Si Effective coefficient of thermal expansion at the coating interface (min) & $\bar{\alpha}_{\substrate, \text{min}}$ & $ 4.4\cdot10^{-6}~\mathrm{K}^{-1}$ \\
    c-Si Diffusion coefficient & $D_{\text{cc}}$ & $4 \cdot 10^{-3}$~$\text{m}^2\, \text{s}^{-1}$~\cite{BrunsPRD20ThermalCharge} \\
    c-Si Mean carrier density & $N_0$ & $<10^{18}~\mathrm{m}^{-3}$~\cite{SiegelPRD23RevisitingThermal}  \\
    
    \midrule

    $\mathrm{Ta}_2\mathrm{O}_5$ Young's modulus & $E$ & 120~GPa~\cite{GranataCQG20AmorphousOptical} \\
    $\mathrm{Ta}_2\mathrm{O}_5$ Poisson ratio & $\nu$ & 0.28~\cite{GranataCQG20AmorphousOptical}  \\
    $\mathrm{Ta}_2\mathrm{O}_5$ Density & $\rho$ & 7300~$\mathrm{kg}\,\mathrm{m}^{-3}$~\cite{GranataCQG20AmorphousOptical}\\
    $\mathrm{Ta}_2\mathrm{O}_5$ Loss angle at 10~MHz & $\phi$ & $1\cdot 10^{-3}$~\cite{GrasPRD18DirectMeasurement}\\
    \ifdcc \else $\mathrm{Ta}_2\mathrm{O}_5$ Index of refraction at $\lambda_{\text{laser}} = 1064$ nm & $n$ & $2.076$~\cite{ChabbraPRD25MeasurementThermooptic} \\ \fi
    $\mathrm{Ta}_2\mathrm{O}_5$ Index of refraction at $\lambda_{\text{laser}} = 1550$ nm & $n$ & 2.03~\cite{GranataCQG20AmorphousOptical} \\
    $\mathrm{Ta}_2\mathrm{O}_5$ Thermal conductivity & $\kappa$ & 0.3~W$\,\mathrm{m}^{-1}\,\mathrm{K}^{-1}$~\cite{GrilliSPTOCS96ThermalConductivity}\footnote{There is considerable disagreement in the literature about this value for a thin coating; see~\cite{WelschIJoT99AbsoluteMeasurement} for example. Please note that the appropriate average coating thermal conductivity is a reciprocal sum. We choose this specific value since an ion-beam-sputtered (IBS) coating was measured. In the same measurement, the thermal conductivity of FS was measured to also be 0.3~W$\,\mathrm{m}^{-1}\,\mathrm{K}^{-1}$ (lower than the nominal substrate value), so we choose the smaller of the reported values for the thermal conductivity of $\mathrm{Ta}_2\mathrm{O}_5$. 33 W$\,\mathrm{m}^{-1}\,\mathrm{K}^{-1}$ is the thermal conductivity of bulk sapphire and, despite its use in previous papers, this value should not be used for the thermal conductivity of $\mathrm{Ta}_2\mathrm{O}_5$.} \\
    $\mathrm{Ta}_2\mathrm{O}_5$ Volumetric heat capacity & $C_V$ & $2.2 \cdot 10^{6}$~J$\,\mathrm{m}^{-3}\,\mathrm{K}^{-1}$~(derived from \cite{FejerPRD04ThermoelasticDissipation}) \\
    $\mathrm{Ta}_2\mathrm{O}_5$ Coefficient of thermal expansion & $\alpha$ & $3.6 \cdot10^{-6}$~$\mathrm{K}^{-1}$~\cite{FejerPRD04ThermoelasticDissipation} \\
    $\mathrm{Ta}_2\mathrm{O}_5$ Thermorefractive coefficient $\partial n/\partial T$ & $\beta$ & $14 \cdot 10^{-6}$~$\mathrm{K}^{-1}$~\cite{EvansPRD08ThermoopticNoise} \\
    
    \midrule \midrule
    Coating effective P-wave modulus & $M_\coating$ & 97~GPa \\
    Coating Poisson ratio (volume average) & $\nu_\coating$ & 0.22 \\
    Coating density (volume average) & $\rho_\coating$ & 4300~$\mathrm{kg}\,\mathrm{m}^{-3}$\\
    Coating effective loss angle at 10~MHz & $\phi_\coating$ &  $3\cdot 10^{-4}$ \\
    Coating average index of refraction at $\lambda_{\text{laser}} = 1064$ nm & $n_{\text{avg}}$ & $1.72$ \\
    Coating average index of refraction at $\lambda_{\text{laser}} = 1550$ nm & $n_{\text{avg}}$ & 1.69 \\
    Coating thermal conductivity & $\kappa_\coating$ &  0.6~W$\,\mathrm{m}^{-1}\,\mathrm{K}^{-1}$\\
    Coating volumetric heat capacity (average) & $C_\coating$ & $1.9\cdot10^6$~J$\,\mathrm{m}^{-3}\,\mathrm{K}^{-1}$ \\
    Coating effective coefficient of thermal expansion (min) & $\bar{\alpha}_{\coating, \text{min}}$ & $ 3.1\cdot10^{-6}~\mathrm{K}^{-1}$ \\
    Coating effective thermorefractive coefficient & $\bar{\beta}_\coating$ & $ 8\cdot10^{-6}$~$\mathrm{K}^{-1}$ \\
    \bottomrule
    \end{tabular}
\end{table*}

Density and volumetric heat capacity are simple volume averages. Thermal conductivity and P-wave modulus are a harmonic mean volume average. The Poisson ratio is a simple volume average because it is similar in $\mathrm{Ta}_2\mathrm{O}_5$ and $\mathrm{SiO}_2$. The effective loss angle is defined in \cref{sec:CMN}. The definition of $\bar{\beta}_\coating$ is found in~\cite{EvansPRD08ThermoopticNoise}. The effective coefficient of thermal expansion at the valley minima is given in \crefCTOAvg. Unless otherwise stated, the bulk and substrate parameters for a single material are assumed to be the same.
\bibliography{HFCNIMICitation}

\end{document}